\documentclass[article,aps,prb,twocolumn,superscriptaddress,english,longbibliography]{revtex4-2}
\usepackage{graphicx}
\usepackage{amssymb}
\usepackage{bm}
\usepackage{natbib}
\usepackage{amsmath}
\usepackage{mathrsfs} 
\usepackage{amstext, upgreek}
\usepackage[english]{babel}
\usepackage[table, svgnames, dvipsnames]{xcolor}
\usepackage[colorlinks=true,citecolor=Cerulean,linkcolor=RubineRed,urlcolor=Cerulean]{hyperref}
\usepackage{orcidlink}
\newcommand{\be}{\begin{eqnarray}}
\newcommand{\ee}{\end{eqnarray}}
\newcommand{\no}{\nonumber}

\begin{document}

\title{Krylov Complexity Under Hamiltonian Deformations and Toda Flows
}

\author{Kazutaka Takahashi\orcidlink{0000-0001-7321-2571}}
\affiliation{Department of Physics and Materials Science, University of Luxembourg, 
L-1511 Luxembourg, Luxembourg}
\affiliation{Department of Physics Engineering, Faculty of Engineering, Mie University, 
Mie 514–8507, Japan}
\affiliation{Institute of Science and Engineering, Kanazawa University, Kanazawa 920-1192, Japan}
\author{Pratik Nandy\orcidlink{0000-0001-5383-2458}}
\affiliation{Theoretische Natuurkunde, Vrije Universiteit Brussel (VUB) and \\The International Solvay Institutes, Pleinlaan 2, B-1050 Brussels, Belgium}
\affiliation{Center for Gravitational Physics and Quantum Information, 
Yukawa Institute for Theoretical Physics, Kyoto University, Kyoto 606-8502, Japan}
\affiliation{RIKEN Center for Interdisciplinary Theoretical and Mathematical Sciences (iTHEMS), Wako, Saitama 351-0198, Japan}
\author{Adolfo del Campo\orcidlink{0000-0003-2219-2851}}
\affiliation{Department of Physics and Materials Science, University of Luxembourg, 
L-1511 Luxembourg, Luxembourg}
\affiliation{Donostia International Physics Center,  E-20018 San Sebasti\'an, Spain}

\begin{abstract}
The quantum dynamics of a complex system can be efficiently described in Krylov space, the minimal subspace in which the dynamics unfolds. We apply the Krylov subspace method to Hamiltonian deformations, which provides a systematic way to construct solvable models from known instances. In doing so, we relate the evolution of deformed and undeformed theories and investigate their complexity.
For a certain class of deformations, the resulting Krylov subspace remains unchanged, and time evolution takes place in a reorganized basis.
The tridiagonal form of the generator in the Krylov space is maintained, 
and we obtain generalized Toda equations as a function of the deformation parameters.
The imaginary-time-like evolutions can be described by real-time unitary ones.
As possible applications, we discuss coherent Gibbs states for thermodynamic systems, for which we analyze the survival probability, spread complexity, Krylov entropy, and associated time-averaged quantities. We further discuss the 
statistical properties of random matrices and supersymmetric systems for quadratic deformations.
\end{abstract}

\maketitle

%%%%%%%%%%%%%%%%%%%%%%%%%%%%%%%%%%%%%%%%%%%%%%%%%%%%%%%%%%%%%%%%%%%%%%%%%%%%%%%%%%%
\section{Introduction}

Nonperturbative methods are crucial for understanding physical phenomena that fall beyond the scope of perturbative expansions, which typically rely on small coupling constants or weak-field limits. Such methods become crucial in regimes where perturbation theory breaks down, e.g.,  due to divergences or strong coupling \cite{Marinyo15}. In two dimensions, exactly solvable deformations of integrable field theories have shed light on quantum field theory and holography \cite{Zamolodchikov04,cavaglia2016,SmirnovZ17,jiang2021}. In this context, Hamiltonian deformations, which map the Hamiltonian to a function of the Hamiltonian, have also been discussed \cite{Gross20-1,Gross20-2,kruthoff2020,rosso21,jiang2022,ebert2022}.
Such deformations can be applied to quantum mechanics both in a Hermitian \cite{Gross20-1,Gross20-2} and non-Hermitian Hamiltonian setting ~\cite{Matsoukas23}.

One may wonder whether theories related by such deformations exhibit different complexity and generate fundamentally different kinds of nonequilibrium dynamics. To address this question, one can use quantum complexity measures that quantify the growth of an operator or quantum state under unitary evolution. Among them, Krylov complexity provides an insightful characterization of nonequilibrium dynamics by tracking the spread of an initial operator or quantum state using an orthonormal basis constructed via the Lanczos algorithm \cite{Viswanath94,Parker19}. This basis, known as the Krylov basis, spans the minimal subspace in which the dynamics unfolds. The use of Krylov subspace methods for quantum dynamics is thus of great interest both to the foundations of physics and applications in quantum science and technology \cite{Nandy25,Baiguera26,Rabinovici25}, ranging from quantum control to quantum optimization algorithms \cite{Takahashi24,Takahashi25,Bhattacharjee23arxiv,Grabarits26,Grabarits2025cd,grabarits2025KQPT}. 

In this study, we investigate how the Krylov complexity evolves under Hamiltonian deformations. For a given Hamiltonian and initial state, the associated Krylov subspace is uniquely defined, and the dynamics within this subspace reduces to a one-dimensional nearest-neighbor hopping model. Since the deformation preserves the Krylov subspace, the resulting effective tridiagonal Hamiltonian depends parametrically on the deformation. We thus observe a continuous flow of the tridiagonal Hamiltonian as the deformation is implemented. This framework is reminiscent of the method of flow equations, where a Hamiltonian evolves, typically in a double-bracket form, toward a diagonal form while preserving its spectrum \cite{Brockett91,GlazekWilson93,GlazekWilson94,Wegner94,Kehrein06}.

Among the possible choices of flow generator, the Toda chain flow is particularly natural in our setting: it preserves the tridiagonal structure of the Hamiltonian throughout the evolution and has been shown to saturate the operator-growth quantum speed limit~\cite{Hornedal23}. The Toda chain is one of the earliest and most celebrated examples of a nonlinear classical integrable
system~\cite{Toda67-1,Toda67-2}, and Toda flows also arise naturally in the description of quantum dynamics~\cite{Okuyama16}. The relevance of the Toda lattice to the Krylov method, and equivalently to the theory of orthogonal polynomials, has been discussed in Ref.~\cite{Ismail05}.

A related connection between Krylov dynamics and the Toda lattice has been analyzed in 
previous works~\cite{Dymarsky20,Kundu23,angelinos25}, 
where the imaginary-time correlation function of an operator generates a Toda flow 
for the associated Lanczos coefficients. 
In those approaches, the flow parameter is identified with the Euclidean time 
appearing in the correlation function, and the Krylov method was applied to operators.
In contrast, in the present work, we apply the Krylov method to state vectors.
The physical time ($t$) evolution of the state vector 
is always governed by the same Hamiltonian $H$, 
and the Toda flow arises with respect to the deformation parameter $\tau$ for the initial state.
The deformation reorganizes the Krylov representation while preserving the underlying Krylov subspace.

The paper is organized as follows. Section~\ref{sec:hdeform} defines the Hamiltonian
deformations considered in this work. The Krylov subspace method is introduced in
Sec.~\ref{sec:krylov}, and Sec.~\ref{sec:todaflows} establishes that the generator
in Krylov space generically satisfies the generalized Toda equations. Properties of
these equations are analyzed in Sec.~\ref{sec:todaeqs}. Sections~\ref{sec:cgs}
and~\ref{sec:rmt} discuss applications: the coherent Gibbs state, used to probe
thermodynamic properties of many-body systems, and random matrix Hamiltonians, used
to study statistical properties of spreading in Krylov space. Possible applications
to supersymmetric systems are discussed in Sec.~\ref{sec:susy}, and
Sec.~\ref{sec:summary} provides a summary and outlook.

%%%%%%%%%%%%%%%%%%%%%%%%%%%%%%%%%%%%%%%%%%%%%%%%%%%%%%%%%%%%%%%%%%%%%%%%%%%%%%%
\section{Hamiltonian deformations}
\label{sec:hdeform}

For a given Hamiltonian $H=\sum_{\mu=1}^DE_\mu|E_\mu\rangle\langle E_\mu|$ 
and an initial state $|\psi_0\rangle$ defined in 
a $D$-dimensional Hilbert space, we consider the unitary time evolution
$|\psi(t)\rangle = e^{-iHt}|\psi_0\rangle$.
A quantum state of broad interest in quantum chaos, black hole physics 
and the study of thermal averages is 
the coherent Gibbs state~\cite{delcampo2018sff,Xu21,Balasubramanian22}
\be
 |\psi_0(\beta)\rangle=\frac{1}{\sqrt{Z(\beta)}}
 \sum_{\mu=1}^De^{-\beta E_\mu/2}|E_\mu\rangle, \label{cgs}
\ee
where $Z(\beta)$ represents the partition function, the ``one-copy'' version 
of the thermofield double state~\cite{Maldacena03,Papadodimas15,Chapman19}. 
This state can be associated with the choice of an initial quantum state, the coherent superposition of all energy eigenstates with equal weight
$|\psi_0\rangle=\sum_\mu|E_\mu\rangle/\sqrt{D}$, subject to the transformation 
\be 
 |\psi_0\rangle\to \frac{e^{-\beta H/2}|\psi_0\rangle}
 {\sqrt{\langle\psi_0|e^{-\beta H}|\psi_0\rangle}} = |\psi_0(\beta)\rangle. 
\ee
In the study of Hamiltonian deformations, we consider generalizing
this transformation as  
\be
 |\psi_0\rangle\to|\psi_0(\tau)\rangle = \frac{e^{-f(H,\tau)/2}|\psi_0\rangle}
 {\sqrt{\langle\psi_0|e^{-f(H,\tau)}|\psi_0\rangle}}. \label{psitau}
\ee
Here, $f(\cdot,\tau)$ is a real function that involves a set of real parameters
$\tau=(\tau_1,\tau_2,\dots)$.
Although the following discussions can essentially be applied without specifying
the explicit form of $f(H,\tau)$, we are interested in a particular choice  
\be
 f(H,\tau)=\tau_1H+\tau_2H^2, \label{f}
\ee
which includes, e.g., the $T\overline{T}$-deformation~\cite{Gross20-2} 
at first order, as well as its generalizations to the non-Hermitian case, 
which arise naturally in the context of energy dephasing 
in the absence of quantum jumps \cite{Matsoukas23,Matsoukas23b}.

The generator of the real-time evolution is kept unchanged and 
the time-evolved state is given by 
\be
 |\psi(t,\tau)\rangle = e^{-iHt}|\psi_0(\tau)\rangle.
\ee
A natural measure to compare the time evolution between the deformed and 
undeformed theories is the survival (or Loschmidt echo) amplitude 
\be
 \langle \psi_0(\tau)|\psi(t,\tau)\rangle
 =\frac{\langle\psi_0|e^{-iHt}e^{-f(H,\tau)}|\psi_0\rangle}
 {\langle\psi_0|e^{-f(H,\tau)}|\psi_0\rangle}. \label{sp}
\ee
The coherent Gibbs state is obtained when $(\tau_1,\tau_2)=(\beta,0)$, 
for which the survival amplitude reduces to the complex partition function, 
$Z(\beta+it)/Z(\beta)$. 
Note that its absolute square value is then the spectral form factor, used in the characterization of quantum chaos and the spectral characterization of complex many-body 
systems~\cite{Haake01,Cotler2017,delcampo2018sff,cornelius2022,Matsoukas24,MartinezAzcona25}.

The survival amplitude can be written as the Fourier transform of 
the local density of states,
\be
 \langle \psi_0(\tau)|\psi(t,\tau)\rangle =\int dE\,e^{-iEt}\rho(E,\tau),
\ee
defined as
\be
 \rho(E,\tau) &=& \langle\psi_0(\tau)|\delta(E-H)|\psi_0(\tau)\rangle \no\\
 &=& \frac{e^{-f(E,\tau)}\rho(E,0)}{\int dE\,e^{-f(E,\tau)}\rho(E,0)}. \label{dos}
\ee
Thus, in this setting, all other quantities are fully determined by 
the local density of states.
Furthermore, the structure of the local density of states implies that only the case of a real-valued function $f$ is physically relevant, 
since any imaginary contribution vanishes when evaluating the overlap 
in Eq.~(\ref{sp}).

The aim of the present work is to understand the relation between
the original time-evolved state $|\psi(t)\rangle$ and
the deformed state $|\psi(t,\tau)\rangle$.
Since both states belong to the space spanned by
${|\psi_0\rangle, H|\psi_0\rangle, H^2|\psi_0\rangle, \dots}$,
the Krylov subspace method provides a natural and effective strategy.

%%%%%%%%%%%%%%%%%%%%%%%%%%%%%%%%%%%%%%%%%%%%%%%%%%%%%%%%%%%%%%%%%%%%%%%%%%%%%%%
\section{Krylov subspace method}
\label{sec:krylov}

To describe the unitary time evolution in Hilbert space,
it is convenient to identify the minimal subspace in which the state evolves.
In the Krylov subspace method, the state space is spanned by the Krylov basis~\cite{Viswanath94}.
Starting with the initial basis $|K_0(\tau)\rangle = |\psi_0(\tau)\rangle$, one constructs an orthonormal basis set through the recurrence relation:
\be
 |K_{n+1}(\tau)\rangle b_{n+1}(\tau)
 &=& H |K_n(\tau)\rangle-|K_n(\tau)\rangle a_n(\tau)
 \no\\
 && -|K_{n-1}(\tau)\rangle b_n(\tau). \label{rec}
\ee
The so-called Lanczos coefficients are determined by imposing the orthonormality  relation
\be
 \langle K_m(\tau)|K_n(\tau)\rangle=\delta_{m,n}, \label{orthonormal}
\ee
and can be identified as
\be
 && a_n(\tau) =\langle K_n(\tau)|H|K_n(\tau)\rangle, \\
 && b_n(\tau) =\langle K_{n-1}(\tau)|H|K_n(\tau)\rangle,
\ee
where $a_n(\tau)$ and $b_n(\tau)$ are real and, in addition, $b_n(\tau)$ is nonnegative.
The recurrence relation is applied until the condition
$H|K_{d-1}(\tau)\rangle - |K_{d-1}(\tau)\rangle a_{d-1}(\tau) - |K_{d-2}(\tau)\rangle b_{d-1}(\tau) = 0$
is satisfied.
The Krylov dimension $d$, representing the total number of basis elements,
is less than or equal to the dimension of the Hilbert space $D$.
In our setting, the initial state at $\tau=0$ is written as 
\be
 |\psi_0\rangle 
 = \sum_{\mu=1}^D \frac{1}{\sqrt{D}}|E_\mu\rangle 
 = \sum_{n=0}^{d-1}\sqrt{\frac{d_n}{D}}|\epsilon_n\rangle, \label{psi0}
\ee
where 
\be
 |\epsilon_n\rangle = \frac{1}{\sqrt{d_n}}\sum_{\mu=1}^D
 |E_\mu\rangle\delta_{E_\mu,\epsilon_n}, \label{stateepsn}
\ee
and the degeneracy $d_n$ represents the number of the eigenstates 
with $E_\mu=\epsilon_n$.
Since the state space of the equienergy $E_\mu=\epsilon_n$ is 
spanned by the single basis $|\epsilon_n\rangle$, 
the Krylov dimension is given by the number of distinct eigenvalues
of the original Hamiltonian.

In the present formulation, since the initial state depends on 
the parameter $\tau$, all other quantities naturally inherit 
a dependence on $\tau$.
The main objective of this work is to characterize the $\tau$-evolution.
As mentioned at the end of the previous section,
the Krylov space spanned by $\{|K_n(\tau)\rangle\}_{n=0}^{d-1}$
must remain independent of $\tau$.
Consequently, the basis vectors $|K_n(\tau)\rangle$ vary continuously with $\tau$,
and the Krylov dimension $d$ remains constant, independent of $\tau$.

An essential property of the Krylov subspace method is that the time evolution can be effectively described as a one-dimensional motion within the Krylov space.
Expanding the state as
\be
|\psi(t,\tau)\rangle = \sum_{n=0}^{d-1} \varphi_n(t,\tau) |K_n(\tau)\rangle, \label{psittau}
\ee
the wave function components $\varphi_n(t,\tau)$ are given by
\be
\varphi_n(t,\tau) = \langle n| e^{-i L(\tau) t} |0\rangle. \label{varphittau}
\ee
Here, we define the fixed orthonormal basis 
$\{|n\rangle\}_{n=0}^{d-1}$ to represent the Krylov subspace.
The time-evolution generator $L(\tau)$, expressed in this basis, takes a tridiagonal form:
\be
 L(\tau) &=& \sum_{n=0}^{d-1} a_n(\tau) |n\rangle\langle n| \no\\
 && +\sum_{n=1}^{d-1} b_n(\tau) (|n-1\rangle\langle n|+|n\rangle\langle n-1|). 
\ee
In the Krylov space representation, the state is initially set to $|0\rangle$
and propagates along the finite or semi-infinite chain by the generator $L(\tau)$.

Generally, the Krylov basis $|K_n\rangle$ and the Lanczos coefficients $a_n$ and $b_n$  can be identified by specifying the Hamiltonian and the initial state.
In other words, the local density of states $\rho(E,\tau)$ defined 
in Eq.~(\ref{dos}) uniquely determines the Krylov expansion.
To see this, we represent the Krylov basis as 
\be
 |K_n(\tau)\rangle = P_n(H,\tau)|K_0(\tau)\rangle.
\ee
$P_n(\cdot,\tau)$ is a real polynomial of the $n$th degree.
The orthonormality of the Krylov basis in Eq.~(\ref{orthonormal}) is written as 
\be
 \int dx\,\rho(x,\tau)P_m(x,\tau)P_n(x,\tau)=\delta_{m,n}. \label{orthonormalp}
\ee
This relation shows that the Krylov basis is closely related to orthogonal polynomials~\cite{Viswanath94,Parker19,Muck22}.
The recurrence relation is written as 
\be
 b_{n+1}(\tau)P_n(x,\tau) &=& xP_n(x,\tau)-a_n(\tau)P_n(x,\tau) \no\\
 && -b_{n}(\tau)P_{n-1}(x,\tau).
\ee
We can also represent the local density of states as  
\be
 \rho(x,\tau)=\langle 0|\delta(x-L(\tau))|0\rangle. \label{dosl}
\ee
Using the transformation operator 
\be
 F(\tau)=\sum_{n=0}^{d-1}|K_n(\tau)\rangle\langle n|, \label{Ftau}
\ee
connecting vectors in the original Hilbert space to those in the Krylov space, 
we can write $|K_0(\tau)\rangle=F(\tau)|0\rangle$ and $F^\dag(\tau)HF(\tau)=L(\tau)$,
which give Eq.~(\ref{dosl})

As mentioned above, since the Krylov expansion depends not only on the Hamiltonian but also
on the choice of the initial state, $L(\tau)$ depends on $\tau$.
The comparison between the original definition of 
the local density of states in Eq.~(\ref{dos})
and the Krylov space representation in Eq.~(\ref{dosl}) implies that 
any eigenvalue of $L(\tau)$ is equal to one of the eigenvalues of $H$.
Equation (\ref{psi0}) shows that the eigenvalues of $L(\tau)$
are given by $\{\epsilon_n\}_{n=0}^{d-1}$.
We note $\epsilon_m\ne\epsilon_n$ for $m\ne n$ by construction.
Using a unitary matrix $V(\tau)$, we can write 
$L(\tau)=V(\tau)L(0)V^\dag(\tau)$ and the equation of motion
\be
 \frac{\partial L(\tau)}{\partial \tau_\mu}=[M_\mu(\tau),L(\tau)], \label{lax}
\ee
where $M_\mu(\tau)$ is a real antisymmetric matrix.
Since $iM_\mu(\tau)$ is Hermitian, this describes a unitary time evolution.
We show this property in the next section.
It is interesting to note that we obtain a unitary time evolution starting 
from the imaginary-time-like evolution in Eq.~(\ref{psitau}).

When we apply the equivalence of Eqs.~(\ref{dos}) and (\ref{dosl}) 
to the coherent Gibbs state $|\psi_0(\tau)\rangle$ with $\tau=(\tau_1,\tau_2)=(\beta,0)$, 
the thermodynamic properties are reflected in the Lanczos coefficients.
Especially, $a_0(\tau)$ represents the thermodynamic energy and 
$b_1(\tau)$ represents the energy variance.
When the system has a thermal phase transition, the energy variance is divergent
and $b_1(\tau)$ displays a singular behavior.
Since the effective Hamiltonian $L(\tau)$ contains information about the state, in addition to that of the Hamiltonian, this behavior is therefore natural.

%%%%%%%%%%%%%%%%%%%%%%%%%%%%%%%%%%%%%%%%%%%%%%%%%%%%%%%%%%%%%%%%%%%%%%%%%%%%%%%
\section{Toda chain flows}
\label{sec:todaflows}

In the following, we consider Eq.~(\ref{f}) for the deformed function $f(x,\tau)$.
To identify how the Krylov basis and the Lanczos coefficients
are dependent on $\tau$, we consider the derivative of the orthonormal relation in 
Eq.~(\ref{orthonormalp})~\cite{Ismail05}.
After some calculations described in Appendix~\ref{sec:todacalc}, we finally obtain 
\be
 && \partial_{\tau_1}a_n = -(b_{n+1}^2-b_n^2), \label{toda1-1}\\
 && \partial_{\tau_1}b_n = -\frac{1}{2}b_n(a_n-a_{n-1}), \label{toda1-2}
\ee
and 
\be
 && \partial_{\tau_2}a_n = 
 -\left[b_{n+1}^2(a_{n+1}+a_n)-b_n^2(a_n+a_{n-1})\right], \label{toda2-1}\\
 && \partial_{\tau_2}b_n = -\frac{1}{2}b_n\left(
 b_{n+1}^2-b_{n-1}^2+a_n^2-a_{n-1}^2\right). \label{toda2-2}
\ee
The first set of differential equations represents the Toda equations~\cite{Toda67-1,Toda67-2}
and the second is a generalization~\cite{Moser75book}.
Correspondingly, Eq.~(\ref{lax}) is known as the Lax form~\cite{Lax68}, which represents a key signature of integrability \cite{Faddeev2007}.
The explicit forms of $M_1(\tau)$ and $M_2(\tau)$ in the present case are given by  
\be
 M_1(\tau) &=& \frac{1}{2}\sum_{n=1}^{d-1}
 b_n(\tau)(-|n-1\rangle\langle n|+|n\rangle\langle n-1|), \\
 M_2(\tau) &=& \frac{1}{2}\sum_{n=1}^{d-1}
 (a_n(\tau)+a_{n-1}(\tau))b_n(\tau)
 \no\\ && \times
 (-|n-1\rangle\langle n|+|n\rangle\langle n-1|) \no\\
 && +\frac{1}{2}\sum_{n=1}^{d-2}b_n(\tau)b_{n+1}(\tau) 
 \no\\ && \times
 (-|n-1\rangle\langle n+1|+|n+1\rangle\langle n-1|).
\ee
$M_1(\tau)$ has a tridiagonal form while $M_2(\tau)$ has a pentadiagonal form.
Since the latter is obtained from the quadratic term of Eq.~(\ref{f}), it is not difficult to imagine generalizations to arbitrary-order polynomials.
There exist corresponding Toda equations~\cite{Moser75book, Kac75, Moser75} and the resulting generator $M_\mu$ takes a band matrix form.

The Hermitian matrices $iM_1(\tau)$ and $iM_2(\tau)$ represent
the generators for the unitary evolution $V(\tau)$
satisfying $\partial_{\tau_\mu} V(\tau)=M_\mu(\tau)V(\tau)$.
We show in Appendix~\ref{sec:todacalc} that
\be
 F(\tau)=F(0)V^\dag(\tau). \label{ftau}
\ee
Since two generators are described by the same unitary matrix $V(\tau)$, they satisfy the zero-curvature condition
\be
 \partial_{\tau_1}M_2(\tau)-\partial_{\tau_2}M_1(\tau)-[M_1(\tau),M_2(\tau)]=0.
\ee
This set of Hamiltonians is an example of
integrable time-dependent Hamiltonians~\cite{Sinitsyn18}.
Arbitrary evolutions represented by the path-ordering operator as 
\be
 |\phi(\tau^{f};\tau^{i})\rangle = \mathrm{P}_\tau \exp\left[
 \sum_\mu\int_{\tau^{i}}^{\tau^{f}} d\tau_\mu M_\mu(\tau)
 \right]|\phi\rangle,
\ee
are determined by specifying the two points and are independent of 
the choice of the path in the parameter space $(\tau_1,\tau_2)$,
provided $M_\mu(\tau)$ is a smooth function.

In this formulation, the $\tau$-evolution can be described
universally by the Toda equations.
System-specific properties such as the Hamiltonian and the initial state are 
incorporated into the initial condition for the $\tau$-evolution.
This universal dynamics comes from the property that
$L(\tau)$ keeps its tridiagonal form for any $\tau$.

Consider the time-evolved state $|\psi(t,\tau)\rangle$, 
given by Eq.~(\ref{psittau}) along with Eq.~(\ref{varphittau}).
Since $|\psi_0(\tau)\rangle=|K_0(\tau)\rangle$, 
the survival amplitude is simply written as 
\be
 \langle \psi_0(\tau)|\psi(t,\tau)\rangle 
 = \langle 0|e^{-iL(\tau)t}|0\rangle. \label{spl}
\ee
Equation (\ref{psittau}) is written in the basis set $\{|K_n(\tau)\rangle\}_{n=0}^{d-1}$.
As mentioned in the previous section, the subspace is independent of $\tau$
and it is possible to write the state with respect to
the basis before the deformation $\{|K_n\rangle=|K_n(0)\rangle\}_{n=0}^{d-1}$.
Using Eq.~(\ref{ftau}), we can write 
\be
 |\psi(t,\tau)\rangle 
 = \sum_{n=0}^{d-1}|K_n\rangle\langle n|e^{-iL(0)t}V^\dag(\tau)|0\rangle.
\ee
Of particular interest is to consider $t=0$, when
\be
 |\psi_0(\tau)\rangle
 = \sum_{n=0}^{d-1}|K_n\rangle\langle n|V^\dag(\tau)|0\rangle.
\ee
This state is originally prepared as an initial state.
Now, the imaginary-time-like evolution is described as a unitary time evolution
in Krylov space.
Note that the anti-time ordering is used in the evolution described by $V^\dag(\tau)$.

%%%%%%%%%%%%%%%%%%%%%%%%%%%%%%%%%%%%%%%%%%%%%%%%%%%%%%%%%%%%%%%%%%%%%%%%%%%%%%%
\section{Toda equations and spread complexity}
\label{sec:todaeqs}

%%%%%%%%%%%%%%%%%%%%%%%%%%%%%%%%%%%%%%%%%%%%%%%%%%%%%%%%%%%%%%%%%%%%%%%%%%%%%%%
\subsection{Exact solutions from the first Toda equations}

To understand the general structure of Hamiltonian deformations in Krylov space, 
it is instructive to analyze explicit solutions of the Toda equations.
In the context of the Krylov algorithm, analytical solutions 
are known in systems with dynamical symmetries~\cite{Caputa22,Hornedal22}.
Here, we show that the Hamiltonian deformations with $\tau=(\tau_1,0)$
preserve the symmetries, admitting analytical solutions.

We consider the first set of the Toda equations in Eqs.~(\ref{toda1-1}) and (\ref{toda1-2}).
As a finite-dimensional system, we can consider SU(2) symmetry.
We choose as an ansatz 
\be
 && a_n(\tau_1)=2\gamma(\tau_1)\left(n-\frac{d-1}{2}\right)+\delta, \\
 && b_n(\tau_1)=\alpha(\tau_1)\sqrt{n(d-n)},
\ee
where the Krylov dimension $d$ takes a finite integer value.
With this choice, each of $L$ and $M$ is represented by a linear combination of spin operators.
Their operators give closed algebra and the time evolution is confined in a two-dimensional space~\cite{Hornedal23}.

As the sum $\alpha^2(\tau_1)+\gamma^2(\tau_1)$ takes a constant value given the Toda equations,  we choose the parametrization
$\gamma(\tau_1)=\gamma_0\cos\theta(\tau_1)$ and $\alpha(\tau_1)=\gamma_0\sin\theta(\tau_1)$.
Then, we obtain 
\be
 \partial_{\tau_1}\theta(\tau_1)=-\gamma_0\sin\theta(\tau_1).
\ee
The general solution is given by 
\be
 \cos\theta(\tau_1)=\frac{\cos\theta(0)+\tanh\gamma_0\tau_1}{1+\cos\theta(0)\tanh\gamma_0\tau_1}.
\ee
This function increases monotonically
from $\cos\theta(0)$ to 1 as $\tau_1$ increases from 0 to $\infty$.
As a result, we find the Toda flow, 
a diagonalizing evolution of the $L$-matrix~\cite{Moser75book}.
Since the $\tau$-evolution keeps the structure of the Lanczos coefficients, 
we can find the dynamical properties at arbitrary $\tau_1$, once we know those at $\tau_1=0$.
For example, consider the spread complexity, which is generally defined as~\cite{Parker19,Balasubramanian22}
\be
 K(t,\tau)=\sum_{n=0}^{d-1}n|\varphi_n(t,\tau)|^2. \label{complexity}
\ee
For systems with SU(2) symmetry, 
by explicitly calculating 
$K(t,\tau)=\langle 0|e^{iL(\tau)t}{\cal K}e^{-iL(\tau)t}|0\rangle$ 
with ${\cal K}=\mathrm{diag}\,(0,1,\dots,d-1)$, we obtain
\be
 K(t,\tau_1)=(d-1)\left(\frac{\sin\theta(0)\sin\gamma_0t}
 {\cosh\gamma_0\tau_1+\cos\theta(0)\sinh\gamma_0\tau_1}\right)^2. \no\\
\ee

The second example is the system with Heisenberg-Weyl symmetry.
The Krylov dimension is infinite, and we put 
\be
 && a_n(\tau_1)=2\gamma_0 n+\delta(\tau_1), \label{HWa} \\
 && b_n(\tau_1)=\alpha(\tau_1)\sqrt{n}. \label{HWb}
\ee
The corresponding $L$-matrix represents the one-dimensional particle 
in a harmonic oscillator potential with translational motion, for example.
The Toda equations give the solution 
\be
 && \alpha(\tau_1)=\alpha(0)e^{-\gamma_0\tau_1}, \\
 && \delta(\tau_1)=\delta(0)-\frac{\alpha^2(0)}{2\gamma_0}(1-e^{-2\gamma_0\tau_1}),
\ee
which also gives a diagonalization flow.
The spread complexity is given by 
\be
 K(t,\tau_1)=\left(\frac{\alpha(0)}{\gamma_0}e^{-\gamma_0\tau_1}\sin\gamma_0t\right)^2.
\ee

The third example is the system with SL(2,R) symmetry given by 
\be
 && a_n(\tau_1)=2\gamma(\tau_1)(n+h)+\delta, \\
 && b_n(\tau_1)=\alpha(\tau_1)\sqrt{n(n+2h-1)}.
\ee
The Toda equations give $\alpha^2(\tau_1)-\gamma^2(\tau_1)=\mathrm{const.}$,
and we can formally study stable, unstable, and critical solutions.
The stable solution is obtained by setting 
$\gamma(\tau_1)=\gamma_0\cosh\theta(\tau_1)$ and
$\alpha(\tau_1)=\gamma_0\sinh\theta(\tau_1)$.
The differential equation for $\theta(\tau_1)$
\be
 \partial_{\tau_1}\theta(\tau_1)=-\gamma_0\sinh\theta(\tau_1),
\ee
is solved as 
\be
 \cosh\theta(\tau_1)=\frac{\cosh\theta(0)+\tanh\gamma_0\tau_1}
 {1+\cosh\theta(0)\tanh\gamma_0\tau_1}.
\ee
The spread complexity is given by
\be
 K(t,\tau_1)=2h\left(\frac{\sinh\theta(0)\sin\gamma_0t}
 {\cosh\gamma_0\tau_1+\cosh\theta(0)\sinh\gamma_0\tau_1}\right)^2. \label{kstable}
\ee
We observe again a diagonalizing flow of the $L$-matrix 
and an exponential decay of the spread complexity.

The unstable solution is obtained when $\gamma(\tau_1)<\alpha(\tau_1)$.
We put $\gamma(\tau_1)=\gamma_0\sinh\theta(\tau_1)$ and 
$\alpha(\tau_1)=\gamma_0\cosh\theta(\tau_1)$ to obtain
\be
 \partial_{\tau_1}\theta(\tau_1)=-\gamma_0\cosh\theta(\tau_1),
\ee
and
\be
 \sinh\theta(\tau_1)=\frac{\sinh\theta(0)-\tan\gamma_0\tau_1}
 {1+\sinh\theta(0)\tan\gamma_0\tau_1}.
\ee
The spread complexity grows without bound as a function of time
\be
 K(t,\tau_1)=2h\left(\frac{\cosh\theta(0)\sinh\gamma_0t}
 {\cos\gamma_0\tau_1+\sinh\theta(0)\sin\gamma_0\tau_1}\right)^2. \label{kunst}
\ee
This solution of the Toda equations was obtained in Ref.~\cite{Dymarsky20}
and was discussed as an onset of chaos.
It is interesting to see that the spread complexity in Eq.~(\ref{kunst}) 
is a nonmonotonic function of $\tau_1$ when $\theta(0)\ne 0$.
It is a decreasing function for small $\tau_1$ as in the other examples,
and turns into an increasing one as $\tau_1$ becomes larger.
This is because $a_n(\tau_1)$ is decreasing in $n$ for large $\tau_1$.
This scenario is reminiscent of systems with an unstable fixed point, 
as the inverted harmonic oscillator.

We can also study the critical point $\gamma(\tau_1)=\alpha(\tau_1)$ 
as the boundary between stable and unstable solutions.
The equation of $\gamma(\tau_1)$, 
\be
 \partial_{\tau_1}\gamma(\tau_1) = -\gamma^2(\tau_1), 
\ee
is easily solved as 
\be
 \gamma(\tau_1)=\frac{\gamma_0}{1+\gamma_0\tau_1}.
\ee
The spread complexity is given by
\be
 K(t,\tau_1)=2h\left(\frac{\gamma_0t}{1+\gamma_0\tau_1}\right)^2. \label{kmarg}
\ee

%%%%%%%%%%%%%%%%%%%%%%%%%%%%%%%%%%%%%%%
\begin{figure*}[t]
\centering\includegraphics[width=.8\linewidth]{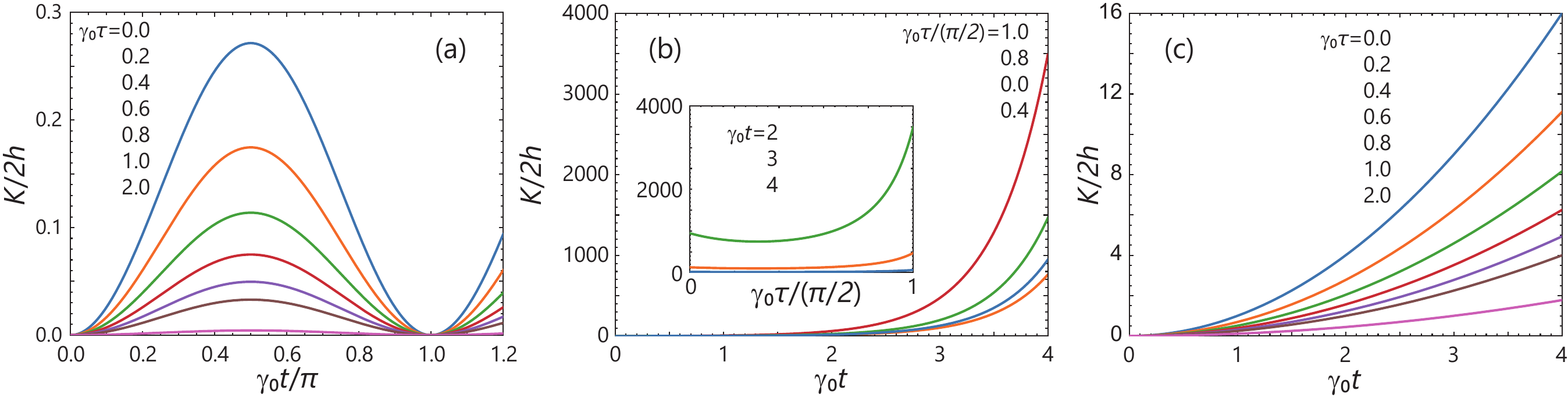}
\caption{The spread complexity for systems with SL(2,R) symmetry.
We set $\theta(0)=0.5$ and show three possible cases: 
(a) stable solution in Eq.~(\ref{kstable}), 
(b) unstable in Eq.~(\ref{kunst}), (c) marginal in Eq.~(\ref{kmarg}).
The $t$-dependence of the spread complexity 
for the SU(2) case and for the Heisenberg-Weyl case
has the same form as the function in panel (a).
}
\label{fig:k1}
\end{figure*}
%%%%%%%%%%%%%%%%%%%%%%%%%%%%%%%%%%%%%%%

We summarize the possible behavior of the spread complexity in Fig.~\ref{fig:k1}.
In all of the results described in this subsection, 
we can find closed complexity algebra
among the diagonal part of $L$, the off-diagonal part of $L$, the Lax pair $M_1$, 
and the complexity operator ${\cal K}=\mathrm{diag}\,(0,1,2,\dots)$~\cite{Caputa22, Hornedal22}.
Since the $\tau_1$ dependence of each matrix appears as an overall factor, 
the algebraic structure is unchanged by the deformation and we can solve the problem analytically, generalizing the known results for the underformed case.

%%%%%%%%%%%%%%%%%%%%%%%%%%%%%%%%%%%%%%%%%%%%%%%%%%%%%%%%%%%%%%%%%%%%%%%%%%%%%%%
\subsection{Exact solutions from the second Toda equations}

We next discuss the second Toda equations~(\ref{toda2-1}) and (\ref{toda2-2}).
In contrast to the first equations, we can put $a_n(\tau_2)=0$ to find a simple but nontrivial solution.
The equation for $b_n$ reads 
\be
 \partial_{\tau_2}b_n^2(\tau_2) = -b_n^2(\tau_2)[b_{n+1}^2(\tau_2)-b_{n-1}^2(\tau_2)]. \label{toda2-a0}
\ee
This equation denotes that the time evolutions of the odd components are 
determined from the even components, and vice versa.
We also see that this equation is solvable when 
$b_{n+1}^2(\tau_2)-b_{n-1}^2(\tau_2)$ is independent of $n$.
We  use the ansatz 
\be
 && b_{2n+1}^2(\tau_2) = (2n+1)\gamma^2(\tau_2), \label{alt-1}\\
 && b_{2n}^2(\tau_2) = 2n\alpha^2(\tau_2). \label{alt-2}
\ee
Given that $\alpha^2(\tau_2)-\gamma^2(\tau_2)$ remains constant as a function of $\tau_2$, 
we choose the parametrizations 
$(\alpha,\gamma)=(\gamma_0\cosh\theta(\tau_2),\gamma_0\sinh\theta(\tau_2))$ and 
$(\alpha,\gamma)=(\gamma_0\sinh\theta(\tau_2),\gamma_0\cosh\theta(\tau_2))$.
Both forms yield the same equation 
\be
 \partial_{\tau_2}\theta(\tau_2)=-\gamma_0^2\sinh\theta(\tau_2)\cosh\theta(\tau_2),
\ee
and we obtain 
\be
 \cosh2\theta(\tau_2)=\frac{\cosh 2\theta(0)+\tanh\gamma_0^2\tau_2}{1+\cosh 2\theta(0)\tanh\gamma_0^2\tau_2}.
\ee
This solution shows that the $L$-matrix is block-diagonalized into many $2\times 2$ matrices. 

We can also consider the marginal solution with 
$\alpha(\tau_2)=\gamma(\tau_2)$.
Then, we obtain 
\be
 \alpha^2(\tau_2)=\frac{\alpha^2(0)}{1+2\alpha^2(0)\tau_2}.
\ee
In this case, the $L$-matrix takes a simple, known form 
with $b_n(\tau_2)=\alpha(\tau_2)\sqrt{n}$ and the spread complexity is obtained as 
\be
 K(t,\tau_2)=\alpha^2(\tau_2)t^2=\frac{\alpha^2(0)t^2}{1+2\alpha^2(0)\tau_2}.
 \label{alt-eq}
\ee

%%%%%%%%%%%%%%%%%%%%%%%%%%%%%%%%%%%%%%%
\begin{figure}[t]
\centering\includegraphics[width=1.\columnwidth]{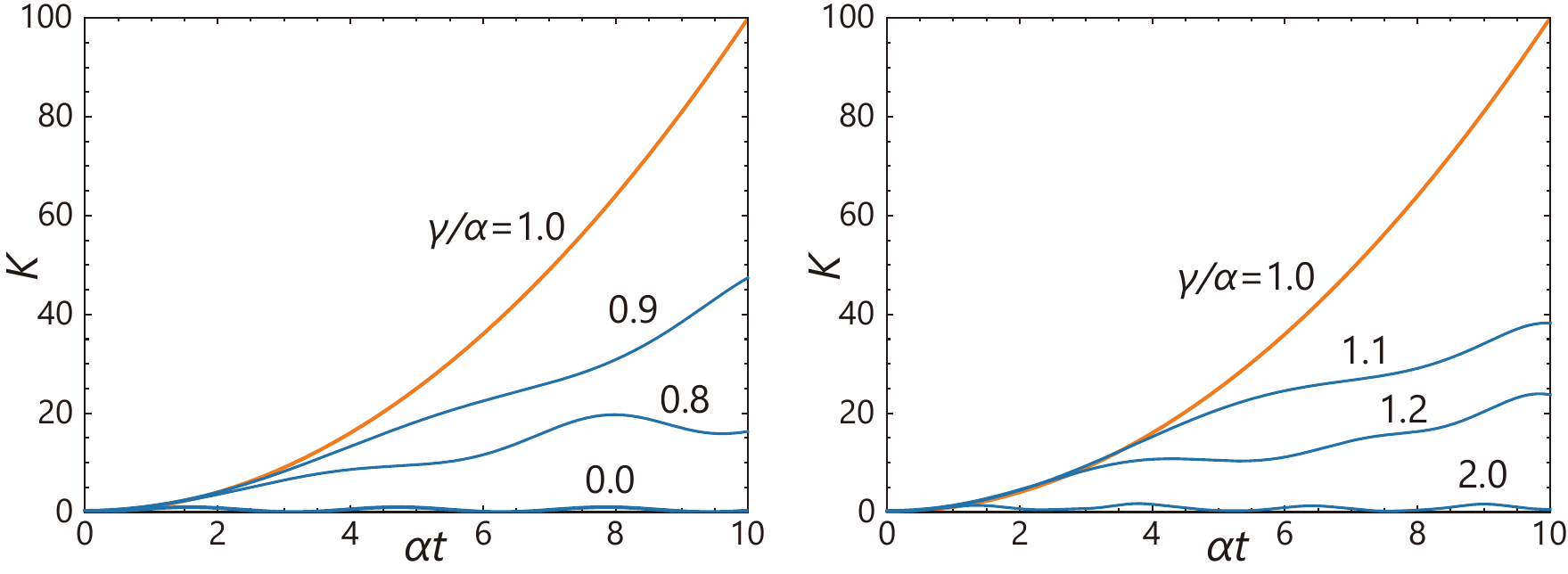}
\caption{The spread complexity for systems described 
by Eqs.~(\ref{alt-1}) and (\ref{alt-2}).
We set $d=100$ and several values of $\gamma(\tau_2)/\alpha(\tau_2)$
to calculate the spread complexity numerically.
The result for $\alpha=\gamma$ at $d\to\infty$ 
is obtained analytically as Eq.~(\ref{alt-eq}).
}
\label{fig:k2}
\end{figure}
%%%%%%%%%%%%%%%%%%%%%%%%%%%%%%%%%%%%%%%

In the case of the second Toda equations, 
finding closed complexity algebra is not a simple task.
Correspondingly, we have no analytical solutions of the spread complexity 
at $\gamma/\alpha\ne 0, 1,\infty$.
We plot the numerical results of the spread complexity in Fig.~\ref{fig:k2}.

The second Toda equations with $a_n=0$ are closely related to the first ones.
In that case, $L^2$ is written as 
\be
 L^2 = \left(\begin{array}{ccccc} b_1^2 & 0  & b_1b_2 &  & \\
 0 & b_1^2+b_2^2 & 0 & b_2b_3 & \cdots\\
 b_1b_2 & 0 & b_2^2+b_3^2 & 0 & \\
 &\vdots && & \ddots
 \end{array}\right).
\ee
We can define two independent tridiagonal matrices from the even and odd sectors.
Identifying each matrix as a tridiagonal ``$L$'', 
we can find that the redefined coefficients, ``$a_n$'' and ``$b_n$'', 
follow the first Toda equations.
Also, in the case of $a_n=0$, 
we can construct the even Krylov series and odd series separately~\cite{Muck22}.
In Sec.~\ref{sec:susy}, we utilize this property for supersymmetric systems.

%%%%%%%%%%%%%%%%%%%%%%%%%%%%%%%%%%%%%%%%%%%%%%%%%%%%%%%%%%%%%%%%%%%%%%%%%%%%%%%
\subsection{Fixed-point solutions and saturation of complexity}

The introduction of the parameter $\tau$ gives 
a Toda flow in the Lanczos-coefficients space 
$(a_0,a_1,\dots;b_1,b_2,\dots)$.
We are interested in the large-$\tau$ regime 
where the flow is influenced by the fixed point.

We first consider the first set of the Toda equations, 
Eqs.~(\ref{toda1-1}) and (\ref{toda1-2}), with $\tau=(\tau_1,0)$.
Setting the left-hand side of each equation to be zero, 
we find $b_n=0$.
Since the off-diagonal components go to zero, 
the $\tau$-evolution represents a diagonalization flow.
The fixed point is given by 
$(\epsilon_0,\epsilon_1,\dots,\epsilon_{d-1};0,\dots)$
where $(\epsilon_0,\epsilon_1,\dots,\epsilon_{d-1})$ 
represent distinct eigenvalues of the original Hamiltonian $H$.
To find the convergence $b_n\to 0$ in Eq.~(\ref{toda1-2}),
we require $a_n>a_{n-1}$, which gives the unique order 
$\epsilon_0<\epsilon_1<\dots<\epsilon_{d-1}$~\cite{Moser75book,Monthus16}.
At large $\tau_1$, we obtain exponential decay
\be
 b_n(\tau_1)\sim\exp\left[-\frac{1}{2}
 (\epsilon_n-\epsilon_{n-1})\tau_1\right]. 
 \label{as1}
\ee

Next, we consider the second set of the Toda equations, 
Eqs.~(\ref{toda2-1}) and (\ref{toda2-2}), with $\tau=(0,\tau_2)$.
In that case, the fixed-point condition is written as 
\be
 && b_n^2(a_{n-1}+a_n)=0 \qquad (n=1,2,\dots,d-1), \\
 && b_{n}b_{n+1}=0 \qquad (n=1,2,\dots, d-2).
\ee
These relations denote two possible kinds of fixed points.

The simplest possible solution is given by $b_n=0$ for all $n$.
From Eq.~(\ref{toda2-2}), we require $a_n^2>a_{n-1}^2$ for this point to be stable.
As in the previous case, we can find 
$(\epsilon_0,\epsilon_1,\dots,\epsilon_{d-1};0,\dots)$ as the fixed point 
and exponential decay
\be
 b_n(\tau_2)\sim\exp\left[-\frac{1}{2}
 (\epsilon_n^2-\epsilon_{n-1}^2)\tau_2\right],
 \label{as2}
\ee
provided that the condition $|\epsilon_0|<|\epsilon_1|< \dots <|\epsilon_{d-1}|$
is satisfied.
In contrast to the previous case, we require the absolute value of each energy level.
This is because $H^2$ is used in the deformed initial state.

Although $\epsilon_m\ne\epsilon_n$ holds for $m\ne n$ by definition,  
there exist possibilities that $|\epsilon_m|=|\epsilon_n|$ holds even for $m\ne n$.
In that case, we can find a different fixed point with $b_n\ne 0$.
For example, when 
$\epsilon_0<\epsilon_1<0<\epsilon_{2}<\epsilon_3$
with $-\epsilon_1=\epsilon_2<-\epsilon_0<\epsilon_3$ at $d=4$, 
the fixed point is given by
$(a_0,-a_0,\epsilon_0,\epsilon_3;b_1,0,0)$
with $\sqrt{a_0^2+b_1^2}=|\epsilon_1|$.
The generalization is straightforward 
and we obtain many block structures 
when there exist many energy levels with $\epsilon_m=-\epsilon_n$.

In certain kinds of systems, we can find a symmetric spectrum as 
$(\pm \epsilon_1, \pm\epsilon_2,\dots,\pm\epsilon_{d/2})$.
Here, we assume that $d$ is even and 
$0<\epsilon_1<\epsilon_2<\dots<\epsilon_{d/2}$.
In this case, the diagonal components $a_n$ are shown to be zero and
the Toda equations are simplified to Eq.~(\ref{toda2-a0}).
The equations have a fixed point
$(0,\dots,0;\epsilon_1,0,\epsilon_2,0,\dots, 0,\epsilon_{d/2})$,
and the convergence to this point takes an exponential form
as Eq.~(\ref{as2}).

We discuss the implications of the Toda flow for the spread complexity written as 
\be
 K(t,\tau)=\langle 0|e^{iL(\tau)t}{\cal K}e^{-iL(\tau)t}|0\rangle,
 \label{complexity2}
\ee
where ${\cal K}=\mathrm{diag}\,(0,1,2,\dots,d-1)$.
It depends on both $t$ and $\tau$, and exhibits a rich behavior.
In chaotic systems, we are often interested in instances that saturate 
the complexity and its growth.
It is convenient to introduce time-averaged quantities.
We define 
\be
 \overline{K}(\tau)=\frac{1}{T}\int_0^T dt\,K(t,\tau).
\ee
When $K(t,\tau)$ is periodic in $t$, the time duration $T$ is set to the period,
while we let $T\to\infty$ in the general nonperiodic case.
In Appendix~\ref{sec:timeav}, we show that
the quantity $\overline{K}(\tau)$ can be written as 
\be
 && \overline{K}(\tau)=\sum_{n=0}^{d-1}
 \frac{e^{-f(\epsilon_n,\tau)+S_n}}{Z(\tau)}K_n(\tau), \label{kav} \\
 && K_n(\tau) = \langle \phi_n(\tau)|{\cal K}|\phi_n(\tau)\rangle, \\
 && |\phi_n(\tau)\rangle = F^\dag(\tau)|\epsilon_n\rangle,\label{phin}
\ee
where $S_n=\ln d_n$, $Z(\tau)=\sum_n e^{-f(\epsilon_n,\tau)+S_n}$, 
and $F(\tau)$ is defined in Eq.~(\ref{Ftau}).
The transformation operator $F(\tau)$ connects the eigenstate 
$|\epsilon_n\rangle$ in the original space to that in the Krylov space.
We see that $|\phi_n(\tau)\rangle$
is an eigenstate of $L(\tau)$ satisfying 
\be
 L(\tau)|\phi_n(\tau)\rangle=\epsilon_n|\phi_n(\tau)\rangle. \label{lphi}
\ee
Since $S_n$ represents the entropy for a given energy $\epsilon_n$,
Eq.~(\ref{kav}) is interpreted as the generalized canonical average 
of $K_n(\tau)$.
Although $K_n(\tau)$ itself depends on $\tau$, 
it takes values between 0 and $d-1$.

When $\tau$ is large enough
and the flow is governed by the diagonal fixed point discussed above, the off-diagonal components of the $L$ matrix are small, and the diagonal components align in the described order.
In that case, we can estimate  
\be
 K_n(\tau) \sim n.
 \label{kavn}
\ee
The saturation value of the complexity is given by
the generalized canonical average of the excitation number
from the ground state or the zero-energy state.
The scale justifying this relation is obtained from 
Eq.~(\ref{as1}) or Eq.~(\ref{as2}). 
We have $\tau_1\gg 1/\min_n (\epsilon_n-\epsilon_{n-1})$
and $\tau_2\gg 1/\min_n (\epsilon_n^2-\epsilon_{n-1}^2)$.
For many-body Hamiltonians with very small energy gaps, 
we require very large $\tau$ to find Eq.~(\ref{kavn}).

%%%%%%%%%%%%%%%%%%%%%%%%%%%%%%%%%%%%%%%%%%%%%%%%%%%%%%%%%%%%%%%%%%%%%%%%%%%%%%%
\section{Coherent Gibbs state for thermodynamic systems}
\label{sec:cgs}

%%%%%%%%%%%%%%%%%%%%%%%%%%%%%%%%%%%%%%%%%%%%%%%%%%%%%%%%%%%%%%%%%%%%%%%%%%%%%%%
\subsection{Lanczos coefficients}

%%%%%%%%%%%%%%%%%%%%%%%%%%%%%%%%%%%%%%%
\begin{figure}[t]
\centering\includegraphics[width=1.\columnwidth]{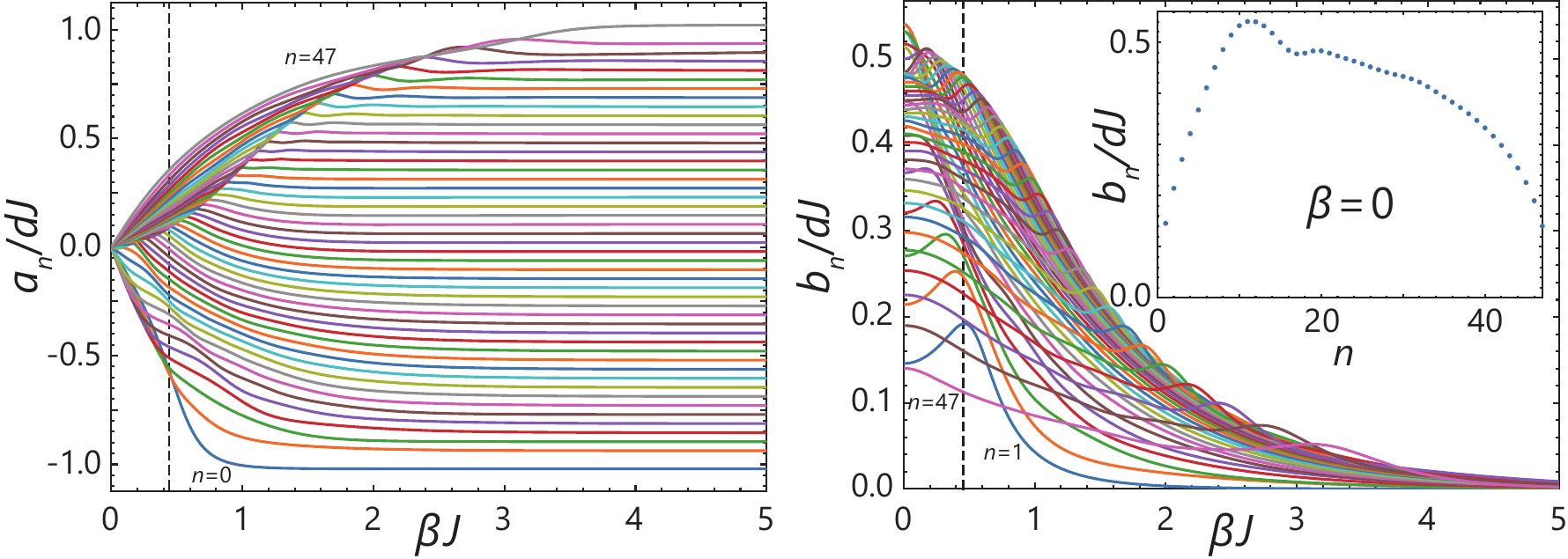}
\caption{The Lanczos coefficients $a_n(\beta)$ and $b_n(\beta)$ 
of the two-dimensional Ising model.
We take a $6\times 5$ lattice with open boundary conditions,  
which gives the Krylov dimension $d=48$. 
At $\beta=0$, the diagonal components $a_n(0)$ are zero and the off-diagonal 
components $\beta_n(0)$ are plotted in the inset of the right panel.
The critical point at the thermodynamic limit is given by 
$\beta J=\frac{1}{2}\ln (\sqrt{2}+1)\approx 0.4407$
and is denoted by the vertical dashed line in each panel.
}
\label{fig:ising2d-ab}
\end{figure}
%%%%%%%%%%%%%%%%%%%%%%%%%%%%%%%%%%%%%%%
%%%%%%%%%%%%%%%%%%%%%%%%%%%%%%%%%%%%%%%
\begin{figure}[t]
\centering\includegraphics[width=1.\columnwidth]{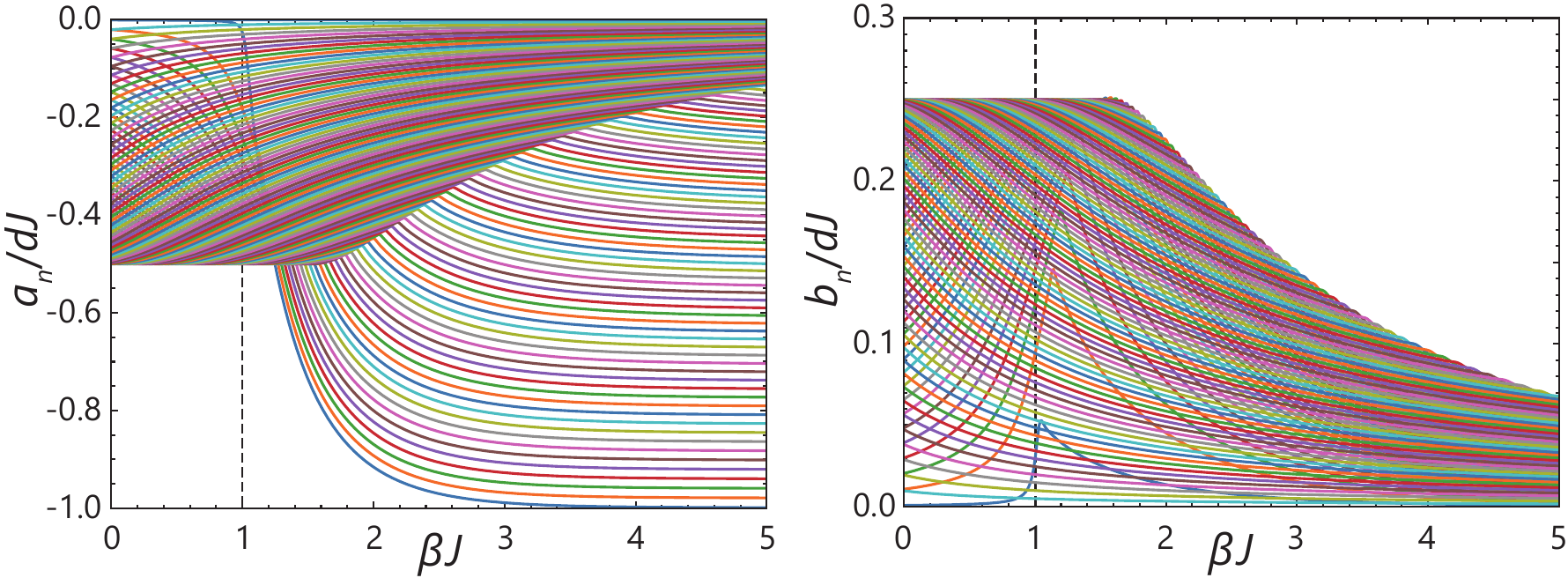}
\caption{The Lanczos coefficients $a_n$ and $b_n$ of the fully-connected Ising model
with $N=2000$.
The Krylov dimension is given by $d=N/2+1=1001$, and 
we plot the coefficients at intervals of 10.
The critical point at the thermodynamic limit is given by $\beta J=1$
and is denoted by the vertical dashed line in each panel.
}
\label{fig:isingfc-ab}
\end{figure}
%%%%%%%%%%%%%%%%%%%%%%%%%%%%%%%%%%%%%%%

In this section, we set $\tau=(\beta,0)$ and use the coherent 
Gibbs state in Eq.~(\ref{cgs}) as an initial state.
When the Hamiltonian represents many-body systems, 
the properties of the dynamics are closely related to
those of the corresponding thermodynamic system.
In fact, once we know the Lanczos coefficients at $\beta=0$, 
we can obtain the thermodynamic properties by solving the Toda equations.
As mentioned in Sec.~\ref{sec:krylov}, 
$a_0(\beta)$ denotes the thermodynamic energy, and 
$b_1(\beta)$ is related to the heat capacity of the system.

In thermodynamic systems, one is often interested in the thermodynamic limit.
In that case, we can find singular behavior for the Lanczos coefficients.
For example, we consider ``classical'' Ising models generally written as 
\be
 H = -\sum_{(i,j)}J_{ij}\sigma_i\sigma_j, \label{ising}
\ee
where $\sigma_i=\pm 1$  represents spin variable at site $i$.
In the standard Ising model on a two-dimensional lattice, where 
$J_{ij}$ takes a finite value $J$ only for nearest neighbor pairs $(i,j)$, 
the Onsager solution gives the critical point 
$\beta J=\frac{1}{2}\ln (\sqrt{2}+1)\approx 0.4407$.
Although the Hilbert space dimension increases exponentially as $2^N$
with the number of spins $N$, 
the number of the distinct eigenvalues, the Krylov dimension, is of the order of $N$.

In the following, we show numerical results 
on the Lanczos coefficients 
and several quantities obtained from the time evolution $e^{-iL(\tau)t}|0\rangle$.
The Lanczos coefficients at $\tau=0$ are calculated from the recurrence relation in Eq.~(\ref{rec}) 
with $|K_0(0)\rangle\propto(\sqrt{d_0},\sqrt{d_1},\dots,\sqrt{d_{d-1}})^\mathrm{T}$, 
and $H=\mathrm{diag}(\epsilon_0,\epsilon_1,\dots,\epsilon_{d-1})$.
We note that $(\epsilon_0,\epsilon_1,\dots,\epsilon_{d-1})$ denotes the set of distinct energy levels
and $(d_0,d_1,\dots,d_{d-1})$ denotes that of the corresponding degeneracies.
After finding all $a_n(0)$ and $b_n(0)$, 
we solve the Toda equations to obtain the $\tau$-dependent coefficients.

We show an example of a $6\times 5$ lattice with open boundary conditions 
in Fig.~\ref{fig:ising2d-ab}.
%\epsilon_n/J= \{-49, -45, -43, -41, -39, -37, -35, -33, -31, -29, -27, -25, -23, -21, -19, -17, -15, -13, -11, -9, -7, -5, -3, -1, 1, 3, 5, 7, 9, 11, 13, 15, 17, 19, 21, 23, 25, 27, 29, 31, 33, 35, 37, 39, 41, 43, 45, 49\}, 
%d_n = \{2, 8, 44, 88, 274, 1012, 2748, 7458, 20848, 54020, 133984, 322128, 738314, 1607784, 3332792, 6542614, 12108512, 21069584, 34287436, 51960806, 73064022, 94821196, 113108464, 123686774, 123686774, 113108464, 94821196, 73064022, 51960806, 34287436, 21069584, 12108512, 6542614, 3332792, 1607784, 738314, 322128, 133984, 54020, 20848, 7458, 2748, 1012, 274, 88, 44, 8, 2\}.
%$d= 48$.
As we expect from the general discussions in Sec.~\ref{sec:todaeqs}, 
the result shows a diagonalization flow.
The diagonal components $a_n(\beta)$ converge from $a_n(0)=0$ at $\beta=0$ 
to the eigenvalues of the Hamiltonian in ascending order at $\beta\to\infty$,
and the off-diagonal components $b_n(\beta)$ decay exponentially 
at large $\beta$.
We observe a sharp peak of $b_1(\beta)$ around the critical point.
It is interesting to see that the other components representing 
higher-order cumulants also show sharp peaks at different points.
Some of the coefficients $a_n$ interchange their order at those peak points,
which can be understood from the Toda equations.

We also study the case of the fully connected Ising model,
where the sum in Eq.~(\ref{ising}) is taken over all pairs and $J_{ij}=J/N$.
The Lanczos coefficients at $N=2000$ are plotted in Fig.~\ref{fig:isingfc-ab}.
In this case, the critical point is given by $\beta J=1$.
We observe a sharp peak of $b_1(\beta)$ and 
a sudden drop of $a_0(\beta)$ at the point.
These results are consistent with the known mean-field solution \cite{NishimoriOrtiz2011}.

%%%%%%%%%%%%%%%%%%%%%%%%%%%%%%%%%%%%%%%%%%%%%%%%%%%%%%%%%%%%%%%%%%%%%%%%%%%%%%%
\subsection{Survival amplitude}

%%%%%%%%%%%%%%%%%%%%%%%%%%%%%%%%%%%%%%%
\begin{figure}[t]
\centering\includegraphics[width=1.\columnwidth]{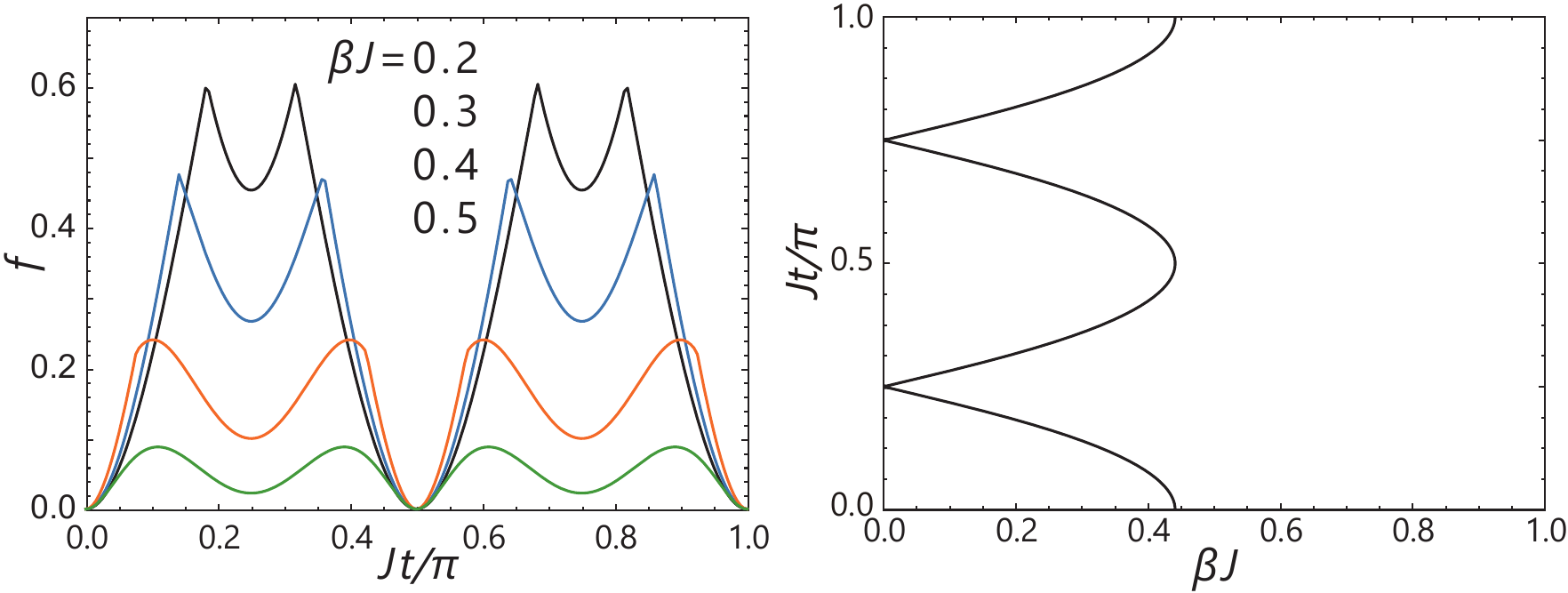}
\caption{The left panel represents the rate function 
of the two-dimensional Ising model at the thermodynamic limit.
The function is periodic in $t$ with the period $Jt=\pi/2$.
The right panel represents the phase diagram obtained from the Lee-Yang zeros.
}
\label{fig:ising2d-surv}
\end{figure}
%%%%%%%%%%%%%%%%%%%%%%%%%%%%%%%%%%%%%%%

As we mention in Sec.~\ref{sec:hdeform}, the survival amplitude is directly 
related to the canonical partition function $Z(\beta)$ as
\be
 \langle\psi_0(\beta)|\psi(t,\beta)\rangle 
 = \langle 0|\varphi(t,\beta)\rangle = \frac{Z(\beta+it)}{Z(\beta)},
\ee
where $|\varphi(t,\beta)\rangle=\sum_{n=0}^{d-1}|n\rangle\varphi_n(t,\beta)$.
In the thermodynamic limit, it is useful to define the rate function 
\be
 f(t,\beta) = -\frac{1}{N}\ln |\langle 0|\varphi(t,\beta)\rangle|^2,
\ee
where $N$ generally represents a system size.
This quantity is represented by using the free energy density.
The rate function was used to establish and characterize  
dynamical quantum phase transitions in quenched systems~\cite{Heyl13,Heyl18}, although it should be noted that it is generally not an intensive quantity \cite{delcampo16,delcampo21}.

In the case of the two-dimensional Ising model, 
the free energy is obtained analytically.
We plot the rate function in Fig.~\ref{fig:ising2d-surv}.
We observe several singular points, which are obtained 
from the Lee-Yang zeros of the partition function~\cite{Yang52,Lee52,Itzykson91} 
\be
 |\sinh [2J(\beta+it)]| = 1.
\ee
The property that the partition function is characterized by zeros
indicates that the singular behavior of the Lanczos coefficients in 
the thermodynamic limit is limited to the critical point.

If we are interested in the overlap of $|\psi(t,\beta)\rangle$ with
the initial state at $\beta=0$, $|\psi_0\rangle$, 
we can use the relations~\cite{Obuchi12}
\be
 \langle\psi_0|\psi(t,\beta)\rangle = \langle 0|V(\beta)|\varphi(t,\beta)\rangle
 = \frac{Z(\beta/2+it)}{\sqrt{Z(0)Z(\beta)}}.
\ee
The singularity in the $\beta$-axis appears at $\beta=2\beta_c$ where $\beta_c$ 
represents the critical point of the corresponding thermodynamic system.

%%%%%%%%%%%%%%%%%%%%%%%%%%%%%%%%%%%%%%%%%%%%%%%%%%%%%%%%%%%%%%%%%%%%%%%%%%%%%%%
\subsection{Spread complexity and Krylov entropy}

%%%%%%%%%%%%%%%%%%%%%%%%%%%%%%%%%%%%%%%
\begin{figure}[t]
\centering\includegraphics[width=1.\columnwidth]{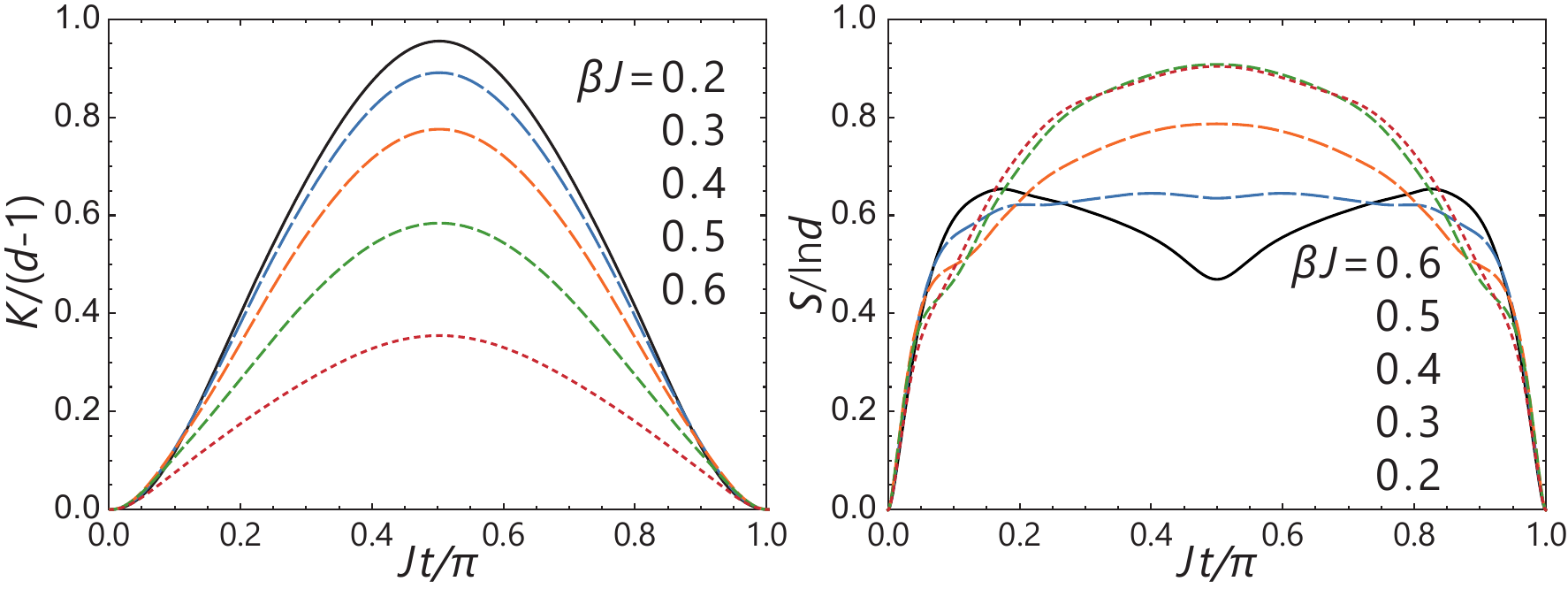}
\caption{The complexity $K$ and the Krylov entropy $S$ of the $6\times 5$ Ising model.
They are periodic in $t$ with the period $Jt=\pi$.
}
\label{fig:ising2d-ks}
\end{figure}
%%%%%%%%%%%%%%%%%%%%%%%%%%%%%%%%%%%%%%%

To study how the thermodynamic and dynamical singularities are incorporated into  other quantities defined in the Krylov space, we numerically calculate 
the spread complexity $K(t,\beta)$, Eq.~(\ref{complexity}), and the Krylov entropy \cite{Rabinovici21} 
\be
 S(t,\beta) = -\sum_{n=0}^{d-1}|\varphi_n(t,\beta)|^2\ln |\varphi_n(t,\beta)|^2.
\ee
We note that the Krylov entropy measures randomness in the Krylov space 
and is different from the thermodynamic entropy.

We plot the result for the $6\times 5$ Ising model 
in Fig.~\ref{fig:ising2d-ks}.
In contrast to the survival amplitude, 
the spread complexity does not show any singularity and changes smoothly.
This is because the spread complexity does not include the contribution
from $\varphi_0(t,\beta)$ and involves an average over the other states.
On the other hand, for the Krylov entropy, 
we observe unstable behavior around the dynamical singular points.
We also find a cusp when $\beta J$ is smaller than the critical value.
A similar result was obtained in a quenched system 
showing dynamical quantum phase transitions~\cite{Bento24,Takahashi25-2}.

When $\beta$ becomes larger, both quantities averaged over $t$ decrease rapidly.
The state at a large $\beta$ represents a low-temperature state, and
the state space explored is reduced to the low-energy states.
In the following, we investigate the dependence of the time-averaged quantities on $\beta$.

%%%%%%%%%%%%%%%%%%%%%%%%%%%%%%%%%%%%%%%%%%%%%%%%%%%%%%%%%%%%%%%%%%%%%%%%%%%%%%%
\subsection{Time-averaged quantities}

%%%%%%%%%%%%%%%%%%%%%%%%%%%%%%%%%%%%%%%
\begin{figure}[t]
\centering\includegraphics[width=1.\columnwidth]{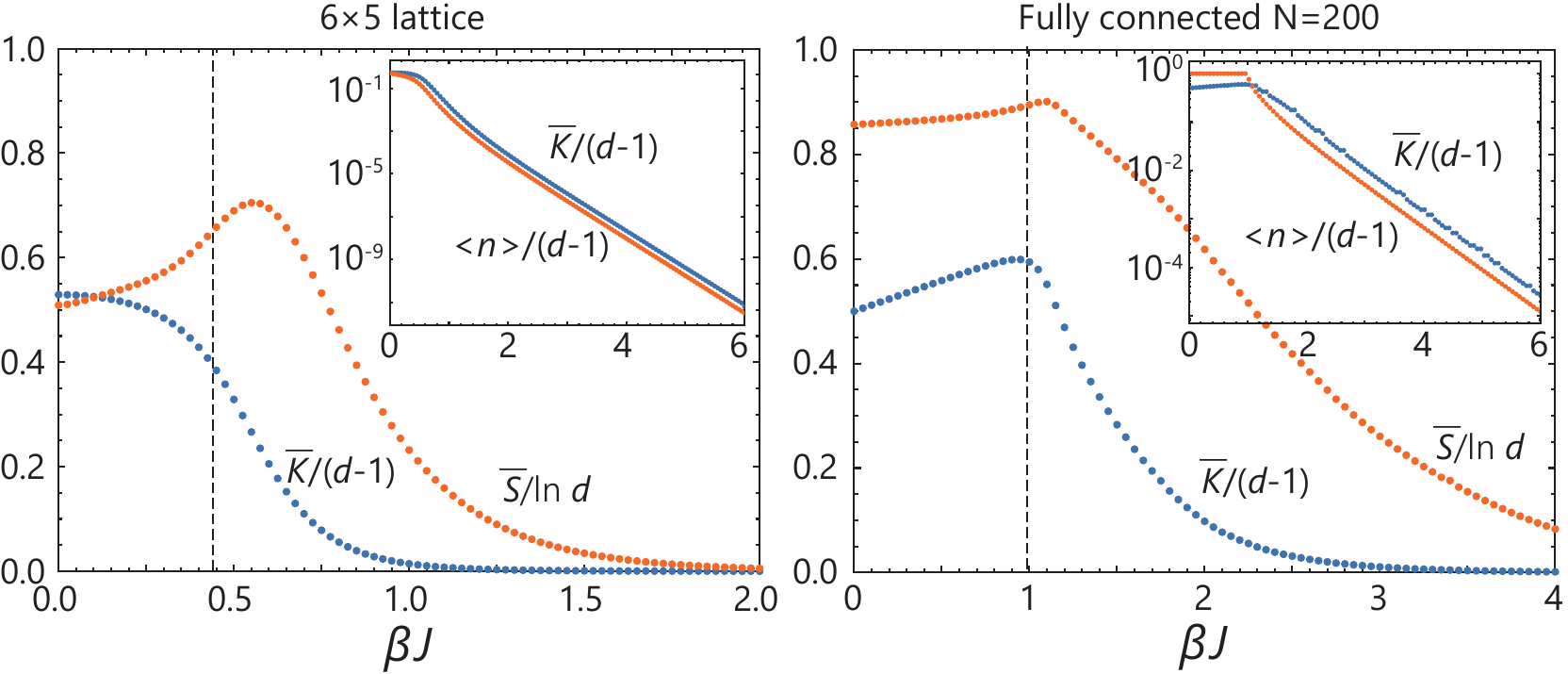}
\caption{The time-averaged complexity and entropy.
The left panel is for the  $6\times 5$ Ising model
and the right is for the fully-connected model with $N=200$.
The vertical dashed line in each panel represents 
the critical point obtained at the thermodynamic limit.
In each panel, the inset denotes the long-time behavior of 
the complexity on a log scale, which is compared to 
$\langle n\rangle=\sum_n n e^{-\beta\epsilon_n+S_n}/Z(\beta)$.
}
\label{fig:ising-tav}
\end{figure}
%%%%%%%%%%%%%%%%%%%%%%%%%%%%%%%%%%%%%%%

In the thermodynamic limit, 
the large deviation principle says that the sum over microscopic states 
in Eq.~(\ref{kav})
is effectively restricted to a single value with 
$\epsilon_n=\epsilon^*(\beta)$.
The value is determined by a competition between minimizing the energy and maximizing the entropy.
Then, the average complexity is written as 
$\overline{K}(\beta)\sim\langle \epsilon^*(\beta)|F(\beta){\cal K}F^\dag(\beta)|\epsilon^*(\beta)\rangle$.
Across the critical point where the choice of $\epsilon^*$ changes qualitatively, 
we expect that the complexity involves a singularity.

In Fig.~\ref{fig:ising-tav}, we plot the averaged complexity and the averaged entropy 
for a two-dimensional Ising model and a fully-connected model.
As expected, the average complexity exhibits a drastic change across the critical point.
Although the entropy cannot be written as a simple expectation value, 
it also shows a similar behavior.

The results in Figs.~\ref{fig:ising2d-ab} and \ref{fig:isingfc-ab} 
indicate that $a_n\sim\epsilon_n$ and $b_n\sim 0$ for $\beta J\gg 1$.
In that case, we can justify Eq.~(\ref{kavn}),
as we see in the inset of each panel. 
 
%%%%%%%%%%%%%%%%%%%%%%%%%%%%%%%%%%%%%%%%%%%%%%%%%%%%%%%%%%%%%%%%%%%%%%%%%%%%%%%
\section{Random matrix quenches}
\label{sec:rmt}

Random matrix theory (RMT) provides an ideal framework for exploring the complexity of deformed Hamiltonians. 
In particular, Ref.~\cite{Balasubramanian22} has extensively analyzed the evolution of complexity in RMT with different Dyson symmetry classes, focusing on evolutions initiated from the standard reference 
states to coherent Gibbs states. Building upon these results, this section considers 
a broader generalization, where the properties of complexity evolution are re-examined using the deformed Hamiltonian. We especially focus on the long-time behavior of the spread complexity.

%%%%%%%%%%%%%%%%%%%%%%%%%%%%%%%%%%%%%%%%%%%%%%%%%%%%%%%%%%%%%%%%%%
\subsection{Two-dimensional example}

Building on Refs.~\cite{Caputa:2024vrn, Nandy:2024mml}, 
we present a two-dimensional random matrix example, 
where explicit computation of Krylov complexity is possible.
The complexity is calculated for a single realization of $2\times 2$ matrix 
and the ensemble average is carried out afterwards.

When the Hamiltonian has eigenvalues 
$\epsilon_0=E-\omega/2$ and $\epsilon_1=E+\omega/2$ with $\omega\ge 0$, the initial state is written as 
\be
 |\psi_0(\tau)\rangle = |0\rangle \sqrt{p(\epsilon_0,\tau)}
 +|1\rangle \sqrt{p(\epsilon_1,\tau)},
\ee
with $p(\epsilon,\tau)\propto e^{-f(\epsilon,\tau)}$ and 
$p(\epsilon_0,\tau)+p(\epsilon_1,\tau)=1$.
It is a simple task to apply the Krylov algorithm.
We obtain 
\be
 && a_0(\tau) =\epsilon_0p(\epsilon_0,\tau)+\epsilon_1p(\epsilon_1,\tau), \\
 && a_1(\tau) =\epsilon_1p(\epsilon_0,\tau)+\epsilon_0p(\epsilon_1,\tau), \\
 && b_1(\tau) = \omega\sqrt{p(\epsilon_0,\tau)p(\epsilon_1,\tau)}.
\ee
It is also straightforward to solve the Schr\"odinger 
equation in the Krylov space as 
\be
 |\varphi(t,\tau)\rangle=\left(\begin{array}{c}
 p(\epsilon_0,\tau)e^{-i\epsilon_0t}+p(\epsilon_1,\tau)e^{-i\epsilon_1t} \\
 \sqrt{p(\epsilon_0,\tau)p(\epsilon_1,\tau)}(
 -e^{-i\epsilon_0t}+e^{-i\epsilon_1t})
 \end{array}\right).\no\\
\ee
Since the dimension of the system is minimal, 
the time dependence of the spread complexity is determined irrespective 
of the choice of the deformation function $f$ as 
\be
 K(t,\tau)= 4p(\epsilon_0,\tau)p(\epsilon_1,\tau)\sin^2\frac{\omega t}{2}.
\ee
This gives the time-averaged complexity 
\be
 \overline{K}(\tau)
 = 2p(\epsilon_0,\tau)p(\epsilon_1,\tau)
 = \frac{2e^{-(f(\epsilon_1,\tau)-f(\epsilon_0,\tau))}}
 {[1+e^{-(f(\epsilon_1,\tau)-f(\epsilon_0,\tau))}]^2}.\no\\ \label{kav-rmt2}
\ee

In the case of the coherent Gibbs state $\tau=(\beta,0)$, 
Eq.~(\ref{kav-rmt2}) is dependent only on $\omega=\epsilon_1-\epsilon_0$, 
and not on $E=(\epsilon_0+\epsilon_1)/2$.
For the Gaussian RMT associated with the Dyson index $\beta_{\mathrm{D}}=1,2,4$, 
the probability distribution of the level spacing is given by~\cite{haake1991quantum} 
\be
 \rho_\omega(\omega) = A_{\beta_\mathrm{D}}
 \frac{\omega^{\beta_\mathrm{D}}}{\Delta^{\beta_\mathrm{D}+1}}
 \exp\left(-B_{\beta_\mathrm{D}}\frac{\omega^2}{\Delta^2}\right),
\ee
where the coefficients $A_{\beta_\mathrm{D}}$ and $B_{\beta_\mathrm{D}}$ 
are determined from the normalization $\int_0^\infty d\omega\rho_\omega(\omega)=1$ and 
the definition of the mean level spacing $\Delta=\int_0^\infty d\omega\rho_\omega(\omega)\omega$.
The average of Eq.~(\ref{kav-rmt2}) is written as 
\be
 \langle \overline{K}(\beta)\rangle &=&
 \frac{2A_{\beta_\mathrm{D}}}{(\beta\Delta)^{\beta_\mathrm{D}+1}}
 \int_0^\infty dz 
 \frac{z^{\beta_\mathrm{D}}}
 {(e^{z/2}+e^{-z/2})^2} 
 \no\\ && \times
 \exp\left[-B_{\beta_\mathrm{D}}\frac{z^2}{(\beta\Delta)^2}\right]. \label{kavav-rmt2}
\ee

\begin{figure}[t]
\centering\includegraphics[width=1.\columnwidth]{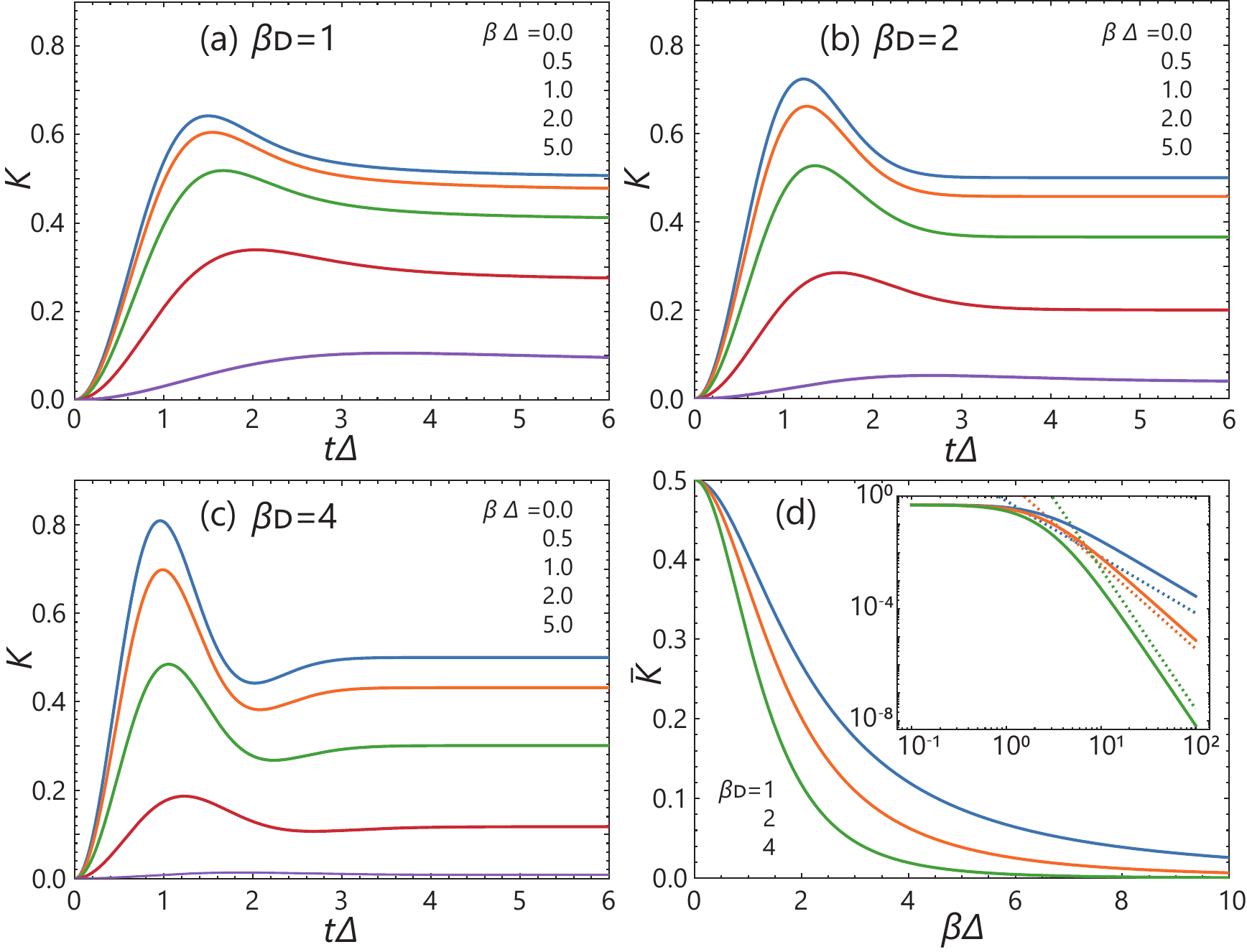}
\caption{The ensemble average of the spread complexity $K$ for 
random matrices with $d=2$ and $\tau=(\beta,0)$.
Panels (a), (b), and (c) represent 
the $t$-dependence for the symmetry class with the Dyson index $\beta_\mathrm{D}=1,2,4$, respectively. Panel (d) represents the average of the saturation value of the complexity.
The inset denotes the log-scale plot. 
The dotted lines denote the asymptotic forms $1/\beta^{\beta_\mathrm{D}+1}$
evaluated from Eq.~(\ref{kavav-rmt2}).
}
\label{fig:2drmt1}
\end{figure}

In Fig.~\ref{fig:2drmt1}, we plot the average of the spread complexity
including 
the $t$-depedence of $\langle K(t,\beta)\rangle$
for several fixed values of $\beta$ 
and the $\beta$-dependence of $\langle \overline{K}(\beta)\rangle$. 
The result for $\beta\Delta = 0$ can be analytically derived 
as Eq.(8) in Ref.\cite{Nandy:2024mml}, 
while for generic values of $\beta$, the analytic expressions are unknown. 
The level repulsion makes the spread complexity large, 
as we see from the small-$t$ behavior.
On the other hand, the asymptotic forms at large $\beta$ denote that 
the complexity decays as 
$\langle \overline{K}(\beta)\rangle\sim 1/\beta^{\beta_\mathrm{D}+1}$,
which is consistent with the expression in Eq.~(\ref{kavav-rmt2}).

\begin{figure}[t]
\centering\includegraphics[width=1.\columnwidth]{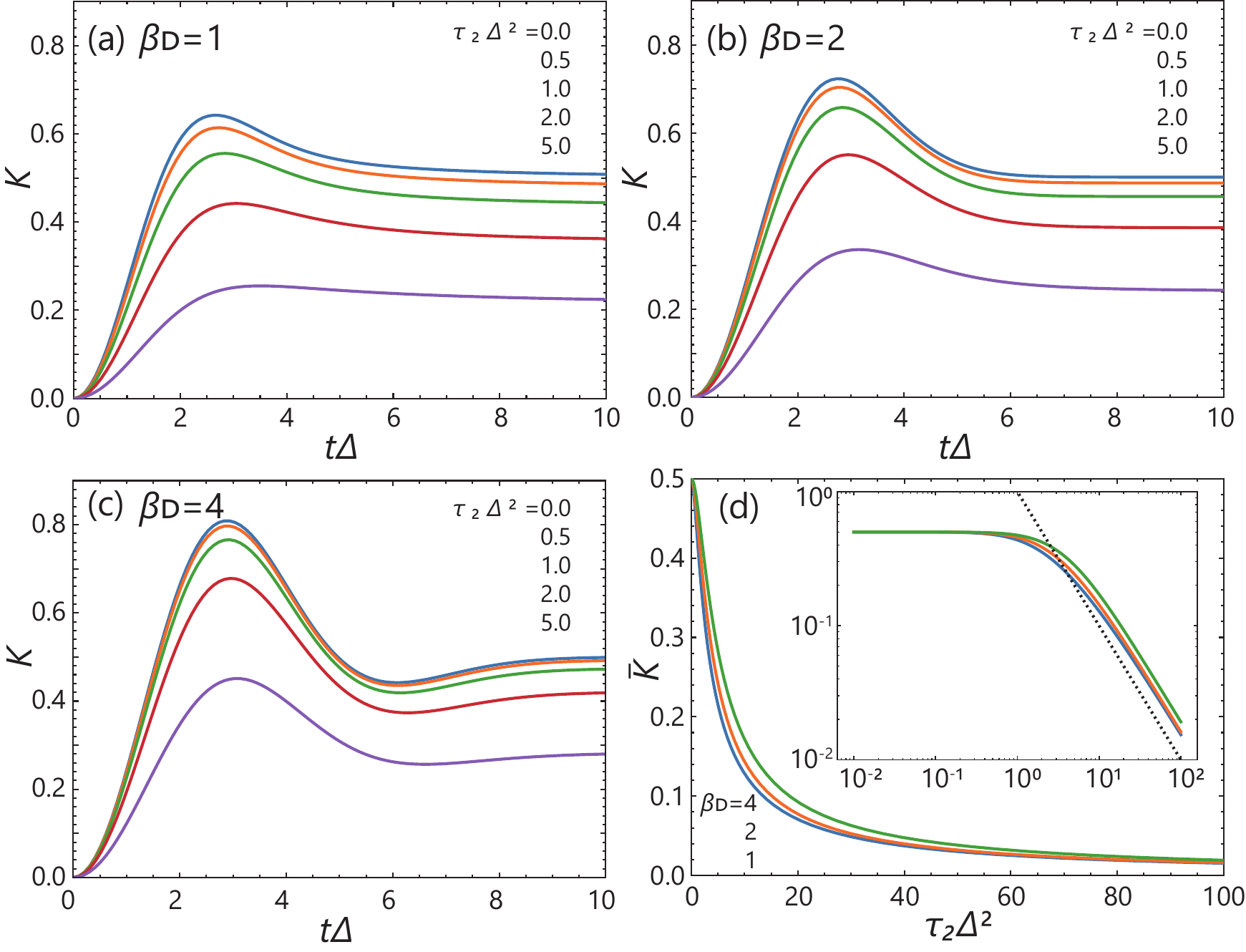}
\caption{
The result for $d=2$ and $\tau=(0,\tau_2)$.
See the caption in Fig.~\ref{fig:2drmt1} for the other details.
The dotted line in panel (d) denotes $1/\tau_2\Delta^2$.
}
\label{fig:2drmt2}
\end{figure}

We also plot the case of $\tau=(0,\tau_2)$ in Fig.~\ref{fig:2drmt2}.
Equation (\ref{kav-rmt2}) is dependent not only on $\omega$ but also on $E$ and 
we take the average by $\rho_E(E)\propto \exp[-4A_{\beta_\mathrm{D}}(E/\Delta)^2]$
as well as by $\rho_\omega(\omega)$.
In this case, the asymptotic form of $\langle\overline{K}(\tau_2)\rangle$
at large $\tau_2$ is $\langle\overline{K}(\tau_2)\rangle\sim 1/\tau_2$
and is independent of $\beta_\mathrm{D}$.

%%%%%%%%%%%%%%%%%%%%%%%%%%%%%%%%%%%%%%%
%%%%%%%%%%%%%%%%%%%%%%%%%%%%%%%%%%%%%%%
%%%%%%%%%%%%%%%%%%%%%%%%%%%%%%%%%%%%%%%
%%%%%%%%%%%%%%%%%%%%%%%%%%%%%%%%%%%%%%%

%%%%%%%%%%%%%%%%%%%%%%%%%%%%%%%%%%%%%%%%%%%%%%%%%%%%%
\subsection{Large dimensional random matrices}

%%%%%%%%%%%%%%%%%%%%%%%%%%%%%%%%%%%%%%%
\begin{figure}[t]
\centering\includegraphics[width=1.\columnwidth]{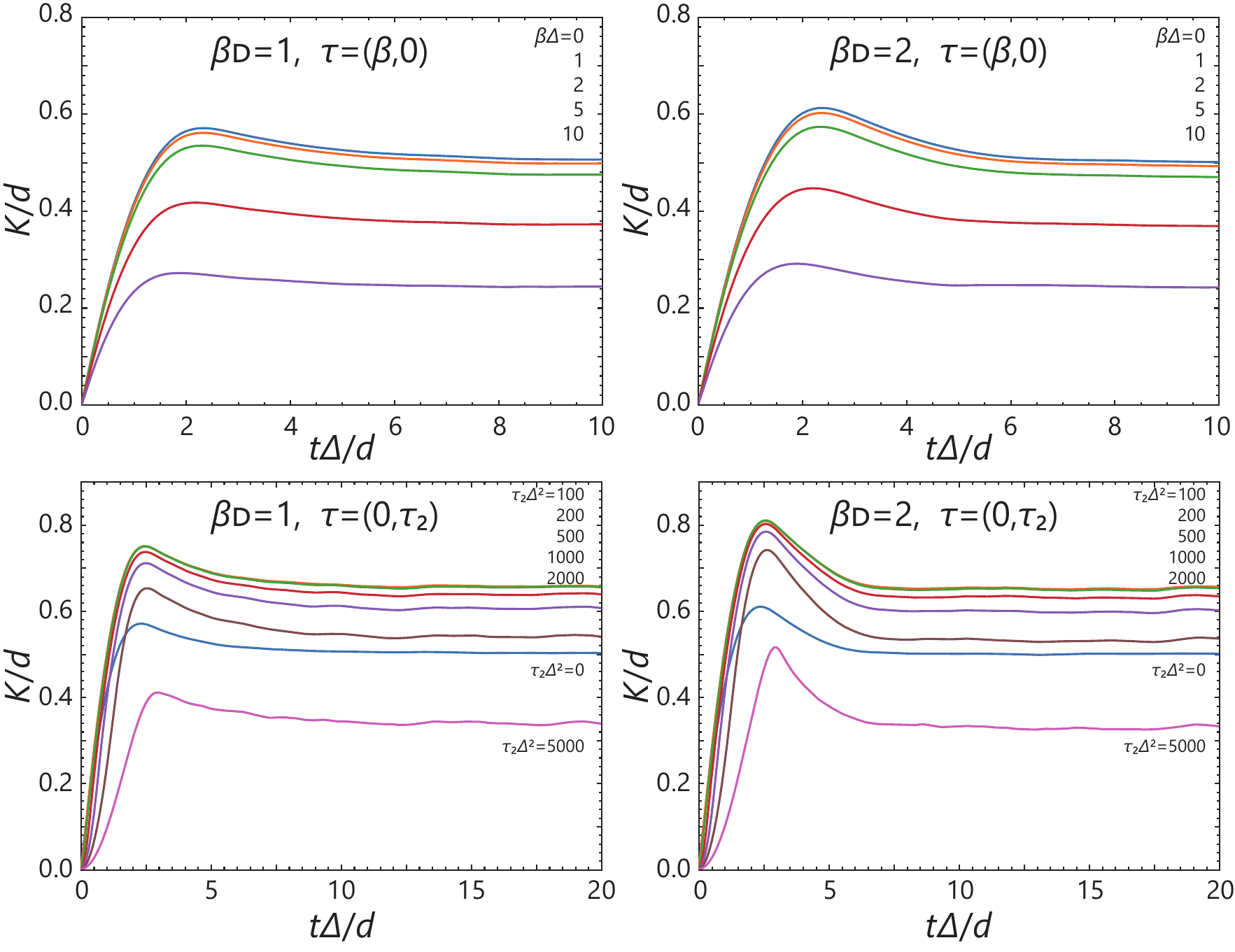}
\caption{
The spread complexity for the random matrix ensembles.
The size of the matrix is $d=1000$ 
and the average is taken over 100 samples.
We use the orthogonal ensemble $\beta_\mathrm{D}=1$ in the left panels
and the unitary ensemble $\beta_\mathrm{D}=2$ in the right panels. 
The upper panels represent $\tau=(\beta,0)$
with several values of $\beta$, 
and the lower panels represent $\tau=(0,\tau_2)$.
}
\label{fig:rmt-k}
\end{figure}
%%%%%%%%%%%%%%%%%%%%%%%%%%%%%%%%%%%%%%%
%%%%%%%%%%%%%%%%%%%%%%%%%%%%%%%%%%%%%%%
\begin{figure}[t]
\centering\includegraphics[width=1.\columnwidth]{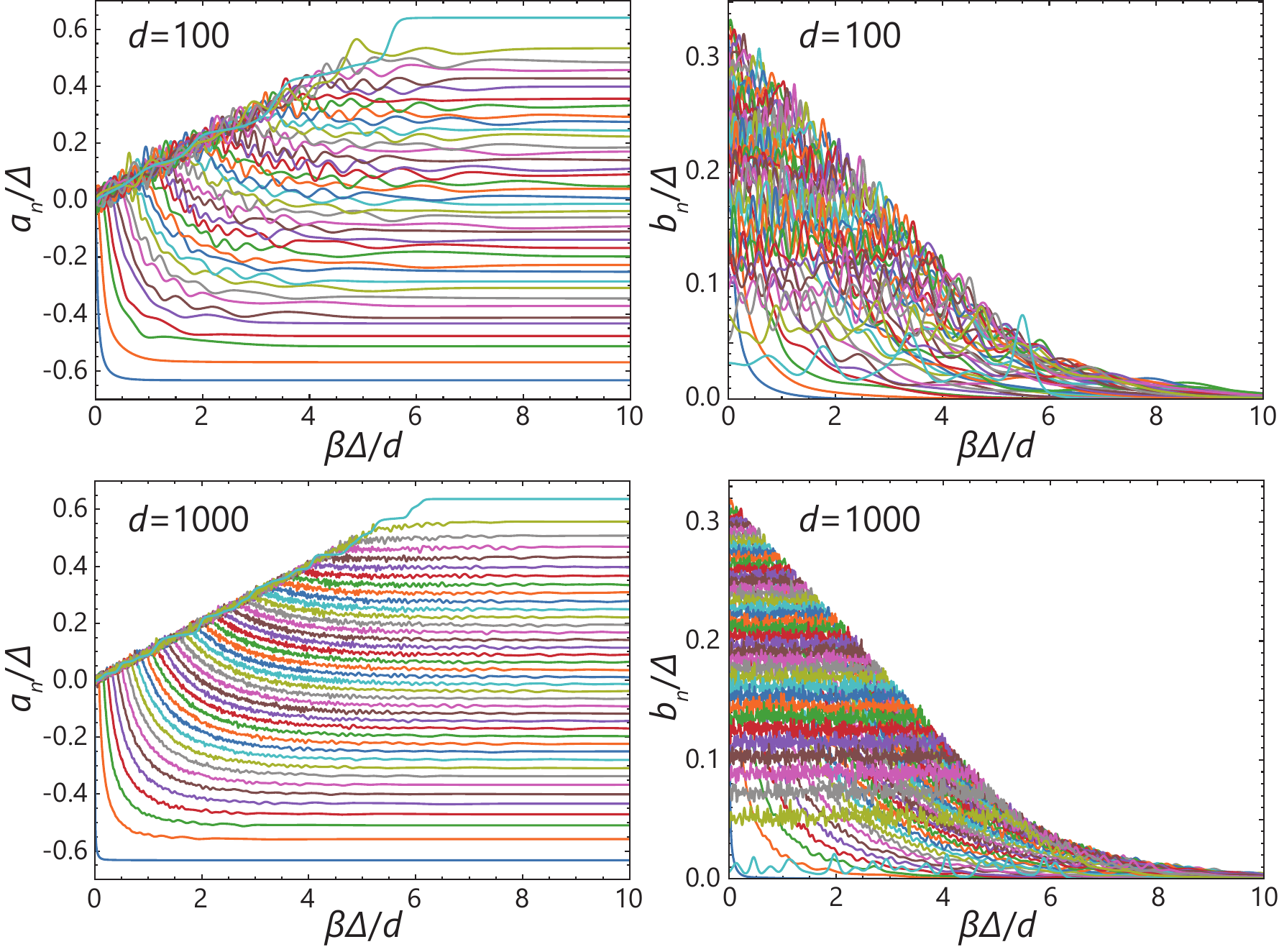}
\caption{
The Lanczos coefficients $a_n(\beta)$ (left panels) and 
$b_n(\beta)$ (right)
for a single realization of the unitary ensemble and $\tau=(\beta,0)$.
We take the matrix size $d=100$ in the upper panels and 
$d=1000$ in the lower panels.
In each panel, we plot the coefficients for 40 values of $n$ 
selected at regular intervals from the range of $n$.
}
\label{fig:rmt1-ab}
\end{figure}
%%%%%%%%%%%%%%%%%%%%%%%%%%%%%%%%%%%%%%%
%%%%%%%%%%%%%%%%%%%%%%%%%%%%%%%%%%%%%%%
\begin{figure}[t]
\centering\includegraphics[width=1.\columnwidth]{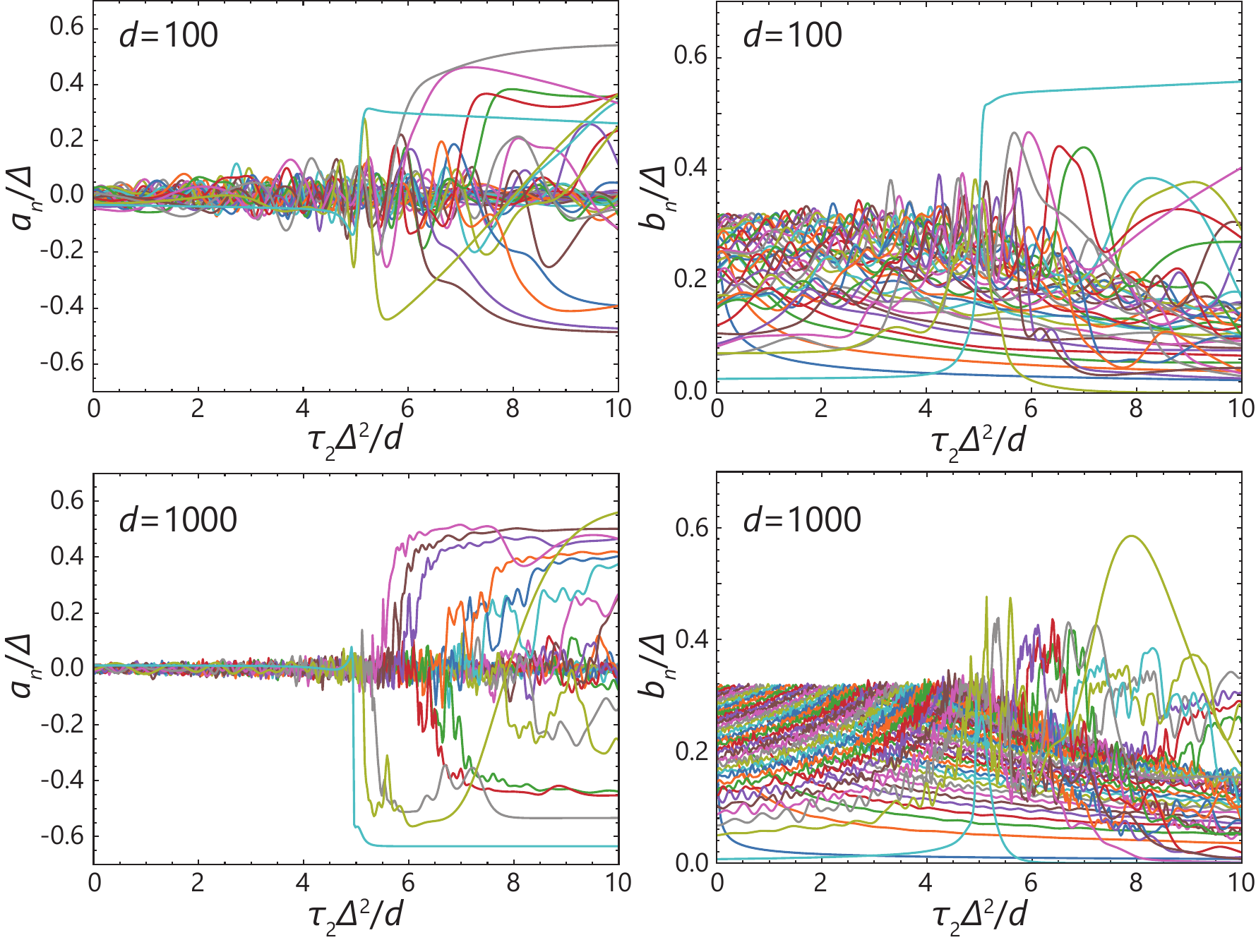}
\caption{
The Lanczos coefficients for $\tau=(0,\tau_2)$
with $\tau_2\Delta^2\sim d$.
See the caption in Fig.~\ref{fig:rmt1-ab} for the other details.
}
\label{fig:rmt2-ab1}
\end{figure}
%%%%%%%%%%%%%%%%%%%%%%%%%%%%%%%%%%%%%%%
%%%%%%%%%%%%%%%%%%%%%%%%%%%%%%%%%%%%%%%
\begin{figure}[t]
\centering\includegraphics[width=1.\columnwidth]{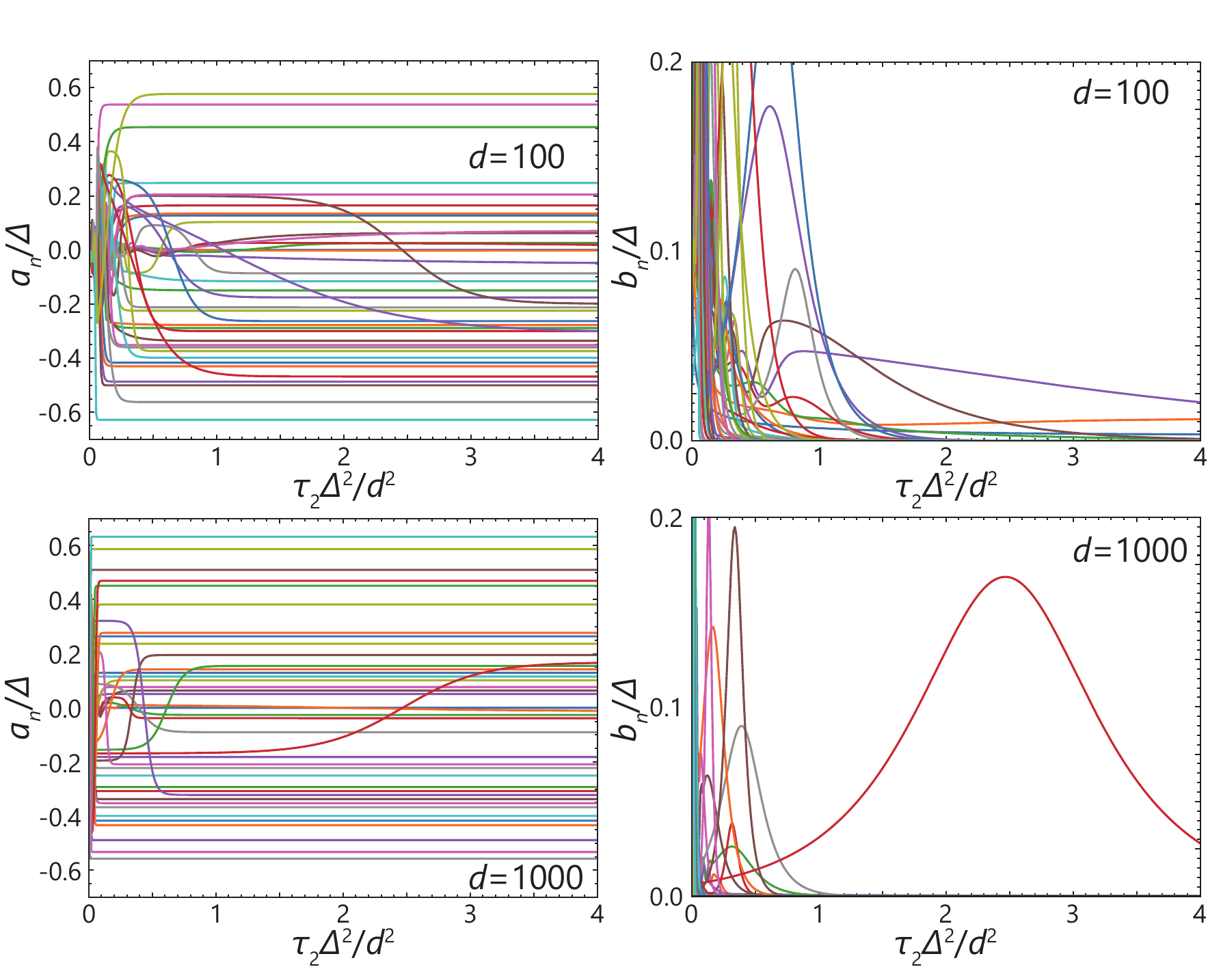}
\caption{
The Lanczos coefficients for $\tau=(0,\tau_2)$
with $\tau_2\Delta^2\sim d^2$.
See the caption in Fig.~\ref{fig:rmt1-ab} for the other details.
}
\label{fig:rmt2-ab2}
\end{figure}
%%%%%%%%%%%%%%%%%%%%%%%%%%%%%%%%%%%%%%%
%%%%%%%%%%%%%%%%%%%%%%%%%%%%%%%%%%%%%%%
\begin{figure}[t]
\centering\includegraphics[width=1.\columnwidth]{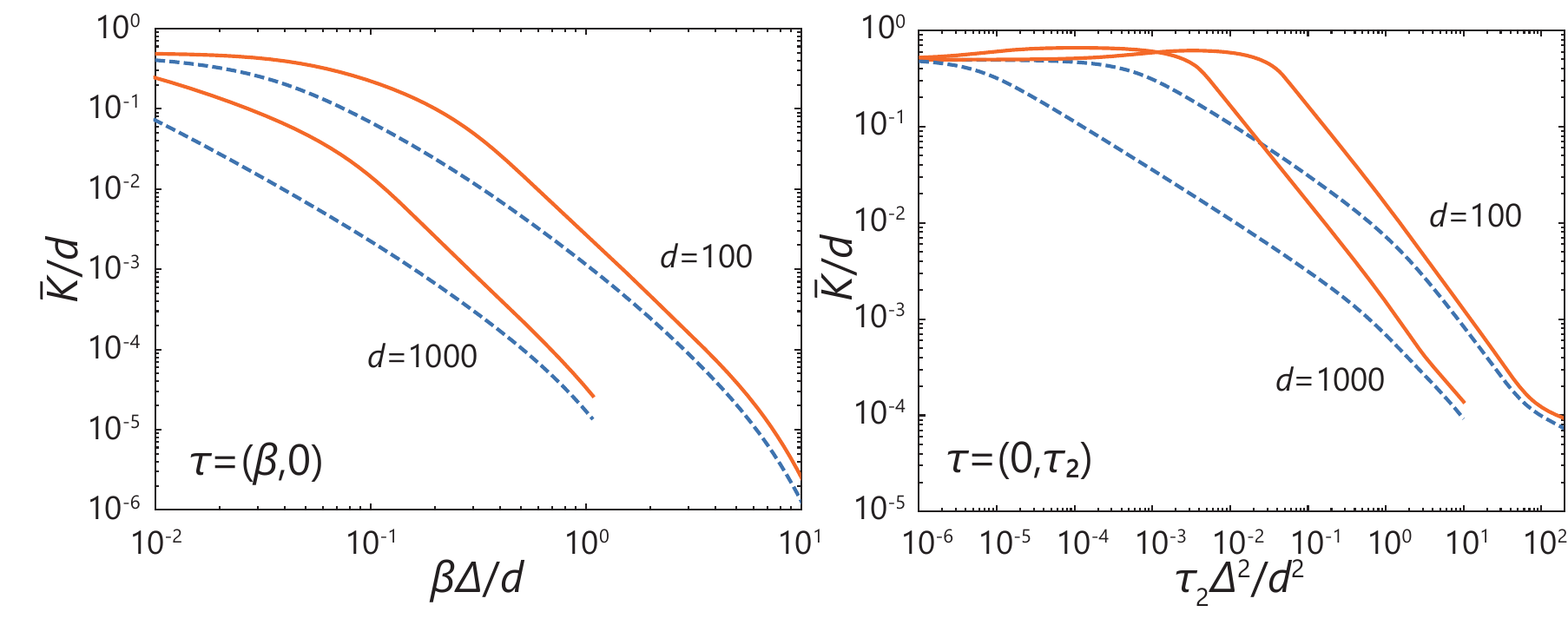}
\caption{
The time-averaged complexity $\overline{K}(\tau)$ (solid lines) and 
$\langle n\rangle=\sum_n n e^{-f(\epsilon_n,\tau)+S_n}/Z(\tau)$ (dashed lines)
for random matrices of the unitary ensemble.
We take $d=100$ and $d=1000$, and each result is averaged over 100 samples.
The left panel denotes the case $\tau=(\beta,0)$
and the right denotes $\tau=(0,\tau_2)$.
}
\label{fig:rmt-tav}
\end{figure}
%%%%%%%%%%%%%%%%%%%%%%%%%%%%%%%%%%%%%%%

Next, we consider large random matrices.
The random-matrix ensemble is produced from the probability distribution 
\be
P(H)\propto \exp\left(-\frac{d\beta_\mathrm{D}\pi^2}
{4\Delta^2}\mathrm{Tr}\,H^2\right). \label{ensemble}
\ee
Here, we consider the orthogonal ensemble ($\beta_\mathrm{D}=1$) and 
the unitary ensemble ($\beta_\mathrm{D}=2$)
and denote the size of $H$ as $d$. For generic RMT, the analytical structure of the Lanczos coefficients in this limit is predicted in \cite{Bhattacharjee:2024yxj}. The asymptotic form of the average density of states at $d\to\infty$ 
is given by 
\be
 \rho(E)=\left<\sum_{n=0}^{d-1}\delta(E-\epsilon_n)\right>
 \to \frac{d}{\Delta}\sqrt{1-\left(\frac{\pi E}{2\Delta}\right)^2},
 \label{dos-rmt}
\ee
denoting that $\Delta/d$ is the mean level spacing 
at $E=0$~\cite{haake1991quantum}.

According to the probability distribution in  Eq.~(\ref{ensemble}),
we numerically produce a $d\times d$ random matrix $H$ and 
calculate the eigenvalue set $\{E_\mu\}_{\mu=1}^{d}$.
Since the random matrix exhibits level repulsion, there is no accidental degeneracy, and we have $D=d$.
For $|\psi_0(\tau)\rangle$, we apply the Krylov algorithm
to construct $L(\tau)$.
Then, the spread complexity $K(t,\tau)$ is obtained from 
the general formula in Eq.~(\ref{complexity}), 
or Eq.~(\ref{complexity2}).
The result is averaged over random instances.
We note that the result is independent of the order of eigenvalues 
$\{E_\mu\}_{\mu=1}^{d}$.
The eigenvalue set $\{\epsilon_n\}_{n=0}^{d-1}$ aligns 
in the described order.

We show our numerical results of the complexity in Fig.~\ref{fig:rmt-k}.
We consider both cases of $\tau=(\beta,0)$ and $\tau=(0,\tau_2)$.
The matrix size is $d=1000$. 
Although the displayed result is averaged over 100 samples, 
we can find a similar behavior for every single realization.
For a fixed $\tau$, we find that the maximum value of $K(t,\tau)$
is dependent on the ensemble and the saturation value 
at $t\to\infty$, $\langle\overline{K}(\tau)\rangle$, is not.
The saturation value decays monotonically as a function of $\beta$
for $\tau=(\beta,0)$, while it gives a maximum 
at an intermediate value of $\tau_2$ for $\tau=(0,\tau_2)$.
Since the saturation value is independent of the ensemble,
we treat the unitary ensemble in the following analysis.

Before studying how $\overline{K}(\tau)$ depends on $d$ and $\tau$, 
we calculate the Lanczos coefficients $a_n(\tau)$ and $b_n(\tau)$
for a single realization of $H$.
The initial values at $\tau=0$ are obtained from the Krylov algorithm 
and the $\tau$-dependence is obtained by solving 
the Toda equations numerically.
The results are plotted in Fig.~\ref{fig:rmt1-ab} for $\tau=(\beta,0)$, 
and in Figs.\ref{fig:rmt2-ab1} and \ref{fig:rmt2-ab2}
for $\tau=(0,\tau_2)$.

Figure \ref{fig:rmt1-ab} for $\tau=(\beta,0)$ shows 
a convergence to the fixed point
$(0,\dots,0;\epsilon_0,\dots,\epsilon_{d-1})$ with
$\epsilon_0<\dots<\epsilon_{d-1}$.
The eigenvalues distribute according to Eq.~(\ref{dos-rmt}) and 
the minimum energy gap is the order of $\Delta/d$.
Equation (\ref{as1}) denotes that the scale of 
the convergence is determined by $\beta\Delta/d$, which is consistent 
with the result in Fig.~\ref{fig:rmt1-ab}.
The global structure of the Lanczos coefficients at large $d$ 
is determined by $\beta\Delta/d$.
Equation (\ref{kavn}) is justified for $\beta\Delta\gg d$.

On the other hand, 
the result for $\tau=(0,\tau_2)$ shows a more complicated behavior.
Figure \ref{fig:rmt2-ab1} represents plots of the Lanczos coefficients
as functions of $\tau_2\Delta^2/d$.
This scale is convenient for finding a drastic structural change of 
the coefficients around $\tau_2\Delta^2/d\sim 5$.

To find the convergence to the fixed point, 
we require $\tau_2\Delta^2\gg d^2$.
Equation~(\ref{as2}) shows that 
$\tau_2$ must be the order of $1/\min_n(\epsilon_n^2-\epsilon_{n-1}^2)$.
For large $\tau_2$, the dominant contribution comes from eigenstates
at the band center $E\sim 0$, where $\epsilon_n^2-\epsilon_{n-1}^2$
is the order of $\Delta^2/d^2$.
This estimation is consistent with the result in Fig.~\ref{fig:rmt2-ab2}.

In Fig.~\ref{fig:rmt-tav}, we plot $\langle\overline{K}(\tau)\rangle$ 
for $d=100$ and $d=1000$.
We consider both cases of $\tau=(\beta,0)$ and $\tau=(0,\tau_2)$.
As we expect from the results of the Lanczos coefficients, 
Eq.~(\ref{kavn}), $K_n(\tau)\sim n$, holds for 
$\beta\Delta\gg d$ or $\tau_2\Delta^2\gg d^2$.
We note that the condition $d\gg 1$ is not required to find
Eq.~(\ref{kavn}).

Each result in Fig.~\ref{fig:rmt-tav} implies a power-law decay
at $\beta\Delta \gg d\gg 1$ or $\tau_2\Delta^2 \gg d^2\gg 1$. 
However, $\overline{K}(\tau)$ is negligibly small in this domain 
and it is difficult to access numerically.
We find that the largest accessible value of $\beta\Delta$ is around 1000.
It represents $\beta\Delta/d\sim 1$ for $d=1000$.
We can find a slower decay for the case of $\tau=(0,\tau_2)$
and the result can be plotted until around $\tau_2\Delta^2\sim 10^7$.

We can find a power-law behavior at large $\tau$ from the analytical expression.
For $\tau=(\beta,0)$, the energy levels align in ascending order.
We use 
\be
 \frac{n}{d-1}\sim
 \frac{1}{d}\int_{E_\mathrm{min}}^E dE'\,\rho(E') \label{rhon}
\ee
to write 
\be
 \frac{\langle\overline{K}(\beta)\rangle}{d-1}
 \sim\frac{\frac{1}{d}
 \int_{E_\mathrm{min}}^{E_\mathrm{max}} dE\,e^{-\beta E}\rho(E)
 \int_{E_\mathrm{min}}^E dE'\,\rho(E')}
 {\int_{E_\mathrm{min}}^{E_\mathrm{max}} dE\,e^{-\beta E}\rho(E)}.
\ee
Here, $E_\mathrm{min}=-2\Delta/\pi$ is the minimum energy 
and $E_\mathrm{max}=2\Delta/\pi$ is the maximum.
For $\beta\Delta\gg 1$, the dominant contribution to the integral 
comes from near the edge of the spectrum $E\sim E_\mathrm{min}$.
Using 
\be
 \frac{\rho(E)}{d}\sim \frac{1}{\Delta}\sqrt{\frac{\pi}{\Delta}
 (E-E_\mathrm{min})},
\ee
we can find 
\be
 \frac{\langle\overline{K}(\beta)\rangle}{d-1}
 &\sim& \frac{\pi^{1/2}}{\Delta^{3/2}}
 \frac{\int_{0}^{\infty} dE\,e^{-\beta E}E^{1/2}
 \int_{0}^{E} dE'\,E'^{1/2}}
 {\int_{0}^{\infty} dE\,e^{-\beta E}E^{1/2}}
 \no\\
 &\propto& \frac{1}{(\beta\Delta)^{3/2}}.
\ee
This behavior is independent of the symmetry class.

A similar calculation is possible for $\tau=(0,\tau_2)$.
The eigenvalues are sorted in ascending order 
with respect to their absolute values.
For large-$\tau_2$, the dominant contribution to the integral over $E$
comes from $E\sim 0$.
We can use 
\be
 \frac{n}{d-1}\sim \frac{1}{d}\int_{-|E|}^{|E|} dE'\,\rho(E'),
\ee
to write 
\be
 \frac{\langle\overline{K}(\tau_2)\rangle}{d-1}
 &\sim& \frac{\frac{2}{d}
 \int_{0}^{E_\mathrm{max}} dE\,e^{-\tau_2 E^2}\rho(E)
 \int_{0}^{E} dE'\,\rho(E')}
 {\int_{0}^{E_\mathrm{max}} dE\,e^{-\tau_2 E^2}\rho(E)} \no\\
 &\sim& \frac{2}{\Delta}\frac{
 \int_{0}^{\infty} dE\,e^{-\tau_2 E^2}E}
 {\int_{0}^{\infty} dE\,e^{-\tau_2 E^2}} \propto\frac{1}{(\tau_2\Delta^2)^{1/2}}.
\ee
Our numerical results show that the range of the plot is not enough to find 
these asymptotic forms.

%%%%%%%%%%%%%%%%%%%%%%%%%%%%%%%%%%%%%%%%%%%%%%%%%%%%%%%%%%%%%%%%%%
\subsection{Random matrices with chiral symmetry}

%%%%%%%%%%%%%%%%%%%%%%%%%%%%%%%%%%%%%%%
\begin{figure}[t]
\centering\includegraphics[width=1.\columnwidth]{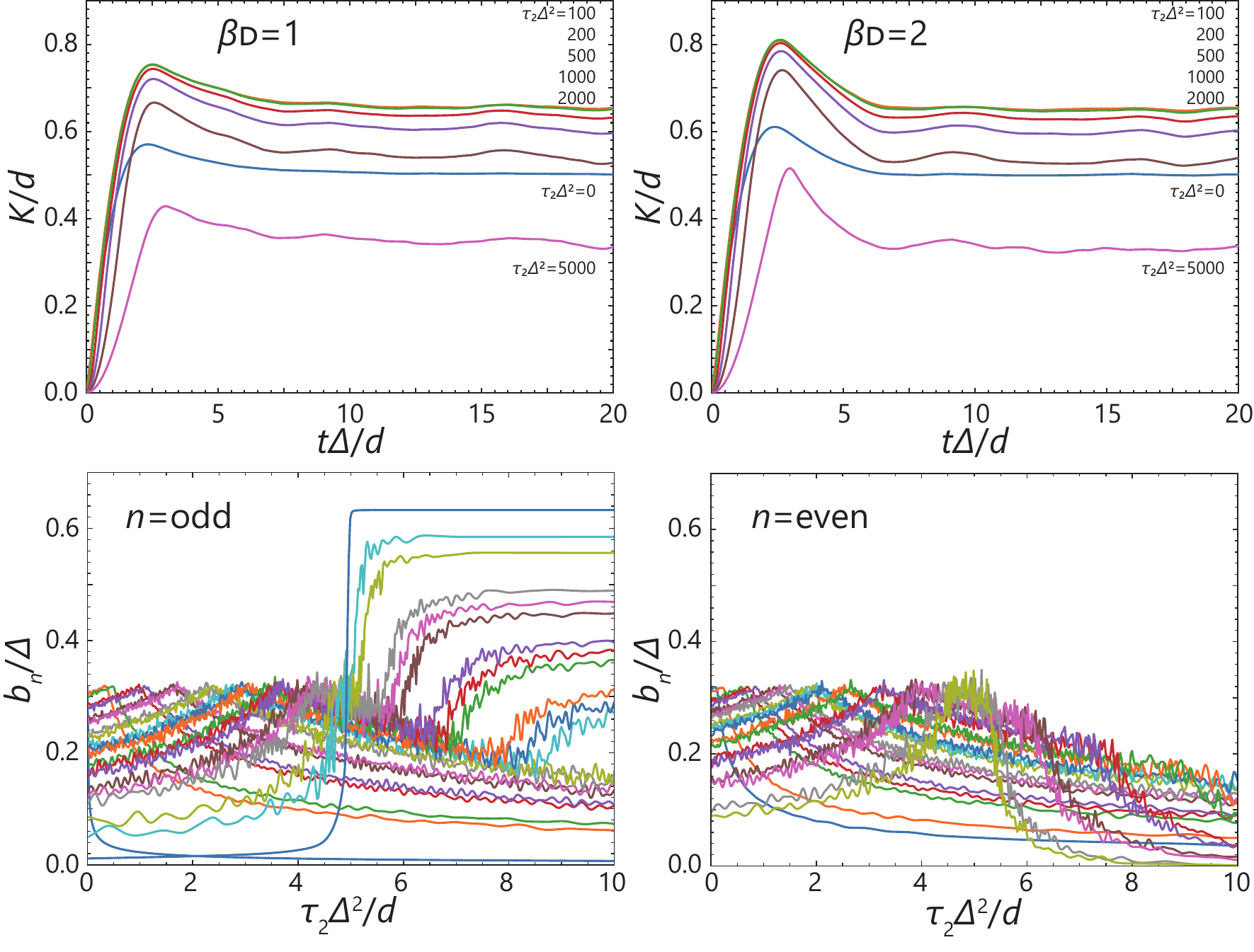}
\caption{
The upper panels represent $\langle K(t,\tau)\rangle$
for the chiral random matrix ensembles.
We take $d=1000$, $\tau=(0,\tau_2)$, and the average is taken over 100 samples.
The lower panels represent the Lanczos coefficients.
We note $a_n=0$ in the present setting.
}
\label{fig:crmt}
\end{figure}
%%%%%%%%%%%%%%%%%%%%%%%%%%%%%%%%%%%%%%%

In the standard choices of the random matrix ensemble,
the eigenvalues satisfy $|\epsilon_m|\ne |\epsilon_n|$ for $m\ne n$.
It is interesting to study random matrices with symmetric spectrum 
for $\tau=(0,\tau_2)$.
For random matrices described by the probability distribution 
in Eq.~(\ref{ensemble}), we impose the form
\be
 H = \left(\begin{array}{cc}
 0 & W \\ W^\dag & 0 \end{array}\right),
\ee
where $W$ represents a complex matrix with the size $d/2\times d/2$ 
for even $d$.
Then, we can find random eigenvalues 
$(\pm\epsilon_1,\pm\epsilon_2,\dots,\pm \epsilon_{d/2})$.
The corresponding Gaussian ensembles are known as 
chiral random matrix theory~\cite{Verbaarschot93, Verbaarschot94}
and belong to the tenfold classification of Hermitian 
random matrices~\cite{Altland97}.

We show our numerical results in Fig.~\ref{fig:crmt}.
Since the density of state in Eq.~(\ref{dos-rmt}) 
is basically unchanged by the symmetrization, 
we find the same results for $K(t,\tau)$ 
as the lower panels in Fig.~\ref{fig:rmt-k}.

Due to the symmetric distribution of the eigenvalues, 
we can find $a_n=0$ and the Toda equations reduce to Eq.~(\ref{toda2-a0}).
The fixed point is given by 
$(0,\dots,0;\epsilon_1,0,\epsilon_2,0,\dots,0,\epsilon_{d/2})$ with 
$0<\epsilon_1<\epsilon_2<\dots<\epsilon_{d/2}$.
The lower panels in Fig.~\ref{fig:crmt} show that 
the structural change of the coefficients is described by the scale
$\tau_2\Delta/d$ as we see in Fig.~\ref{fig:rmt2-ab1},
while the convergence to the fixed point is described by $\tau_2\Delta/d^2$.
We can also confirm that $\overline{K}(\tau)$ shows a similar behavior.
The only difference is that $\overline{K}(\tau)\to 1/2$ at $\tau_2\to\infty$. 
This is due to the structure of the fixed point, where many $2\times 2$ matrices
are obtained.
Numerically, $\overline{K}(\tau)/(d-1)$ for $\tau_2\Delta\gg d^2\gg 1$
is small and the difference from the standard random matrix is negligible.

It is known in the chiral random matrix theory 
that the density of states at $E=0$ has 
a universal structure~\cite{Verbaarschot93, Verbaarschot94}.
However, we require a very large $\tau_2$ and 
the complexity is not an ideal quantity for finding such a structure.

%%%%%%%%%%%%%%%%%%%%%%%%%%%%%%%%%%%%%%%%%%%%%%%%%%%%%%%%%%%%%%%%%%%%%%%%%%%%%%%
\section{Supersymmetry for quadratic deformations}
\label{sec:susy}

%%%%%%%%%%%%%%%%%%%%%%%%%%%%%%%%%%%%%%%%%%%%%%%%%%%%%%%%%%%%%%%%%%%%%%%%%%%%%%%
\subsection{Supersymmetric quantum mechanics}

The quadratic deformation $f(H,\tau)=\tau_2 H^2$ fits well with
the squared form of the Hamiltonian.
In supersymmetric quantum mechanics~\cite{Witten81,Cooper83,Cooper95}, 
we introduce a set of Hamiltonians $H_+ = A^\dag A$ and $H_- = AA^\dag$ 
for any operator $A$.
These Hamiltonians can be written as 
\be
 Q^2 = \left(\begin{array}{cc} H_+ & 0 \\ 0 & H_- \end{array}\right), 
\ee
where
\be
 Q = \left(\begin{array}{cc} 0 & A^\dag \\ A & 0 \end{array}\right).
\ee
When $H_+$ has an energy eigenstate $|n_+\rangle$ with a positive eigenvalue $E_n>0$,
$A|n_+\rangle$ represent an energy eigenstate of $H_-$ with the same eigenvalue $E_n$.
For the Hamiltonian $H_+$ and an initial state $|\psi_{0+}\rangle$, 
we consider the time evolution $|\psi_+(t)\rangle = e^{-iH_+t}|\psi_{0+}\rangle$.
Then, we introduce 
\be
 |\psi_{0-}\rangle = \frac{A|\psi_{0+}\rangle}{\sqrt{\langle\psi_{0+}|H_+|\psi_{0+}\rangle}}.
\ee
The time evolution of this state with $H_-$, 
$|\psi_-(t)\rangle = e^{-iH_-t}|\psi_{0-}\rangle$, is given by 
\be
 |\psi_{-}(t)\rangle = \frac{A|\psi_{+}(t)\rangle}{\sqrt{\langle\psi_{0+}|H_+|\psi_{0+}\rangle}}.
\ee

The state $|\psi_{0-}\rangle$ is taken as the first Krylov vector.
We prepare the initial state
\be
 |K_0\rangle = \left(\begin{array}{c} |\psi_{0+}\rangle \\ 0 \end{array}\right),
\ee
and construct the Krylov series 
\be
 |K_{n+1}\rangle b_{n+1} = Q|K_n\rangle - |K_{n-1}\rangle b_{n}.
\ee
We note that $a_n=0$ in this case.
The first Krylov basis represents the initial state for $H_-$, that is, 
\be
 |K_1\rangle = \left(\begin{array}{c} 0 \\ |\psi_{0-}\rangle \end{array}\right).
\ee
For a given initial state $|\Psi_0\rangle$, the time-evolved state is constructed as 
\be
 |\Psi(t)\rangle = e^{-iQ^2t}|\Psi_0\rangle.
\ee
We have 
\be
 |\Psi(t)\rangle = \left\{\begin{array}{cc}
 \left(\begin{array}{c} |\psi_{+}(t)\rangle \\ 0 \end{array}\right) & 
 \mbox{for}\quad |\Psi_0\rangle=|K_0\rangle, \\
 \left(\begin{array}{c} 0 \\ |\psi_{-}(t)\rangle \end{array}\right) & 
 \mbox{for}\quad |\Psi_0\rangle=|K_1\rangle.
 \end{array}\right.
\ee
We note that $|\psi_+(t)\rangle$ is represented by the even-order vectors of the Krylov expansion 
and $|\psi_-(t)\rangle$ by the odd-order vectors.
We expand the time-evolved state by $F=\sum_n |K_n\rangle\langle n|$ as 
$|\Psi(t)\rangle = F|\Phi(t)\rangle$ to define the vector $|\Phi(t)\rangle$ in the Krylov space.
We obtain 
\be
 i\partial_t |\Phi(t)\rangle = L^2|\Phi(t)\rangle.
\ee
The initial condition is given by 
$|\Psi(0)\rangle = (1,0,0,\dots)^\mathrm{T}$ or
$|\Psi(0)\rangle = (0,1,0,\dots)^\mathrm{T}$.
Since the diagonal components of the tridiagonal $L$-matrix are all equal to zero,
$L^2$ is written as 
\be
 L^2 = P_\mathrm{even}X^\dag X P_\mathrm{even}+P_\mathrm{odd}XX^\dag P_\mathrm{odd},
\ee
where $P_\mathrm{even}$ is the projection operator onto even-order space ($0,2,4,\dots$), 
$P_\mathrm{odd}$ is the projection onto odd space ($1,3,5,\dots$), and 
\be
 X = \left(\begin{array}{cccc}
 b_1 & b_2 & 0 & \\ 0 & b_3 & b_4 & \\
 0 & 0 & b_5 & \\ &&& \ddots \end{array}\right).
\ee
This upper tridiagonal matrix $X$ is nothing but the Krylov transformation of the operator $A$ 
introduced as $AF=FX$.
Each time the evolution of $\pm$ systems is transformed to that in the Krylov space 
\be
 && |\varphi_+(t)\rangle = e^{-iX^\dag Xt}|0\rangle, \\
 && |\varphi_-(t)\rangle = e^{-iXX^\dag t}|0\rangle.
\ee
These states are related to each other as 
\be
 |\varphi_-(t)\rangle = \frac{1}{b_1}X|\varphi_+(t)\rangle. \label{varphipm}
\ee

We note that the Krylov dimension is determined by the number of eigenstates 
included in the initial state.
The spectrum of $H_+$ equals that of $H_-$, 
except for the zero-energy of the $H_+$ system.
Since the zero-energy state is defined as $A|0\rangle=0$, it cannot be transformed to the $H_-$ system.
Thus, when $|\psi_{0+}\rangle$ involves $d_+$ eigenstates with $E_n>0$, 
the total Krylov dimension is given by $d=2d_+$.
When $|\psi_{0+}\rangle$ involves $d_++1$ eigenstates including  $|0\rangle$, 
we obtain $d=2d_++1$.
In the latter case, $X$ is a rectangular matrix $d_+\times (d_++1)$.

Since the matrix $X$ does not commute with
the complexity operator ${\cal K}=\mathrm{diag}\,(0,1,\dots)$, 
we see from Eq.~(\ref{varphipm}) that the spread complexity of the $H_+$ system
is different from that of the $H_-$ system.
It is known that certain kinds of systems have the property
\be
 H_-(\alpha)=H_+(\alpha_1)+\epsilon(\alpha_1),
\ee
where $\alpha$ represents a set of parameters,
$\alpha_1=f(\alpha)$ is a function of $\alpha$, and $\epsilon(\alpha_1)$ is a positive function.
This shape invariance allows us to obtain the complete spectrum of the system~\cite{Gendenshtein83,Cooper95}.
In the Krylov subspace method, the tridiagonal matrix $L$ representing the generator of the time evolution depends not only on the original Hamiltonian
but also on the initial state,
which indicates that we require additional conditions on the state. 
To find the invariance, we impose the condition 
\be
 |\psi_{0-}(\alpha)\rangle \propto |\psi_{0+}(\alpha_1)\rangle. \label{psi0pm}
\ee
Then, we obtain 
\be
 X(\alpha)X^\dag(\alpha)=X^\dag(\alpha_1)X(\alpha_1)+\epsilon(\alpha_1),
\ee
which is also written as 
\be
 && b_{2n+2}^2(\alpha)+b_{2n+1}^2(\alpha)=b_{2n+1}^2(\alpha_1)+b_{2n}^2(\alpha_1)+\epsilon(\alpha_1), \no\\ \\
 && b_{2n+1}(\alpha)b_{2n}(\alpha)=b_{2n}(\alpha_1)b_{2n-1}(\alpha_1).
\ee
The time-evolved state in the Krylov subspace is written as 
\be
 |\varphi_-(t,\alpha)\rangle = e^{-i\epsilon(\alpha_1)t}|\varphi_+(t,\alpha_1)\rangle.
 \label{varphipm-si}
\ee
This relation shows that the spread complexity of the $H_-$ system with $\alpha$ 
is equal to that of the $H_+$ system with $\alpha_1$:
\be
 K_-(t,\alpha)=K_+(t,\alpha_1).
\ee

We can also show that $X$ and $X^\dag$ represent lowering and raising operators.
Combining Eq.~(\ref{varphipm}) with Eq.~(\ref{varphipm-si}), we can write 
\be
 \frac{1}{b_1(\alpha)}X(\alpha)|\varphi_+(t,\alpha)\rangle
 = e^{-i\epsilon(\alpha_1)t}|\varphi_+(t,\alpha_1)\rangle.
\ee
For the eigenstates of $X^\dag (\alpha)X(\alpha)$, $|\phi_n(\alpha)\rangle$, satisfying 
$X^\dag(\alpha)X(\alpha)|\phi_n(\alpha)\rangle=|\phi_n(\alpha)\rangle\epsilon_n(\alpha)$,
we obtain 
\be
 && X(\alpha)|\phi_n(\alpha)\rangle = |\phi_{n-1}(\alpha_1)\rangle \sqrt{\epsilon_n(\alpha)}, \\
 && X^\dag(\alpha)|\phi_n(\alpha_1)\rangle = |\phi_{n+1}(\alpha)\rangle \sqrt{\epsilon_{n+1}(\alpha)}.
\ee

One example that exhibits shape invariance is the harmonic oscillator Hamiltonian.
When choosing the initial state to be a coherent state, we obtain Eq.~(\ref{psi0pm}).
However, in that case, we obtain $\alpha_1=\alpha$ and $H_+-H_-=\mathrm{const.}$.
The Lanczos coefficients are given by Eqs.~(\ref{HWa}) and (\ref{HWb}), and 
we can calculate everything analytically.
Many nontrivial examples of shape invariant Hamiltonians are known~\cite{Cooper95}.
It would be an interesting problem to find nontrivial examples
that give implementable forms of Eq.~(\ref{psi0pm}).

%%%%%%%%%%%%%%%%%%%%%%%%%%%%%%%%%%%%%%%%%%%%%%%%%%%%%%%%%%%%%%%%%%%%%%%%%%%%%%%
\subsection{Deformations}

We apply our deformation scheme to the supersymmetric system.
With the deformed initial state 
\be
 |\psi_{0\pm}(\tau_2)\rangle = \frac{e^{-H_\pm \tau_2/2}|\psi_{0\pm}\rangle}
 {\sqrt{\langle\psi_{0\pm}|e^{-H_\pm \tau_2}|\psi_{0\pm}\rangle}},
\ee
the time-evolved state is written as
$|\psi_{\pm}(t,\tau_2)\rangle = e^{-iH_\pm t}|\psi_{0\pm}(\tau_2)\rangle$. 
This is a quadratic deformation, and we can find the second Toda equations, 
Eq.~(\ref{toda2-2}), with $a_n=0$.
The Lax equation reads
\be
 && \partial_{\tau_2}(X^\dag(\tau_2)X(\tau_2)) = [M_+(\tau_2),X^\dag(\tau_2)X(\tau_2)], \\
 && \partial_{\tau_2}(X(\tau_2)X^\dag(\tau_2)) = [M_-(\tau_2),X(\tau_2)X^\dag(\tau_2)],
\ee
where 
\be
 && M_+ = \frac{1}{2}\left(\begin{array}{cccc}
 0 & -b_1b_2 & 0 & \\
 b_1b_2 & 0 & -b_3b_4 & \\
 0 & b_3b_4 & 0 & \\
 &&& \ddots
 \end{array}\right), \\
 && M_- = \frac{1}{2}\left(\begin{array}{cccc}
 0 & -b_2b_3 & 0 & \\
 b_2b_3 & 0 & -b_4b_5 & \\
 0 & b_4b_5 & 0 & \\
 &&& \ddots
 \end{array}\right).
\ee
These relations are reduced to the equation for $X$ as 
\be
 \partial_{\tau_2}X(\tau_2) = M_-(\tau_2)X(\tau_2)-X(\tau_2)M_+(\tau_2).
\ee

In the case of the shape invariant systems, 
we obtain $M_-(\tau_2,\alpha)=M_+(\tau_2,\alpha_1)$.

%%%%%%%%%%%%%%%%%%%%%%%%%%%%%%%%%%%%%%%%%%%%%%%%%%%%%%%%%%%%%%%%%%%%%%%%%%%%%%%
\subsection{Closed complexity algebra and exact solutions}

When we use Eqs.~(\ref{alt-1}) and (\ref{alt-2}) for the Lanczos coefficients, we can find closed complexity algebra and 
the spread complexity for the $\pm$-systems can be calculated analytically.

For ${\cal K}_\pm=\mathrm{diag}\,(0,1,2,\dots)$, we obtain 
\be
 && [X^\dag X, {\cal K}_+]= -2M_+, \\
 && [X X^\dag, {\cal K}_-]= -2M_-.
\ee
Taking the commutators again, we obtain 
\be
 [X^\dag X, M_+] &=& 2\alpha^2\gamma^2+8\alpha^2\gamma^2 {\cal K}_+
 \no\\ && 
 +(\alpha^2+\gamma^2)P_\mathrm{even}(X^\dag X)_\mathrm{od}P_\mathrm{even},
\ee
and 
\be
 [X X^\dag, M_-] &=& 6\alpha^2\gamma^2+8\alpha^2\gamma^2 {\cal K}_-
 \no\\ && 
 +(\alpha^2+\gamma^2)P_\mathrm{odd}(X X^\dag)_\mathrm{od}P_\mathrm{odd}.
\ee
Here, $Z_\mathrm{od}$ represents the matrix that takes the diagonal components of $Z$
to be zero.
We further obtain 
\be
 && [X^\dag X, [X^\dag X, M_+]] = 4(\alpha^2-\gamma^2)^2 M_+,\\
 && [X X^\dag, [X X^\dag, M_-]] = 4(\alpha^2-\gamma^2)^2 M_-.
\ee
Thus, taking the commutators twice, $M_\pm$ go back to the original matrix.
By using this property, we can calculate the spread complexity analytically.
For $+$-system, we obtain 
\be
 K_+(t,\tau_2)=2\alpha^2(\tau_2)\gamma^2(\tau_2)
 \left[\frac{\sin(\alpha^2(\tau_2)-\gamma^2(\tau_2))t}
 {\alpha^2(\tau_2)-\gamma^2(\tau_2)}\right]^2,
 \no\\
\ee
and for $-$-system 
\be
 K_-(t,\tau_2) = 3K_+(t,\tau_2).
\ee
The difference of the factor 3 comes from the relations  
$b_1^2b_2^2=2\alpha^2\gamma^2$ and $b_2^2b_3^2=6\alpha^2\gamma^2$.

%%%%%%%%%%%%%%%%%%%%%%%%%%%%%%%%%%%%%%%%%%%%%%%%%%%%%%%%%%%%%%%%%%%%%%%%%%%%%%%
\section{Summary}
\label{sec:summary}

We have investigated the dynamical evolution of deformed initial states 
using the Krylov subspace method. In our formulation, the deformation 
preserves the eigenvalues of the tridiagonal generator in Krylov space, 
and the resulting flow is always described by Eq.~(\ref{lax}), which takes 
the form of a Lax equation for classical nonlinear integrable systems. 
When the deformation is specified by Eq.~(\ref{f}) with a power-law function $f$, 
the Lanczos coefficients satisfy the generalized Toda equations.

Previous works have shown that particular classes of deformations lead to 
Toda equations for the Lanczos coefficients~\cite{Dymarsky20,Kundu23,angelinos25}. 
Here, we derive similar results from a different perspective and within a 
different setting, which allows for more direct physical applications.

In principle, if the Lanczos algorithm can be applied directly to the 
deformed state, one may obtain the solution without resorting to the Toda 
equations. However, the present framework becomes particularly useful when 
comparing the dynamics associated with different deformed states in a 
systematic manner. In this approach, several properties of the dynamics can 
be understood from the general structure of the Toda equations.

For the coherent Gibbs state, the Lanczos coefficients encode thermodynamic 
properties of the corresponding many-body system. In particular, we have 
shown that the saturation behavior of the spread complexity is closely 
related to thermodynamic quantities.

The results obtained for random matrix ensembles further demonstrate that 
the behavior at large deformation parameter $\tau$ is governed by the 
fixed-point structure of the Toda flow. This structure determines the 
ordering of the eigenvalues and leads to the relation in Eq.~(\ref{kavn}). 
The Toda equations also determine the characteristic scales controlling 
the $\tau$-evolution.

An interesting application of quadratic deformations arises in the Krylov expansion of Heisenberg operators. In this case, one introduces an operator inner product parametrized by a real variable, which generates a flow in parameter space. Because the diagonal elements of the tridiagonal matrix $L$ vanish, the first Toda equations do not contribute, and the second set of Toda equations provides the simplest description of the resulting flow.

Another promising direction is the extension to non-Hermitian Hamiltonians. While the Krylov method is naturally formulated for Hermitian generators, non-Hermitian deformations have been studied in other contexts~\cite{Matsoukas23}. 
It would therefore be interesting to generalize the deformation introduced in Eq.~(\ref{dos}) to the non-Hermitian case.

More broadly, the connection established 
here between Hamiltonian deformations, Krylov dynamics, and integrable 
Toda flows suggest that tools from integrable systems may provide a 
systematic framework for understanding operator growth and complexity in 
quantum many-body systems. Exploring these connections further may open 
new avenues for studying quantum dynamics, complexity, and nonequilibrium 
phenomena from a unified perspective.

{\it Acknowledgments.} 
We acknowledge the financial support 
from the Luxembourg National Research Fund (FNR Grant No. 16434093). 
This project has received funding from the QuantERA II Joint Programme with co-funding from the European Union’s Horizon 2020 research and innovation programme. 
KT further acknowledges support from JSPS KAKENHI Grant No. JP24K00547. 
The work of P.N. is supported by the JSPS Grant-in-Aid for Transformative Research Areas (A) ``Extreme Universe'' No. 21H05190, FWO-Vlaanderen Project No. G012222N, and the VUB Research Council through the Strategic Research Program High-Energy Physics.

%%%%%%%%%%%%%%%%%%%%%%%%%%%%%%%%%%%%%%%%%%%%%%%%%%%%%%%%%%%%%%%%%%%%%%%%%%%%%%%
\appendix
%%%%%%%%%%%%%%%%%%%%%%%%%%%%%%%%%%%%%%%%%%%%%%%%%%%%%%%%%%%%%%%%%%%%%%%%%%%%%%%
\section{Derivation of the Toda equations}
\label{sec:todacalc}

We derive the Toda equations, 
the first set in Eqs.~(\ref{toda1-1}) and (\ref{toda1-2}), 
and the second set in Eqs.~(\ref{toda2-1}) and (\ref{toda2-2}).
We follow the method in Ref.~\cite{Ismail05} where the simplest case $f(H,\tau)=\tau H$ is discussed.
The generalization to the quadratic form in Eq.~(\ref{f}) is a straightforward task.

We first introduce the monic polynomial  
\be
 \tilde{P}_n(x,\tau)=b_1(\tau)\cdots b_n(\tau)P_n(x,\tau). \label{monico}
\ee
Instead of losing the normalization, it keeps 
the coefficient of the highest degree of $\tilde{P}_n(x,\tau)$ to unity.
The recurrence relation is given by 
\be
 \tilde{P}_{n+1}(x,\tau)=(x-a_n(\tau))\tilde{P}_n(x,\tau)-b_n^2(\tau)\tilde{P}_{n-1}(x,\tau), \no\\
\ee
with $\tilde{P}_0(x,\tau)=1$.
The orthonormal relation is changed to
\be
 \int dx\,\rho(x,\tau)\tilde{P}_m(x,\tau)\tilde{P}_n(x,\tau)=h_n(\tau)\delta_{m,n},\label{orthomp}
\ee
where $h_n(\tau)= b_1^2(\tau)b_2^2(\tau)\cdots b_n^2(\tau)$.

The Toda equations can be derived by differentiating Eq.~(\ref{orthomp}) with respect to $\tau$.
Since it is enough to consider the case $m\le n$, we put $m=n-k$ with $k=0,1,\dots, n$.
We use the relations 
\be
 && \partial_{\tau_1}\rho(x,\tau) = -(x-a_0(\tau))\rho(x,\tau), \\
 && \partial_{\tau_2}\rho(x,\tau) = -[x^2-(b_1^2(\tau)+a_0^2(\tau))]\rho(x,\tau),
\ee
and the property that $\partial_\tau\tilde{P}_n(x,\tau)$ is a polynomial of degree at most $n-1$.
Using the recurrence relation and the orthonormal relation, we obtain 
\be
 && \int dx\,\rho(x,\tau)\tilde{P}_{n-k}(x,\tau)\partial_{\tau_1}\tilde{P}_n(x,\tau) \no\\
 &=& \delta_{k,0}[\partial_{\tau_1}h_n+h_n(a_n-a_0)]+\delta_{k,1}h_{n-1}b^2_{n}, \label{eq1} \\
 && \int dx\,\rho(x,\tau)\tilde{P}_{n-k}(x,\tau)\partial_{\tau_2}\tilde{P}_n(x,\tau) \no\\
 &=& \delta_{k,0}[\partial_{\tau_2}h_n+h_n(b_{n+1}^2+b_n^2+a_n^2-b_1^2-a_0^2)] \no\\
 && +\delta_{k,1}h_{n-1}b^2_{n}(a_n+a_{n-1})+\delta_{k,2}h_{n-2}b_n^2b_{n-1}^2. \label{eq2}
\ee
Taking $k=0$ for these relations gives 
\be
 && \partial_{\tau_1}h_n+h_n(a_n-a_0)=0, \\
 && \partial_{\tau_2}h_n+h_n(b_{n+1}^2+b_n^2+a_n^2-b_1^2-a_0^2)=0.
\ee
The first equation gives Eq.~(\ref{toda1-2}) and the second Eq.~(\ref{toda2-2}).
Equations (\ref{eq1}) and (\ref{eq2}) also give 
\be
 && \partial_{\tau_1}\tilde{P}_n =b^2_n\tilde{P}_{n-1}, \\
 && \partial_{\tau_2}\tilde{P}_n =b_n^2(a_n+a_{n-1})\tilde{P}_{n-1}
 +b_n^2b_{n-1}^2\tilde{P}_{n-2}. 
\ee
In a similar way, Eqs.~(\ref{toda1-1}) and (\ref{toda2-1})
can be derived from the derivative of the relation 
\be
 a_n(\tau)h_n(\tau)=\int dx\,\rho(x,\tau) x\tilde{P}_n^2(x,\tau).
\ee

We can also find that the $\tau$-evolution of 
$F(\tau)=\sum_{n=0}^{d-1}|K_n(\tau)\rangle\langle n|$
is described by $V^\dag(\tau)$.
To find the result, we examine the derivative of the Krylov basis.
At the zeroth order, we have 
\be
 && \partial_{\tau_1}|K_0(\tau)\rangle =
 -\frac{1}{2}(H-a_0(\tau))|K_0(\tau)\rangle, \\
 && \partial_{\tau_2}|K_0(\tau)\rangle =
 -\frac{1}{2}[H^2-(b_1^2(\tau)+a_0^2(\tau))]|K_0(\tau)\rangle. \no\\
\ee
We note that $|K_0(\tau)\rangle=|\psi_0(\tau)\rangle$.
The higher-order terms can be calculated from
$|K_n(\tau)\rangle=\tilde{P}_n(H,\tau)|K_0(\tau)\rangle/\sqrt{h_n(\tau)}$.
We obtain 
\be
 && \partial_{\tau_1} |K_n(\tau)\rangle
 = -\frac{1}{2}b_{n+1}(\tau)|K_{n+1}(\tau)\rangle
 +\frac{1}{2}b_n(\tau)|K_{n-1}(\tau)\rangle,
 \no\\ \\ 
 && \partial_{\tau_2} |K_n(\tau)\rangle
 = -\frac{1}{2}(a_n(\tau)+a_{n+1}(\tau))b_{n+1}(\tau)|K_{n+1}(\tau)\rangle
 \no\\ &&
 +\frac{1}{2}(a_{n-1}(\tau)+a_n(\tau))b_n(\tau)|K_{n-1}(\tau)\rangle
 \no\\ &&
 -\frac{1}{2}b_{n+1}(\tau)b_{n+2}(\tau)|K_{n+2}(\tau)\rangle
 +\frac{1}{2}b_{n-1}(\tau)b_{n}(\tau)|K_{n-2}(\tau)\rangle.
 \no\\ 
\ee
These relations give
\be
 \partial_{\tau_\mu} F(\tau) =-F(\tau)M_\mu(\tau),
\ee
and Eq.~(\ref{ftau}).

%%%%%%%%%%%%%%%%%%%%%%%%%%%%%%%%%%%%%%%%%%%%%%%%%%%%%%%%%%%%%%%%%%%%%%%%%%%%%%%
\section{Time average of expectation values in Krylov space}
\label{sec:timeav}

In this section, we derive Eq.~(\ref{kav}) for the coherent Gibbs state.
To broaden the applicability, 
we discuss the derivation in a more general setting.

For an operator ${\cal X}$ in Krylov space,
we consider the expectation value
\be
 X(t,\tau)=\langle\varphi(t,\tau)|{\cal X}|\varphi(t,\tau)\rangle,
\ee
and its time average 
\be
 \overline{X}(\tau) = \frac{1}{T}\int_0^T dt\,X(t,\tau).
\ee

Noting $|\varphi(t,\tau)\rangle=e^{-iL(\tau)t}|0\rangle$,
we introduce the eigenstates of $L(\tau)$ defined from Eq.~(\ref{lphi}).
The eigenvalue $\epsilon_n$ is given by one of the eigenvalues of the original Hamiltonian
and is independent of $\tau$.
The Lax equation (\ref{lax}) indicates that we can write 
\be
 \partial_{\tau_\mu}|\phi_n(\tau)\rangle =M_\mu(\tau)|\phi_n(\tau)\rangle. \label{mphi}
\ee
We can confirm that $|\phi_n(\tau)\rangle$ satisfying Eqs.~(\ref{lphi}) and (\ref{mphi}) is given by Eq.~(\ref{phin}).

Using the spectral decomposition, we have
\be
 X(t,\tau)&=&\sum_{m,n}e^{i(\epsilon_m-\epsilon_n)t}
 \langle 0|\phi_m(\tau)\rangle
 \no\\ && \times\langle\phi_m(\tau)|
 {\cal X}|\phi_n(\tau)\rangle\langle\phi_n(\tau)
 |0\rangle.
\ee
Taking the average over $t$ removes contributions with $m\ne n$.
We obtain 
\be
 \overline{X}(\tau)=\sum_{n}
 |\langle 0|\phi_n(\tau)\rangle|^2\langle\phi_n(\tau)|
 {\cal X}|\phi_n(\tau)\rangle.
\ee

We consider the case that $|\psi_0\rangle$ is given by
the uniform linear combination of the eigenstates of $H$.
By applying $\langle 0|$ to Eq.~(\ref{phin}) from the left
and using Eq.~(\ref{stateepsn}), we obtain 
\be
 \langle 0|\phi_n(\tau)\rangle
 =\langle \psi_0(\tau)|\epsilon_n\rangle 
 =\sqrt{\frac{d_n e^{-f(\epsilon_n,\tau)}}{\sum_n d_n e^{-f(\epsilon_n,\tau)}}}.
\ee
Thus, we find Eq.~(\ref{kav}).

\bibliography{bib-v2}

\begin{thebibliography}{81}%
\makeatletter
\providecommand \@ifxundefined [1]{%
 \@ifx{#1\undefined}
}%
\providecommand \@ifnum [1]{%
 \ifnum #1\expandafter \@firstoftwo
 \else \expandafter \@secondoftwo
 \fi
}%
\providecommand \@ifx [1]{%
 \ifx #1\expandafter \@firstoftwo
 \else \expandafter \@secondoftwo
 \fi
}%
\providecommand \natexlab [1]{#1}%
\providecommand \enquote  [1]{``#1''}%
\providecommand \bibnamefont  [1]{#1}%
\providecommand \bibfnamefont [1]{#1}%
\providecommand \citenamefont [1]{#1}%
\providecommand \href@noop [0]{\@secondoftwo}%
\providecommand \href [0]{\begingroup \@sanitize@url \@href}%
\providecommand \@href[1]{\@@startlink{#1}\@@href}%
\providecommand \@@href[1]{\endgroup#1\@@endlink}%
\providecommand \@sanitize@url [0]{\catcode `\\12\catcode `\$12\catcode `\&12\catcode `\#12\catcode `\^12\catcode `\_12\catcode `\%12\relax}%
\providecommand \@@startlink[1]{}%
\providecommand \@@endlink[0]{}%
\providecommand \url  [0]{\begingroup\@sanitize@url \@url }%
\providecommand \@url [1]{\endgroup\@href {#1}{\urlprefix }}%
\providecommand \urlprefix  [0]{URL }%
\providecommand \Eprint [0]{\href }%
\providecommand \doibase [0]{https://doi.org/}%
\providecommand \selectlanguage [0]{\@gobble}%
\providecommand \bibinfo  [0]{\@secondoftwo}%
\providecommand \bibfield  [0]{\@secondoftwo}%
\providecommand \translation [1]{[#1]}%
\providecommand \BibitemOpen [0]{}%
\providecommand \bibitemStop [0]{}%
\providecommand \bibitemNoStop [0]{.\EOS\space}%
\providecommand \EOS [0]{\spacefactor3000\relax}%
\providecommand \BibitemShut  [1]{\csname bibitem#1\endcsname}%
\let\auto@bib@innerbib\@empty
%</preamble>
\bibitem [{\citenamefont {Mari{\~n}o}(2015)}]{Marinyo15}%
  \BibitemOpen
  \bibfield  {author} {\bibinfo {author} {\bibfnamefont {M.}~\bibnamefont {Mari{\~n}o}},\ }\href {https://books.google.lu/books?id=vGugCgAAQBAJ} {\emph {\bibinfo {title} {Instantons and Large N: An Introduction to Non-Perturbative Methods in Quantum Field Theory}}}\ (\bibinfo  {publisher} {Cambridge University Press},\ \bibinfo {year} {2015})\BibitemShut {NoStop}%
\bibitem [{\citenamefont {Zamolodchikov}(2004)}]{Zamolodchikov04}%
  \BibitemOpen
  \bibfield  {author} {\bibinfo {author} {\bibfnamefont {A.~B.}\ \bibnamefont {Zamolodchikov}},\ }\bibfield  {title} {\bibinfo {title} {Expectation value of composite field $ t \bar t $ in two-dimensional quantum field theory}\ }\href {https://doi.org/10.48550/arXiv.hep-th/0401146} {10.48550/arXiv.hep-th/0401146} (\bibinfo {year} {2004})\BibitemShut {NoStop}%
\bibitem [{\citenamefont {Cavaglià}\ \emph {et~al.}(2016)\citenamefont {Cavaglià}, \citenamefont {Negro}, \citenamefont {Szécsényi},\ and\ \citenamefont {Tateo}}]{cavaglia2016}%
  \BibitemOpen
  \bibfield  {author} {\bibinfo {author} {\bibfnamefont {A.}~\bibnamefont {Cavaglià}}, \bibinfo {author} {\bibfnamefont {S.}~\bibnamefont {Negro}}, \bibinfo {author} {\bibfnamefont {I.~M.}\ \bibnamefont {Szécsényi}},\ and\ \bibinfo {author} {\bibfnamefont {R.}~\bibnamefont {Tateo}},\ }\bibfield  {title} {\bibinfo {title} {{$T\overline{T}$-deformed {2D} quantum field theories}},\ }\href {https://doi.org/10.1007/JHEP10(2016)112} {\bibfield  {journal} {\bibinfo  {journal} {Journal of High Energy Physics}\ }\textbf {\bibinfo {volume} {2016}},\ \bibinfo {pages} {112} (\bibinfo {year} {2016})}\BibitemShut {NoStop}%
\bibitem [{\citenamefont {Smirnov}\ and\ \citenamefont {Zamolodchikov}(2017)}]{SmirnovZ17}%
  \BibitemOpen
  \bibfield  {author} {\bibinfo {author} {\bibfnamefont {F.}~\bibnamefont {Smirnov}}\ and\ \bibinfo {author} {\bibfnamefont {A.}~\bibnamefont {Zamolodchikov}},\ }\bibfield  {title} {\bibinfo {title} {On space of integrable quantum field theories},\ }\href {https://doi.org/https://doi.org/10.1016/j.nuclphysb.2016.12.014} {\bibfield  {journal} {\bibinfo  {journal} {Nuclear Physics B}\ }\textbf {\bibinfo {volume} {915}},\ \bibinfo {pages} {363} (\bibinfo {year} {2017})}\BibitemShut {NoStop}%
\bibitem [{\citenamefont {Jiang}(2021)}]{jiang2021}%
  \BibitemOpen
  \bibfield  {author} {\bibinfo {author} {\bibfnamefont {Y.}~\bibnamefont {Jiang}},\ }\bibfield  {title} {\bibinfo {title} {A pedagogical review on solvable irrelevant deformations of 2d quantum field theory},\ }\href {https://doi.org/10.1088/1572-9494/abe4c9} {\bibfield  {journal} {\bibinfo  {journal} {Communications in Theoretical Physics}\ }\textbf {\bibinfo {volume} {73}},\ \bibinfo {pages} {057201} (\bibinfo {year} {2021})}\BibitemShut {NoStop}%
\bibitem [{\citenamefont {Gross}\ \emph {et~al.}(2020{\natexlab{a}})\citenamefont {Gross}, \citenamefont {Kruthoff}, \citenamefont {Rolph},\ and\ \citenamefont {Shaghoulian}}]{Gross20-1}%
  \BibitemOpen
  \bibfield  {author} {\bibinfo {author} {\bibfnamefont {D.~J.}\ \bibnamefont {Gross}}, \bibinfo {author} {\bibfnamefont {J.}~\bibnamefont {Kruthoff}}, \bibinfo {author} {\bibfnamefont {A.}~\bibnamefont {Rolph}},\ and\ \bibinfo {author} {\bibfnamefont {E.}~\bibnamefont {Shaghoulian}},\ }\bibfield  {title} {\bibinfo {title} {{$T\overline{T}$ in ${\mathrm{AdS_{2}}}$ and quantum mechanics}},\ }\href {https://doi.org/10.1103/PhysRevD.101.026011} {\bibfield  {journal} {\bibinfo  {journal} {Phys. Rev. D}\ }\textbf {\bibinfo {volume} {101}},\ \bibinfo {pages} {026011} (\bibinfo {year} {2020}{\natexlab{a}})}\BibitemShut {NoStop}%
\bibitem [{\citenamefont {Gross}\ \emph {et~al.}(2020{\natexlab{b}})\citenamefont {Gross}, \citenamefont {Kruthoff}, \citenamefont {Rolph},\ and\ \citenamefont {Shaghoulian}}]{Gross20-2}%
  \BibitemOpen
  \bibfield  {author} {\bibinfo {author} {\bibfnamefont {D.~J.}\ \bibnamefont {Gross}}, \bibinfo {author} {\bibfnamefont {J.}~\bibnamefont {Kruthoff}}, \bibinfo {author} {\bibfnamefont {A.}~\bibnamefont {Rolph}},\ and\ \bibinfo {author} {\bibfnamefont {E.}~\bibnamefont {Shaghoulian}},\ }\bibfield  {title} {\bibinfo {title} {{Hamiltonian deformations in quantum mechanics, $T\overline{T}$, and the SYK model}},\ }\href {https://doi.org/10.1103/PhysRevD.102.046019} {\bibfield  {journal} {\bibinfo  {journal} {Phys. Rev. D}\ }\textbf {\bibinfo {volume} {102}},\ \bibinfo {pages} {046019} (\bibinfo {year} {2020}{\natexlab{b}})}\BibitemShut {NoStop}%
\bibitem [{\citenamefont {Kruthoff}\ and\ \citenamefont {Parrikar}(2020)}]{kruthoff2020}%
  \BibitemOpen
  \bibfield  {author} {\bibinfo {author} {\bibfnamefont {J.}~\bibnamefont {Kruthoff}}\ and\ \bibinfo {author} {\bibfnamefont {O.}~\bibnamefont {Parrikar}},\ }\bibfield  {title} {\bibinfo {title} {{On the flow of states under $T\overline{T}$}},\ }\href {https://doi.org/10.21468/SciPostPhys.9.5.078} {\bibfield  {journal} {\bibinfo  {journal} {SciPost Phys.}\ }\textbf {\bibinfo {volume} {9}},\ \bibinfo {pages} {078} (\bibinfo {year} {2020})}\BibitemShut {NoStop}%
\bibitem [{\citenamefont {Rosso}(2021)}]{rosso21}%
  \BibitemOpen
  \bibfield  {author} {\bibinfo {author} {\bibfnamefont {F.}~\bibnamefont {Rosso}},\ }\bibfield  {title} {\bibinfo {title} {{$T\overline{T}$ deformation of random matrices}},\ }\href {https://doi.org/10.1103/PhysRevD.103.126017} {\bibfield  {journal} {\bibinfo  {journal} {Phys. Rev. D}\ }\textbf {\bibinfo {volume} {103}},\ \bibinfo {pages} {126017} (\bibinfo {year} {2021})}\BibitemShut {NoStop}%
\bibitem [{\citenamefont {Jiang}(2022)}]{jiang2022}%
  \BibitemOpen
  \bibfield  {author} {\bibinfo {author} {\bibfnamefont {Y.}~\bibnamefont {Jiang}},\ }\bibfield  {title} {\bibinfo {title} {{$T\overline{T}$-deformed 1d Bose gas}},\ }\href {https://doi.org/10.21468/SciPostPhys.12.6.191} {\bibfield  {journal} {\bibinfo  {journal} {SciPost Phys.}\ }\textbf {\bibinfo {volume} {12}},\ \bibinfo {pages} {191} (\bibinfo {year} {2022})}\BibitemShut {NoStop}%
\bibitem [{\citenamefont {Ebert}\ \emph {et~al.}(2022)\citenamefont {Ebert}, \citenamefont {Ferko}, \citenamefont {Sun},\ and\ \citenamefont {Sun}}]{ebert2022}%
  \BibitemOpen
  \bibfield  {author} {\bibinfo {author} {\bibfnamefont {S.}~\bibnamefont {Ebert}}, \bibinfo {author} {\bibfnamefont {C.}~\bibnamefont {Ferko}}, \bibinfo {author} {\bibfnamefont {H.-Y.}\ \bibnamefont {Sun}},\ and\ \bibinfo {author} {\bibfnamefont {Z.}~\bibnamefont {Sun}},\ }\bibfield  {title} {\bibinfo {title} {{$T\overline{T}$ deformations of supersymmetric quantum mechanics}},\ }\href {https://doi.org/10.1007/JHEP08(2022)121} {\bibfield  {journal} {\bibinfo  {journal} {Journal of High Energy Physics}\ }\textbf {\bibinfo {volume} {2022}},\ \bibinfo {pages} {121} (\bibinfo {year} {2022})}\BibitemShut {NoStop}%
\bibitem [{\citenamefont {Matsoukas-Roubeas}\ \emph {et~al.}(2023{\natexlab{a}})\citenamefont {Matsoukas-Roubeas}, \citenamefont {Roccati}, \citenamefont {Cornelius}, \citenamefont {Xu}, \citenamefont {Chenu},\ and\ \citenamefont {del Campo}}]{Matsoukas23}%
  \BibitemOpen
  \bibfield  {author} {\bibinfo {author} {\bibfnamefont {A.}~\bibnamefont {Matsoukas-Roubeas}}, \bibinfo {author} {\bibfnamefont {F.}~\bibnamefont {Roccati}}, \bibinfo {author} {\bibfnamefont {J.}~\bibnamefont {Cornelius}}, \bibinfo {author} {\bibfnamefont {Z.}~\bibnamefont {Xu}}, \bibinfo {author} {\bibfnamefont {A.}~\bibnamefont {Chenu}},\ and\ \bibinfo {author} {\bibfnamefont {A.}~\bibnamefont {del Campo}},\ }\bibfield  {title} {\bibinfo {title} {Non-hermitian hamiltonian deformations in quantum mechanics},\ }\href {https://doi.org/10.1007/JHEP01(2023)060} {\bibfield  {journal} {\bibinfo  {journal} {Journal of High Energy Physics}\ }\textbf {\bibinfo {volume} {2023}},\ \bibinfo {pages} {60} (\bibinfo {year} {2023}{\natexlab{a}})}\BibitemShut {NoStop}%
\bibitem [{\citenamefont {Viswanath}\ and\ \citenamefont {M\"uller}(1994)}]{Viswanath94}%
  \BibitemOpen
  \bibfield  {author} {\bibinfo {author} {\bibfnamefont {V.}~\bibnamefont {Viswanath}}\ and\ \bibinfo {author} {\bibfnamefont {G.}~\bibnamefont {M\"uller}},\ }\href@noop {} {\emph {\bibinfo {title} {The Recursion Method: Application to Many-Body Dynamics}}}\ (\bibinfo  {publisher} {Springer-Verlag},\ \bibinfo {year} {1994})\BibitemShut {NoStop}%
\bibitem [{\citenamefont {Parker}\ \emph {et~al.}(2019)\citenamefont {Parker}, \citenamefont {Cao}, \citenamefont {Avdoshkin}, \citenamefont {Scaffidi},\ and\ \citenamefont {Altman}}]{Parker19}%
  \BibitemOpen
  \bibfield  {author} {\bibinfo {author} {\bibfnamefont {D.~E.}\ \bibnamefont {Parker}}, \bibinfo {author} {\bibfnamefont {X.}~\bibnamefont {Cao}}, \bibinfo {author} {\bibfnamefont {A.}~\bibnamefont {Avdoshkin}}, \bibinfo {author} {\bibfnamefont {T.}~\bibnamefont {Scaffidi}},\ and\ \bibinfo {author} {\bibfnamefont {E.}~\bibnamefont {Altman}},\ }\bibfield  {title} {\bibinfo {title} {A universal operator growth hypothesis},\ }\href {https://doi.org/10.1103/PhysRevX.9.041017} {\bibfield  {journal} {\bibinfo  {journal} {Phys. Rev. X}\ }\textbf {\bibinfo {volume} {9}},\ \bibinfo {pages} {041017} (\bibinfo {year} {2019})}\BibitemShut {NoStop}%
\bibitem [{\citenamefont {Nandy}\ \emph {et~al.}(2025{\natexlab{a}})\citenamefont {Nandy}, \citenamefont {Matsoukas-Roubeas}, \citenamefont {Martinez-Azcona}, \citenamefont {Dymarsky},\ and\ \citenamefont {del Campo}}]{Nandy25}%
  \BibitemOpen
  \bibfield  {author} {\bibinfo {author} {\bibfnamefont {P.}~\bibnamefont {Nandy}}, \bibinfo {author} {\bibfnamefont {A.~S.}\ \bibnamefont {Matsoukas-Roubeas}}, \bibinfo {author} {\bibfnamefont {P.}~\bibnamefont {Martinez-Azcona}}, \bibinfo {author} {\bibfnamefont {A.}~\bibnamefont {Dymarsky}},\ and\ \bibinfo {author} {\bibfnamefont {A.}~\bibnamefont {del Campo}},\ }\bibfield  {title} {\bibinfo {title} {Quantum dynamics in krylov space: Methods and applications},\ }\href {https://doi.org/https://doi.org/10.1016/j.physrep.2025.05.001} {\bibfield  {journal} {\bibinfo  {journal} {Physics Reports}\ }\textbf {\bibinfo {volume} {1125-1128}},\ \bibinfo {pages} {1} (\bibinfo {year} {2025}{\natexlab{a}})}\BibitemShut {NoStop}%
\bibitem [{\citenamefont {Baiguera}\ \emph {et~al.}(2026)\citenamefont {Baiguera}, \citenamefont {Balasubramanian}, \citenamefont {Caputa}, \citenamefont {Chapman}, \citenamefont {Haferkamp}, \citenamefont {Heller},\ and\ \citenamefont {Halpern}}]{Baiguera26}%
  \BibitemOpen
  \bibfield  {author} {\bibinfo {author} {\bibfnamefont {S.}~\bibnamefont {Baiguera}}, \bibinfo {author} {\bibfnamefont {V.}~\bibnamefont {Balasubramanian}}, \bibinfo {author} {\bibfnamefont {P.}~\bibnamefont {Caputa}}, \bibinfo {author} {\bibfnamefont {S.}~\bibnamefont {Chapman}}, \bibinfo {author} {\bibfnamefont {J.}~\bibnamefont {Haferkamp}}, \bibinfo {author} {\bibfnamefont {M.~P.}\ \bibnamefont {Heller}},\ and\ \bibinfo {author} {\bibfnamefont {N.~Y.}\ \bibnamefont {Halpern}},\ }\bibfield  {title} {\bibinfo {title} {Quantum complexity in gravity, quantum field theory, and quantum information science},\ }\href {https://doi.org/https://doi.org/10.1016/j.physrep.2025.11.001} {\bibfield  {journal} {\bibinfo  {journal} {Physics Reports}\ }\textbf {\bibinfo {volume} {1159}},\ \bibinfo {pages} {1} (\bibinfo {year} {2026})}\BibitemShut {NoStop}%
\bibitem [{\citenamefont {Rabinovici}\ \emph {et~al.}(2025)\citenamefont {Rabinovici}, \citenamefont {Sánchez-Garrido}, \citenamefont {Shir},\ and\ \citenamefont {Sonner}}]{Rabinovici25}%
  \BibitemOpen
  \bibfield  {author} {\bibinfo {author} {\bibfnamefont {E.}~\bibnamefont {Rabinovici}}, \bibinfo {author} {\bibfnamefont {A.}~\bibnamefont {Sánchez-Garrido}}, \bibinfo {author} {\bibfnamefont {R.}~\bibnamefont {Shir}},\ and\ \bibinfo {author} {\bibfnamefont {J.}~\bibnamefont {Sonner}},\ }\href {https://arxiv.org/abs/2507.06286} {\bibinfo {title} {Krylov complexity}} (\bibinfo {year} {2025}),\ \Eprint {https://arxiv.org/abs/2507.06286} {arXiv:2507.06286 [hep-th]} \BibitemShut {NoStop}%
\bibitem [{\citenamefont {Takahashi}\ and\ \citenamefont {del Campo}(2024)}]{Takahashi24}%
  \BibitemOpen
  \bibfield  {author} {\bibinfo {author} {\bibfnamefont {K.}~\bibnamefont {Takahashi}}\ and\ \bibinfo {author} {\bibfnamefont {A.}~\bibnamefont {del Campo}},\ }\bibfield  {title} {\bibinfo {title} {Shortcuts to adiabaticity in krylov space},\ }\href {https://doi.org/10.1103/PhysRevX.14.011032} {\bibfield  {journal} {\bibinfo  {journal} {Phys. Rev. X}\ }\textbf {\bibinfo {volume} {14}},\ \bibinfo {pages} {011032} (\bibinfo {year} {2024})}\BibitemShut {NoStop}%
\bibitem [{\citenamefont {Takahashi}\ and\ \citenamefont {del Campo}(2025)}]{Takahashi25}%
  \BibitemOpen
  \bibfield  {author} {\bibinfo {author} {\bibfnamefont {K.}~\bibnamefont {Takahashi}}\ and\ \bibinfo {author} {\bibfnamefont {A.}~\bibnamefont {del Campo}},\ }\bibfield  {title} {\bibinfo {title} {Krylov subspace methods for quantum dynamics with time-dependent generators},\ }\href {https://doi.org/10.1103/PhysRevLett.134.030401} {\bibfield  {journal} {\bibinfo  {journal} {Phys. Rev. Lett.}\ }\textbf {\bibinfo {volume} {134}},\ \bibinfo {pages} {030401} (\bibinfo {year} {2025})}\BibitemShut {NoStop}%
\bibitem [{\citenamefont {Bhattacharjee}(2023)}]{Bhattacharjee23arxiv}%
  \BibitemOpen
  \bibfield  {author} {\bibinfo {author} {\bibfnamefont {B.}~\bibnamefont {Bhattacharjee}},\ }\href@noop {} {\bibinfo {title} {A lanczos approach to the adiabatic gauge potential}} (\bibinfo {year} {2023}),\ \Eprint {https://arxiv.org/abs/2302.07228} {arXiv:2302.07228 [quant-ph]} \BibitemShut {NoStop}%
\bibitem [{\citenamefont {Grabarits}\ \emph {et~al.}(2026)\citenamefont {Grabarits}, \citenamefont {Balducci},\ and\ \citenamefont {del Campo}}]{Grabarits26}%
  \BibitemOpen
  \bibfield  {author} {\bibinfo {author} {\bibfnamefont {A.}~\bibnamefont {Grabarits}}, \bibinfo {author} {\bibfnamefont {F.}~\bibnamefont {Balducci}},\ and\ \bibinfo {author} {\bibfnamefont {A.}~\bibnamefont {del Campo}},\ }\bibfield  {title} {\bibinfo {title} {Fighting exponentially small gaps by counterdiabatic driving},\ }\href {https://doi.org/10.1103/tgzt-dy3h} {\bibfield  {journal} {\bibinfo  {journal} {PRX Quantum}\ }\textbf {\bibinfo {volume} {7}},\ \bibinfo {pages} {010322} (\bibinfo {year} {2026})}\BibitemShut {NoStop}%
\bibitem [{\citenamefont {Grabarits}\ and\ \citenamefont {del Campo}(2025{\natexlab{a}})}]{Grabarits2025cd}%
  \BibitemOpen
  \bibfield  {author} {\bibinfo {author} {\bibfnamefont {A.}~\bibnamefont {Grabarits}}\ and\ \bibinfo {author} {\bibfnamefont {A.}~\bibnamefont {del Campo}},\ }\href {https://arxiv.org/abs/2503.22212} {\bibinfo {title} {Universal defect statistics in counterdiabatic quantum critical dynamics}} (\bibinfo {year} {2025}{\natexlab{a}}),\ \Eprint {https://arxiv.org/abs/2503.22212} {arXiv:2503.22212 [quant-ph]} \BibitemShut {NoStop}%
\bibitem [{\citenamefont {Grabarits}\ and\ \citenamefont {del Campo}(2025{\natexlab{b}})}]{grabarits2025KQPT}%
  \BibitemOpen
  \bibfield  {author} {\bibinfo {author} {\bibfnamefont {A.}~\bibnamefont {Grabarits}}\ and\ \bibinfo {author} {\bibfnamefont {A.}~\bibnamefont {del Campo}},\ }\href {https://arxiv.org/abs/2510.13947} {\bibinfo {title} {Universal growth of krylov complexity across a quantum phase transition}} (\bibinfo {year} {2025}{\natexlab{b}}),\ \Eprint {https://arxiv.org/abs/2510.13947} {arXiv:2510.13947 [quant-ph]} \BibitemShut {NoStop}%
\bibitem [{\citenamefont {Brockett}(1991)}]{Brockett91}%
  \BibitemOpen
  \bibfield  {author} {\bibinfo {author} {\bibfnamefont {R.~W.}\ \bibnamefont {Brockett}},\ }\bibfield  {title} {\bibinfo {title} {Dynamical systems that sort lists, diagonalize matrices, and solve linear programming problems},\ }\href {https://doi.org/https://doi.org/10.1016/0024-3795(91)90021-N} {\bibfield  {journal} {\bibinfo  {journal} {Linear Algebra and its Applications}\ }\textbf {\bibinfo {volume} {146}},\ \bibinfo {pages} {79} (\bibinfo {year} {1991})}\BibitemShut {NoStop}%
\bibitem [{\citenamefont {G\l{}azek}\ and\ \citenamefont {Wilson}(1993)}]{GlazekWilson93}%
  \BibitemOpen
  \bibfield  {author} {\bibinfo {author} {\bibfnamefont {S.~D.}\ \bibnamefont {G\l{}azek}}\ and\ \bibinfo {author} {\bibfnamefont {K.~G.}\ \bibnamefont {Wilson}},\ }\bibfield  {title} {\bibinfo {title} {Renormalization of hamiltonians},\ }\href {https://doi.org/10.1103/PhysRevD.48.5863} {\bibfield  {journal} {\bibinfo  {journal} {Phys. Rev. D}\ }\textbf {\bibinfo {volume} {48}},\ \bibinfo {pages} {5863} (\bibinfo {year} {1993})}\BibitemShut {NoStop}%
\bibitem [{\citenamefont {Glazek}\ and\ \citenamefont {Wilson}(1994)}]{GlazekWilson94}%
  \BibitemOpen
  \bibfield  {author} {\bibinfo {author} {\bibfnamefont {S.~D.}\ \bibnamefont {Glazek}}\ and\ \bibinfo {author} {\bibfnamefont {K.~G.}\ \bibnamefont {Wilson}},\ }\bibfield  {title} {\bibinfo {title} {Perturbative renormalization group for hamiltonians},\ }\href {https://doi.org/10.1103/PhysRevD.49.4214} {\bibfield  {journal} {\bibinfo  {journal} {Phys. Rev. D}\ }\textbf {\bibinfo {volume} {49}},\ \bibinfo {pages} {4214} (\bibinfo {year} {1994})}\BibitemShut {NoStop}%
\bibitem [{\citenamefont {Wegner}(1994)}]{Wegner94}%
  \BibitemOpen
  \bibfield  {author} {\bibinfo {author} {\bibfnamefont {F.}~\bibnamefont {Wegner}},\ }\bibfield  {title} {\bibinfo {title} {Flow-equations for hamiltonians},\ }\href {https://doi.org/https://doi.org/10.1002/andp.19945060203} {\bibfield  {journal} {\bibinfo  {journal} {Annalen der Physik}\ }\textbf {\bibinfo {volume} {506}},\ \bibinfo {pages} {77} (\bibinfo {year} {1994})}\BibitemShut {NoStop}%
\bibitem [{\citenamefont {Kehrein}(2006)}]{Kehrein06}%
  \BibitemOpen
  \bibfield  {author} {\bibinfo {author} {\bibfnamefont {S.}~\bibnamefont {Kehrein}},\ }\href {https://doi.org/10.1007/3-540-34068-8} {\emph {\bibinfo {title} {The Flow Equation Approach to Many-Particle Systems}}},\ Springer Tracts in Modern Physics\ (\bibinfo  {publisher} {Springer},\ \bibinfo {year} {2006})\BibitemShut {NoStop}%
\bibitem [{\citenamefont {H{\"{o}}rnedal}\ \emph {et~al.}(2023)\citenamefont {H{\"{o}}rnedal}, \citenamefont {Carabba}, \citenamefont {Takahashi},\ and\ \citenamefont {del Campo}}]{Hornedal23}%
  \BibitemOpen
  \bibfield  {author} {\bibinfo {author} {\bibfnamefont {N.}~\bibnamefont {H{\"{o}}rnedal}}, \bibinfo {author} {\bibfnamefont {N.}~\bibnamefont {Carabba}}, \bibinfo {author} {\bibfnamefont {K.}~\bibnamefont {Takahashi}},\ and\ \bibinfo {author} {\bibfnamefont {A.}~\bibnamefont {del Campo}},\ }\bibfield  {title} {\bibinfo {title} {Geometric {O}perator {Q}uantum {S}peed {L}imit, {W}egner {H}amiltonian {F}low and {O}perator {G}rowth},\ }\href {https://doi.org/10.22331/q-2023-07-11-1055} {\bibfield  {journal} {\bibinfo  {journal} {{Quantum}}\ }\textbf {\bibinfo {volume} {7}},\ \bibinfo {pages} {1055} (\bibinfo {year} {2023})}\BibitemShut {NoStop}%
\bibitem [{\citenamefont {Toda}(1967{\natexlab{a}})}]{Toda67-1}%
  \BibitemOpen
  \bibfield  {author} {\bibinfo {author} {\bibfnamefont {M.}~\bibnamefont {Toda}},\ }\bibfield  {title} {\bibinfo {title} {Vibration of a chain with nonlinear interaction},\ }\href {https://doi.org/10.1143/JPSJ.22.431} {\bibfield  {journal} {\bibinfo  {journal} {Journal of the Physical Society of Japan}\ }\textbf {\bibinfo {volume} {22}},\ \bibinfo {pages} {431} (\bibinfo {year} {1967}{\natexlab{a}})}\BibitemShut {NoStop}%
\bibitem [{\citenamefont {Toda}(1967{\natexlab{b}})}]{Toda67-2}%
  \BibitemOpen
  \bibfield  {author} {\bibinfo {author} {\bibfnamefont {M.}~\bibnamefont {Toda}},\ }\bibfield  {title} {\bibinfo {title} {Wave propagation in anharmonic lattices},\ }\href {https://doi.org/10.1143/JPSJ.23.501} {\bibfield  {journal} {\bibinfo  {journal} {Journal of the Physical Society of Japan}\ }\textbf {\bibinfo {volume} {23}},\ \bibinfo {pages} {501} (\bibinfo {year} {1967}{\natexlab{b}})}\BibitemShut {NoStop}%
\bibitem [{\citenamefont {Okuyama}\ and\ \citenamefont {Takahashi}(2016)}]{Okuyama16}%
  \BibitemOpen
  \bibfield  {author} {\bibinfo {author} {\bibfnamefont {M.}~\bibnamefont {Okuyama}}\ and\ \bibinfo {author} {\bibfnamefont {K.}~\bibnamefont {Takahashi}},\ }\bibfield  {title} {\bibinfo {title} {From classical nonlinear integrable systems to quantum shortcuts to adiabaticity},\ }\href {https://doi.org/10.1103/PhysRevLett.117.070401} {\bibfield  {journal} {\bibinfo  {journal} {Phys. Rev. Lett.}\ }\textbf {\bibinfo {volume} {117}},\ \bibinfo {pages} {070401} (\bibinfo {year} {2016})}\BibitemShut {NoStop}%
\bibitem [{\citenamefont {Ismail}(2005)}]{Ismail05}%
  \BibitemOpen
  \bibfield  {author} {\bibinfo {author} {\bibfnamefont {M.~E.~H.}\ \bibnamefont {Ismail}},\ }\href {https://doi.org/10.1017/CBO9781107325982} {\emph {\bibinfo {title} {Classical and Quantum Orthogonal Polynomials in One Variable}}},\ Encyclopedia of Mathematics and its Applications\ (\bibinfo  {publisher} {Cambridge University Press},\ \bibinfo {year} {2005})\BibitemShut {NoStop}%
\bibitem [{\citenamefont {Dymarsky}\ and\ \citenamefont {Gorsky}(2020)}]{Dymarsky20}%
  \BibitemOpen
  \bibfield  {author} {\bibinfo {author} {\bibfnamefont {A.}~\bibnamefont {Dymarsky}}\ and\ \bibinfo {author} {\bibfnamefont {A.}~\bibnamefont {Gorsky}},\ }\bibfield  {title} {\bibinfo {title} {Quantum chaos as delocalization in krylov space},\ }\href {https://doi.org/10.1103/PhysRevB.102.085137} {\bibfield  {journal} {\bibinfo  {journal} {Phys. Rev. B}\ }\textbf {\bibinfo {volume} {102}},\ \bibinfo {pages} {085137} (\bibinfo {year} {2020})}\BibitemShut {NoStop}%
\bibitem [{\citenamefont {Kundu}\ \emph {et~al.}(2023)\citenamefont {Kundu}, \citenamefont {Malvimat},\ and\ \citenamefont {Sinha}}]{Kundu23}%
  \BibitemOpen
  \bibfield  {author} {\bibinfo {author} {\bibfnamefont {A.}~\bibnamefont {Kundu}}, \bibinfo {author} {\bibfnamefont {V.}~\bibnamefont {Malvimat}},\ and\ \bibinfo {author} {\bibfnamefont {R.}~\bibnamefont {Sinha}},\ }\bibfield  {title} {\bibinfo {title} {{State dependence of Krylov complexity in 2d CFTs}},\ }\href {https://doi.org/10.1007/JHEP09(2023)011} {\bibfield  {journal} {\bibinfo  {journal} {JHEP}\ }\textbf {\bibinfo {volume} {09}},\ \bibinfo {pages} {011}},\ \Eprint {https://arxiv.org/abs/2303.03426} {arXiv:2303.03426 [hep-th]} \BibitemShut {NoStop}%
\bibitem [{\citenamefont {Angelinos}\ \emph {et~al.}(2025)\citenamefont {Angelinos}, \citenamefont {Chakraborty},\ and\ \citenamefont {Dymarsky}}]{angelinos25}%
  \BibitemOpen
  \bibfield  {author} {\bibinfo {author} {\bibfnamefont {N.}~\bibnamefont {Angelinos}}, \bibinfo {author} {\bibfnamefont {D.}~\bibnamefont {Chakraborty}},\ and\ \bibinfo {author} {\bibfnamefont {A.}~\bibnamefont {Dymarsky}},\ }\href {https://arxiv.org/abs/2508.19233} {\bibinfo {title} {Temperature dependence in krylov space}} (\bibinfo {year} {2025}),\ \Eprint {https://arxiv.org/abs/2508.19233} {arXiv:2508.19233 [hep-th]} \BibitemShut {NoStop}%
\bibitem [{\citenamefont {del Campo}\ \emph {et~al.}(2017)\citenamefont {del Campo}, \citenamefont {Molina-Vilaplana},\ and\ \citenamefont {Sonner}}]{delcampo2018sff}%
  \BibitemOpen
  \bibfield  {author} {\bibinfo {author} {\bibfnamefont {A.}~\bibnamefont {del Campo}}, \bibinfo {author} {\bibfnamefont {J.}~\bibnamefont {Molina-Vilaplana}},\ and\ \bibinfo {author} {\bibfnamefont {J.}~\bibnamefont {Sonner}},\ }\bibfield  {title} {\bibinfo {title} {Scrambling the spectral form factor: Unitarity constraints and exact results},\ }\href {https://doi.org/10.1103/PhysRevD.95.126008} {\bibfield  {journal} {\bibinfo  {journal} {Phys. Rev. D}\ }\textbf {\bibinfo {volume} {95}},\ \bibinfo {pages} {126008} (\bibinfo {year} {2017})}\BibitemShut {NoStop}%
\bibitem [{\citenamefont {Xu}\ \emph {et~al.}(2021)\citenamefont {Xu}, \citenamefont {Chenu}, \citenamefont {Prosen},\ and\ \citenamefont {del Campo}}]{Xu21}%
  \BibitemOpen
  \bibfield  {author} {\bibinfo {author} {\bibfnamefont {Z.}~\bibnamefont {Xu}}, \bibinfo {author} {\bibfnamefont {A.}~\bibnamefont {Chenu}}, \bibinfo {author} {\bibfnamefont {T.}~\bibnamefont {Prosen}},\ and\ \bibinfo {author} {\bibfnamefont {A.}~\bibnamefont {del Campo}},\ }\bibfield  {title} {\bibinfo {title} {Thermofield dynamics: Quantum chaos versus decoherence},\ }\href {https://doi.org/10.1103/PhysRevB.103.064309} {\bibfield  {journal} {\bibinfo  {journal} {Phys. Rev. B}\ }\textbf {\bibinfo {volume} {103}},\ \bibinfo {pages} {064309} (\bibinfo {year} {2021})}\BibitemShut {NoStop}%
\bibitem [{\citenamefont {Balasubramanian}\ \emph {et~al.}(2022)\citenamefont {Balasubramanian}, \citenamefont {Caputa}, \citenamefont {Magan},\ and\ \citenamefont {Wu}}]{Balasubramanian22}%
  \BibitemOpen
  \bibfield  {author} {\bibinfo {author} {\bibfnamefont {V.}~\bibnamefont {Balasubramanian}}, \bibinfo {author} {\bibfnamefont {P.}~\bibnamefont {Caputa}}, \bibinfo {author} {\bibfnamefont {J.~M.}\ \bibnamefont {Magan}},\ and\ \bibinfo {author} {\bibfnamefont {Q.}~\bibnamefont {Wu}},\ }\bibfield  {title} {\bibinfo {title} {Quantum chaos and the complexity of spread of states},\ }\href {https://doi.org/10.1103/PhysRevD.106.046007} {\bibfield  {journal} {\bibinfo  {journal} {Phys. Rev. D}\ }\textbf {\bibinfo {volume} {106}},\ \bibinfo {pages} {046007} (\bibinfo {year} {2022})}\BibitemShut {NoStop}%
\bibitem [{\citenamefont {Maldacena}(2003)}]{Maldacena03}%
  \BibitemOpen
  \bibfield  {author} {\bibinfo {author} {\bibfnamefont {J.}~\bibnamefont {Maldacena}},\ }\bibfield  {title} {\bibinfo {title} {Eternal black holes in anti-de sitter},\ }\href {https://doi.org/10.1088/1126-6708/2003/04/021} {\bibfield  {journal} {\bibinfo  {journal} {Journal of High Energy Physics}\ }\textbf {\bibinfo {volume} {2003}},\ \bibinfo {pages} {021} (\bibinfo {year} {2003})}\BibitemShut {NoStop}%
\bibitem [{\citenamefont {Papadodimas}\ and\ \citenamefont {Raju}(2015)}]{Papadodimas15}%
  \BibitemOpen
  \bibfield  {author} {\bibinfo {author} {\bibfnamefont {K.}~\bibnamefont {Papadodimas}}\ and\ \bibinfo {author} {\bibfnamefont {S.}~\bibnamefont {Raju}},\ }\bibfield  {title} {\bibinfo {title} {Local operators in the eternal black hole},\ }\href {https://doi.org/10.1103/PhysRevLett.115.211601} {\bibfield  {journal} {\bibinfo  {journal} {Phys. Rev. Lett.}\ }\textbf {\bibinfo {volume} {115}},\ \bibinfo {pages} {211601} (\bibinfo {year} {2015})}\BibitemShut {NoStop}%
\bibitem [{\citenamefont {Chapman}\ \emph {et~al.}(2019)\citenamefont {Chapman}, \citenamefont {Eisert}, \citenamefont {Hackl}, \citenamefont {Heller}, \citenamefont {Jefferson}, \citenamefont {Marrochio},\ and\ \citenamefont {Myers}}]{Chapman19}%
  \BibitemOpen
  \bibfield  {author} {\bibinfo {author} {\bibfnamefont {S.}~\bibnamefont {Chapman}}, \bibinfo {author} {\bibfnamefont {J.}~\bibnamefont {Eisert}}, \bibinfo {author} {\bibfnamefont {L.}~\bibnamefont {Hackl}}, \bibinfo {author} {\bibfnamefont {M.~P.}\ \bibnamefont {Heller}}, \bibinfo {author} {\bibfnamefont {R.}~\bibnamefont {Jefferson}}, \bibinfo {author} {\bibfnamefont {H.}~\bibnamefont {Marrochio}},\ and\ \bibinfo {author} {\bibfnamefont {R.~C.}\ \bibnamefont {Myers}},\ }\bibfield  {title} {\bibinfo {title} {{Complexity and entanglement for thermofield double states}},\ }\href {https://doi.org/10.21468/SciPostPhys.6.3.034} {\bibfield  {journal} {\bibinfo  {journal} {SciPost Phys.}\ }\textbf {\bibinfo {volume} {6}},\ \bibinfo {pages} {034} (\bibinfo {year} {2019})}\BibitemShut {NoStop}%
\bibitem [{\citenamefont {Matsoukas-Roubeas}\ \emph {et~al.}(2023{\natexlab{b}})\citenamefont {Matsoukas-Roubeas}, \citenamefont {Beau}, \citenamefont {Santos},\ and\ \citenamefont {del Campo}}]{Matsoukas23b}%
  \BibitemOpen
  \bibfield  {author} {\bibinfo {author} {\bibfnamefont {A.~S.}\ \bibnamefont {Matsoukas-Roubeas}}, \bibinfo {author} {\bibfnamefont {M.}~\bibnamefont {Beau}}, \bibinfo {author} {\bibfnamefont {L.~F.}\ \bibnamefont {Santos}},\ and\ \bibinfo {author} {\bibfnamefont {A.}~\bibnamefont {del Campo}},\ }\bibfield  {title} {\bibinfo {title} {Unitarity breaking in self-averaging spectral form factors},\ }\href {https://doi.org/10.1103/PhysRevA.108.062201} {\bibfield  {journal} {\bibinfo  {journal} {Phys. Rev. A}\ }\textbf {\bibinfo {volume} {108}},\ \bibinfo {pages} {062201} (\bibinfo {year} {2023}{\natexlab{b}})}\BibitemShut {NoStop}%
\bibitem [{\citenamefont {Haake}(2001)}]{Haake01}%
  \BibitemOpen
  \bibfield  {author} {\bibinfo {author} {\bibfnamefont {F.}~\bibnamefont {Haake}},\ }\href {https://doi.org/10.1007/978-3-642-05428-0} {\emph {\bibinfo {title} {Quantum Signatures of Chaos}}},\ Physics and astronomy online library\ (\bibinfo  {publisher} {Springer},\ \bibinfo {year} {2001})\BibitemShut {NoStop}%
\bibitem [{\citenamefont {Cotler}\ \emph {et~al.}(2017)\citenamefont {Cotler}, \citenamefont {Gur-Ari}, \citenamefont {Hanada}, \citenamefont {Polchinski}, \citenamefont {Saad}, \citenamefont {Shenker}, \citenamefont {Stanford}, \citenamefont {Streicher},\ and\ \citenamefont {Tezuka}}]{Cotler2017}%
  \BibitemOpen
  \bibfield  {author} {\bibinfo {author} {\bibfnamefont {J.~S.}\ \bibnamefont {Cotler}}, \bibinfo {author} {\bibfnamefont {G.}~\bibnamefont {Gur-Ari}}, \bibinfo {author} {\bibfnamefont {M.}~\bibnamefont {Hanada}}, \bibinfo {author} {\bibfnamefont {J.}~\bibnamefont {Polchinski}}, \bibinfo {author} {\bibfnamefont {P.}~\bibnamefont {Saad}}, \bibinfo {author} {\bibfnamefont {S.~H.}\ \bibnamefont {Shenker}}, \bibinfo {author} {\bibfnamefont {D.}~\bibnamefont {Stanford}}, \bibinfo {author} {\bibfnamefont {A.}~\bibnamefont {Streicher}},\ and\ \bibinfo {author} {\bibfnamefont {M.}~\bibnamefont {Tezuka}},\ }\bibfield  {title} {\bibinfo {title} {Black holes and random matrices},\ }\href {https://doi.org/10.1007/JHEP05(2017)118} {\bibfield  {journal} {\bibinfo  {journal} {J. High Energy Phys.}\ }\textbf {\bibinfo {volume} {2017}}\bibinfo  {number} { (5)},\ \bibinfo {pages} {118}}\BibitemShut {NoStop}%
\bibitem [{\citenamefont {Cornelius}\ \emph {et~al.}(2022)\citenamefont {Cornelius}, \citenamefont {Xu}, \citenamefont {Saxena}, \citenamefont {Chenu},\ and\ \citenamefont {del Campo}}]{cornelius2022}%
  \BibitemOpen
\bibfield  {number} {  }\bibfield  {author} {\bibinfo {author} {\bibfnamefont {J.}~\bibnamefont {Cornelius}}, \bibinfo {author} {\bibfnamefont {Z.}~\bibnamefont {Xu}}, \bibinfo {author} {\bibfnamefont {A.}~\bibnamefont {Saxena}}, \bibinfo {author} {\bibfnamefont {A.}~\bibnamefont {Chenu}},\ and\ \bibinfo {author} {\bibfnamefont {A.}~\bibnamefont {del Campo}},\ }\bibfield  {title} {\bibinfo {title} {Spectral filtering induced by non-hermitian evolution with balanced gain and loss: Enhancing quantum chaos},\ }\href {https://doi.org/10.1103/PhysRevLett.128.190402} {\bibfield  {journal} {\bibinfo  {journal} {Phys. Rev. Lett.}\ }\textbf {\bibinfo {volume} {128}},\ \bibinfo {pages} {190402} (\bibinfo {year} {2022})}\BibitemShut {NoStop}%
\bibitem [{\citenamefont {Matsoukas-Roubeas}\ \emph {et~al.}(2024)\citenamefont {Matsoukas-Roubeas}, \citenamefont {Prosen},\ and\ \citenamefont {Campo}}]{Matsoukas24}%
  \BibitemOpen
  \bibfield  {author} {\bibinfo {author} {\bibfnamefont {A.~S.}\ \bibnamefont {Matsoukas-Roubeas}}, \bibinfo {author} {\bibfnamefont {T.}~\bibnamefont {Prosen}},\ and\ \bibinfo {author} {\bibfnamefont {A.~d.}\ \bibnamefont {Campo}},\ }\bibfield  {title} {\bibinfo {title} {Quantum {C}haos and {C}oherence: {R}andom {P}arametric {Q}uantum {C}hannels},\ }\href {https://doi.org/10.22331/q-2024-08-27-1446} {\bibfield  {journal} {\bibinfo  {journal} {{Quantum}}\ }\textbf {\bibinfo {volume} {8}},\ \bibinfo {pages} {1446} (\bibinfo {year} {2024})}\BibitemShut {NoStop}%
\bibitem [{\citenamefont {Martinez-Azcona}\ \emph {et~al.}(2025)\citenamefont {Martinez-Azcona}, \citenamefont {Shir},\ and\ \citenamefont {Chenu}}]{MartinezAzcona25}%
  \BibitemOpen
  \bibfield  {author} {\bibinfo {author} {\bibfnamefont {P.}~\bibnamefont {Martinez-Azcona}}, \bibinfo {author} {\bibfnamefont {R.}~\bibnamefont {Shir}},\ and\ \bibinfo {author} {\bibfnamefont {A.}~\bibnamefont {Chenu}},\ }\bibfield  {title} {\bibinfo {title} {Decomposing the spectral form factor},\ }\href {https://doi.org/10.1103/PhysRevB.111.165108} {\bibfield  {journal} {\bibinfo  {journal} {Phys. Rev. B}\ }\textbf {\bibinfo {volume} {111}},\ \bibinfo {pages} {165108} (\bibinfo {year} {2025})}\BibitemShut {NoStop}%
\bibitem [{\citenamefont {Mück}\ and\ \citenamefont {Yang}(2022)}]{Muck22}%
  \BibitemOpen
  \bibfield  {author} {\bibinfo {author} {\bibfnamefont {W.}~\bibnamefont {Mück}}\ and\ \bibinfo {author} {\bibfnamefont {Y.}~\bibnamefont {Yang}},\ }\bibfield  {title} {\bibinfo {title} {Krylov complexity and orthogonal polynomials},\ }\href {https://doi.org/https://doi.org/10.1016/j.nuclphysb.2022.115948} {\bibfield  {journal} {\bibinfo  {journal} {Nuclear Physics B}\ }\textbf {\bibinfo {volume} {984}},\ \bibinfo {pages} {115948} (\bibinfo {year} {2022})}\BibitemShut {NoStop}%
\bibitem [{\citenamefont {Moser}(1975{\natexlab{a}})}]{Moser75book}%
  \BibitemOpen
  \bibfield  {author} {\bibinfo {author} {\bibfnamefont {J.}~\bibnamefont {Moser}},\ }\bibinfo {title} {Finitely many mass points on the line under the influence of an exponential potential -- an integrable system},\ in\ \href {https://doi.org/10.1007/3-540-07171-7_12} {\emph {\bibinfo {booktitle} {Dynamical Systems, Theory and Applications: Battelle Seattle 1974 Rencontres}}},\ \bibinfo {editor} {edited by\ \bibinfo {editor} {\bibfnamefont {J.}~\bibnamefont {Moser}}}\ (\bibinfo  {publisher} {Springer Berlin Heidelberg},\ \bibinfo {address} {Berlin, Heidelberg},\ \bibinfo {year} {1975})\ pp.\ \bibinfo {pages} {467--497}\BibitemShut {NoStop}%
\bibitem [{\citenamefont {Lax}(1968)}]{Lax68}%
  \BibitemOpen
  \bibfield  {author} {\bibinfo {author} {\bibfnamefont {P.~D.}\ \bibnamefont {Lax}},\ }\bibfield  {title} {\bibinfo {title} {Integrals of nonlinear equations of evolution and solitary waves},\ }\href {https://doi.org/https://doi.org/10.1002/cpa.3160210503} {\bibfield  {journal} {\bibinfo  {journal} {Communications on Pure and Applied Mathematics}\ }\textbf {\bibinfo {volume} {21}},\ \bibinfo {pages} {467} (\bibinfo {year} {1968})}\BibitemShut {NoStop}%
\bibitem [{\citenamefont {Faddeev}\ and\ \citenamefont {Takhtajan}(2007)}]{Faddeev2007}%
  \BibitemOpen
  \bibfield  {author} {\bibinfo {author} {\bibfnamefont {L.~D.}\ \bibnamefont {Faddeev}}\ and\ \bibinfo {author} {\bibfnamefont {L.~A.}\ \bibnamefont {Takhtajan}},\ }\href {https://doi.org/10.1007/978-3-540-69969-9} {\emph {\bibinfo {title} {Hamiltonian Methods in the Theory of Solitons}}},\ Classics in Mathematics\ (\bibinfo  {publisher} {Springer},\ \bibinfo {address} {Berlin, Heidelberg},\ \bibinfo {year} {2007})\BibitemShut {NoStop}%
\bibitem [{\citenamefont {Kac}\ and\ \citenamefont {{van Moerbeke}}(1975)}]{Kac75}%
  \BibitemOpen
  \bibfield  {author} {\bibinfo {author} {\bibfnamefont {M.}~\bibnamefont {Kac}}\ and\ \bibinfo {author} {\bibfnamefont {P.}~\bibnamefont {{van Moerbeke}}},\ }\bibfield  {title} {\bibinfo {title} {On an explicitly soluble system of nonlinear differential equations related to certain toda lattices},\ }\href {https://doi.org/https://doi.org/10.1016/0001-8708(75)90148-6} {\bibfield  {journal} {\bibinfo  {journal} {Advances in Mathematics}\ }\textbf {\bibinfo {volume} {16}},\ \bibinfo {pages} {160} (\bibinfo {year} {1975})}\BibitemShut {NoStop}%
\bibitem [{\citenamefont {Moser}(1975{\natexlab{b}})}]{Moser75}%
  \BibitemOpen
  \bibfield  {author} {\bibinfo {author} {\bibfnamefont {J.}~\bibnamefont {Moser}},\ }\bibfield  {title} {\bibinfo {title} {Three integrable hamiltonian systems connected with isospectral deformations},\ }\href {https://doi.org/https://doi.org/10.1016/0001-8708(75)90151-6} {\bibfield  {journal} {\bibinfo  {journal} {Advances in Mathematics}\ }\textbf {\bibinfo {volume} {16}},\ \bibinfo {pages} {197} (\bibinfo {year} {1975}{\natexlab{b}})}\BibitemShut {NoStop}%
\bibitem [{\citenamefont {Sinitsyn}\ \emph {et~al.}(2018)\citenamefont {Sinitsyn}, \citenamefont {Yuzbashyan}, \citenamefont {Chernyak}, \citenamefont {Patra},\ and\ \citenamefont {Sun}}]{Sinitsyn18}%
  \BibitemOpen
  \bibfield  {author} {\bibinfo {author} {\bibfnamefont {N.~A.}\ \bibnamefont {Sinitsyn}}, \bibinfo {author} {\bibfnamefont {E.~A.}\ \bibnamefont {Yuzbashyan}}, \bibinfo {author} {\bibfnamefont {V.~Y.}\ \bibnamefont {Chernyak}}, \bibinfo {author} {\bibfnamefont {A.}~\bibnamefont {Patra}},\ and\ \bibinfo {author} {\bibfnamefont {C.}~\bibnamefont {Sun}},\ }\bibfield  {title} {\bibinfo {title} {Integrable time-dependent quantum hamiltonians},\ }\href {https://doi.org/10.1103/PhysRevLett.120.190402} {\bibfield  {journal} {\bibinfo  {journal} {Phys. Rev. Lett.}\ }\textbf {\bibinfo {volume} {120}},\ \bibinfo {pages} {190402} (\bibinfo {year} {2018})}\BibitemShut {NoStop}%
\bibitem [{\citenamefont {Caputa}\ \emph {et~al.}(2022)\citenamefont {Caputa}, \citenamefont {Magan},\ and\ \citenamefont {Patramanis}}]{Caputa22}%
  \BibitemOpen
  \bibfield  {author} {\bibinfo {author} {\bibfnamefont {P.}~\bibnamefont {Caputa}}, \bibinfo {author} {\bibfnamefont {J.~M.}\ \bibnamefont {Magan}},\ and\ \bibinfo {author} {\bibfnamefont {D.}~\bibnamefont {Patramanis}},\ }\bibfield  {title} {\bibinfo {title} {Geometry of krylov complexity},\ }\href {https://doi.org/10.1103/PhysRevResearch.4.013041} {\bibfield  {journal} {\bibinfo  {journal} {Phys. Rev. Research}\ }\textbf {\bibinfo {volume} {4}},\ \bibinfo {pages} {013041} (\bibinfo {year} {2022})}\BibitemShut {NoStop}%
\bibitem [{\citenamefont {H{\"o}rnedal}\ \emph {et~al.}(2022)\citenamefont {H{\"o}rnedal}, \citenamefont {Carabba}, \citenamefont {Matsoukas-Roubeas},\ and\ \citenamefont {del Campo}}]{Hornedal22}%
  \BibitemOpen
  \bibfield  {author} {\bibinfo {author} {\bibfnamefont {N.}~\bibnamefont {H{\"o}rnedal}}, \bibinfo {author} {\bibfnamefont {N.}~\bibnamefont {Carabba}}, \bibinfo {author} {\bibfnamefont {A.~S.}\ \bibnamefont {Matsoukas-Roubeas}},\ and\ \bibinfo {author} {\bibfnamefont {A.}~\bibnamefont {del Campo}},\ }\bibfield  {title} {\bibinfo {title} {Ultimate speed limits to the growth of operator complexity},\ }\href {https://doi.org/10.1038/s42005-022-00985-1} {\bibfield  {journal} {\bibinfo  {journal} {Communications Physics}\ }\textbf {\bibinfo {volume} {5}},\ \bibinfo {pages} {207} (\bibinfo {year} {2022})}\BibitemShut {NoStop}%
\bibitem [{\citenamefont {Monthus}(2016)}]{Monthus16}%
  \BibitemOpen
  \bibfield  {author} {\bibinfo {author} {\bibfnamefont {C.}~\bibnamefont {Monthus}},\ }\bibfield  {title} {\bibinfo {title} {Flow towards diagonalization for many-body-localization models: adaptation of the toda matrix differential flow to random quantum spin chains},\ }\href {https://doi.org/10.1088/1751-8113/49/30/305002} {\bibfield  {journal} {\bibinfo  {journal} {Journal of Physics A: Mathematical and Theoretical}\ }\textbf {\bibinfo {volume} {49}},\ \bibinfo {pages} {305002} (\bibinfo {year} {2016})}\BibitemShut {NoStop}%
\bibitem [{\citenamefont {Nishimori}\ and\ \citenamefont {Ortiz}(2011)}]{NishimoriOrtiz2011}%
  \BibitemOpen
  \bibfield  {author} {\bibinfo {author} {\bibfnamefont {H.}~\bibnamefont {Nishimori}}\ and\ \bibinfo {author} {\bibfnamefont {G.}~\bibnamefont {Ortiz}},\ }\href {https://doi.org/10.1093/acprof:oso/9780199577224.001.0001} {\emph {\bibinfo {title} {Elements of Phase Transitions and Critical Phenomena}}},\ \bibinfo {series} {Oxford Graduate Texts}, Vol.\ \bibinfo {volume} {XIII}\ (\bibinfo  {publisher} {Oxford University Press},\ \bibinfo {address} {Oxford},\ \bibinfo {year} {2011})\ p.\ \bibinfo {pages} {358}\BibitemShut {NoStop}%
\bibitem [{\citenamefont {Heyl}\ \emph {et~al.}(2013)\citenamefont {Heyl}, \citenamefont {Polkovnikov},\ and\ \citenamefont {Kehrein}}]{Heyl13}%
  \BibitemOpen
  \bibfield  {author} {\bibinfo {author} {\bibfnamefont {M.}~\bibnamefont {Heyl}}, \bibinfo {author} {\bibfnamefont {A.}~\bibnamefont {Polkovnikov}},\ and\ \bibinfo {author} {\bibfnamefont {S.}~\bibnamefont {Kehrein}},\ }\bibfield  {title} {\bibinfo {title} {Dynamical quantum phase transitions in the transverse-field ising model},\ }\href {https://doi.org/10.1103/PhysRevLett.110.135704} {\bibfield  {journal} {\bibinfo  {journal} {Phys. Rev. Lett.}\ }\textbf {\bibinfo {volume} {110}},\ \bibinfo {pages} {135704} (\bibinfo {year} {2013})}\BibitemShut {NoStop}%
\bibitem [{\citenamefont {Heyl}(2018)}]{Heyl18}%
  \BibitemOpen
  \bibfield  {author} {\bibinfo {author} {\bibfnamefont {M.}~\bibnamefont {Heyl}},\ }\bibfield  {title} {\bibinfo {title} {Dynamical quantum phase transitions: a review},\ }\href {https://doi.org/10.1088/1361-6633/aaaf9a} {\bibfield  {journal} {\bibinfo  {journal} {Reports on Progress in Physics}\ }\textbf {\bibinfo {volume} {81}},\ \bibinfo {pages} {054001} (\bibinfo {year} {2018})}\BibitemShut {NoStop}%
\bibitem [{\citenamefont {del Campo}(2016)}]{delcampo16}%
  \BibitemOpen
  \bibfield  {author} {\bibinfo {author} {\bibfnamefont {A.}~\bibnamefont {del Campo}},\ }\bibfield  {title} {\bibinfo {title} {Exact quantum decay of an interacting many-particle system: the {Calogero–Sutherland} model},\ }\href {https://doi.org/10.1088/1367-2630/18/1/015014} {\bibfield  {journal} {\bibinfo  {journal} {New Journal of Physics}\ }\textbf {\bibinfo {volume} {18}},\ \bibinfo {pages} {015014} (\bibinfo {year} {2016})}\BibitemShut {NoStop}%
\bibitem [{\citenamefont {del Campo}(2021)}]{delcampo21}%
  \BibitemOpen
  \bibfield  {author} {\bibinfo {author} {\bibfnamefont {A.}~\bibnamefont {del Campo}},\ }\bibfield  {title} {\bibinfo {title} {Probing quantum speed limits with ultracold gases},\ }\href {https://doi.org/10.1103/PhysRevLett.126.180603} {\bibfield  {journal} {\bibinfo  {journal} {Phys. Rev. Lett.}\ }\textbf {\bibinfo {volume} {126}},\ \bibinfo {pages} {180603} (\bibinfo {year} {2021})}\BibitemShut {NoStop}%
\bibitem [{\citenamefont {Yang}\ and\ \citenamefont {Lee}(1952)}]{Yang52}%
  \BibitemOpen
  \bibfield  {author} {\bibinfo {author} {\bibfnamefont {C.~N.}\ \bibnamefont {Yang}}\ and\ \bibinfo {author} {\bibfnamefont {T.~D.}\ \bibnamefont {Lee}},\ }\bibfield  {title} {\bibinfo {title} {Statistical theory of equations of state and phase transitions. i. theory of condensation},\ }\href {https://doi.org/10.1103/PhysRev.87.404} {\bibfield  {journal} {\bibinfo  {journal} {Phys. Rev.}\ }\textbf {\bibinfo {volume} {87}},\ \bibinfo {pages} {404} (\bibinfo {year} {1952})}\BibitemShut {NoStop}%
\bibitem [{\citenamefont {Lee}\ and\ \citenamefont {Yang}(1952)}]{Lee52}%
  \BibitemOpen
  \bibfield  {author} {\bibinfo {author} {\bibfnamefont {T.~D.}\ \bibnamefont {Lee}}\ and\ \bibinfo {author} {\bibfnamefont {C.~N.}\ \bibnamefont {Yang}},\ }\bibfield  {title} {\bibinfo {title} {Statistical theory of equations of state and phase transitions. ii. lattice gas and ising model},\ }\href {https://doi.org/10.1103/PhysRev.87.410} {\bibfield  {journal} {\bibinfo  {journal} {Phys. Rev.}\ }\textbf {\bibinfo {volume} {87}},\ \bibinfo {pages} {410} (\bibinfo {year} {1952})}\BibitemShut {NoStop}%
\bibitem [{\citenamefont {Itzykson}\ and\ \citenamefont {Drouffe}(1991)}]{Itzykson91}%
  \BibitemOpen
  \bibfield  {author} {\bibinfo {author} {\bibfnamefont {C.}~\bibnamefont {Itzykson}}\ and\ \bibinfo {author} {\bibfnamefont {J.}~\bibnamefont {Drouffe}},\ }\href {https://books.google.lu/books?id=cKdTrJPBAyUC} {\emph {\bibinfo {title} {Statistical Field Theory}}},\ \bibinfo {series} {Cambridge monographs on mathematical physics}\ No.\ \bibinfo {number} {Bd. 1}\ (\bibinfo  {publisher} {Cambridge University Press},\ \bibinfo {year} {1991})\BibitemShut {NoStop}%
\bibitem [{\citenamefont {Obuchi}\ and\ \citenamefont {Takahashi}(2012)}]{Obuchi12}%
  \BibitemOpen
  \bibfield  {author} {\bibinfo {author} {\bibfnamefont {T.}~\bibnamefont {Obuchi}}\ and\ \bibinfo {author} {\bibfnamefont {K.}~\bibnamefont {Takahashi}},\ }\bibfield  {title} {\bibinfo {title} {Dynamical singularities of glassy systems in a quantum quench},\ }\href {https://doi.org/10.1103/PhysRevE.86.051125} {\bibfield  {journal} {\bibinfo  {journal} {Phys. Rev. E}\ }\textbf {\bibinfo {volume} {86}},\ \bibinfo {pages} {051125} (\bibinfo {year} {2012})}\BibitemShut {NoStop}%
\bibitem [{\citenamefont {Rabinovici}\ \emph {et~al.}(2021)\citenamefont {Rabinovici}, \citenamefont {S{\'a}nchez-Garrido}, \citenamefont {Shir},\ and\ \citenamefont {Sonner}}]{Rabinovici21}%
  \BibitemOpen
  \bibfield  {author} {\bibinfo {author} {\bibfnamefont {E.}~\bibnamefont {Rabinovici}}, \bibinfo {author} {\bibfnamefont {A.}~\bibnamefont {S{\'a}nchez-Garrido}}, \bibinfo {author} {\bibfnamefont {R.}~\bibnamefont {Shir}},\ and\ \bibinfo {author} {\bibfnamefont {J.}~\bibnamefont {Sonner}},\ }\bibfield  {title} {\bibinfo {title} {{Operator complexity: a journey to the edge of Krylov space}},\ }\href {https://doi.org/10.1007/JHEP06(2021)062} {\bibfield  {journal} {\bibinfo  {journal} {Journal of High Energy Physics}\ }\textbf {\bibinfo {volume} {2021}},\ \bibinfo {pages} {62} (\bibinfo {year} {2021})}\BibitemShut {NoStop}%
\bibitem [{\citenamefont {Bento}\ \emph {et~al.}(2024)\citenamefont {Bento}, \citenamefont {del Campo},\ and\ \citenamefont {C\'eleri}}]{Bento24}%
  \BibitemOpen
  \bibfield  {author} {\bibinfo {author} {\bibfnamefont {P.~H.~S.}\ \bibnamefont {Bento}}, \bibinfo {author} {\bibfnamefont {A.}~\bibnamefont {del Campo}},\ and\ \bibinfo {author} {\bibfnamefont {L.~C.}\ \bibnamefont {C\'eleri}},\ }\bibfield  {title} {\bibinfo {title} {{Krylov complexity and dynamical phase transition in the quenched Lipkin-Meshkov-Glick model}},\ }\href {https://doi.org/10.1103/PhysRevB.109.224304} {\bibfield  {journal} {\bibinfo  {journal} {Phys. Rev. B}\ }\textbf {\bibinfo {volume} {109}},\ \bibinfo {pages} {224304} (\bibinfo {year} {2024})}\BibitemShut {NoStop}%
\bibitem [{\citenamefont {Takahashi}(2025)}]{Takahashi25-2}%
  \BibitemOpen
  \bibfield  {author} {\bibinfo {author} {\bibfnamefont {K.}~\bibnamefont {Takahashi}},\ }\bibfield  {title} {\bibinfo {title} {Dynamical quantum phase transition, metastable state, and dimensionality reduction: Krylov analysis of fully connected spin models},\ }\href {https://doi.org/10.1103/m4jf-7svp} {\bibfield  {journal} {\bibinfo  {journal} {Phys. Rev. B}\ }\textbf {\bibinfo {volume} {112}},\ \bibinfo {pages} {054312} (\bibinfo {year} {2025})}\BibitemShut {NoStop}%
\bibitem [{\citenamefont {Caputa}\ \emph {et~al.}(2024)\citenamefont {Caputa}, \citenamefont {Jeong}, \citenamefont {Liu}, \citenamefont {Pedraza},\ and\ \citenamefont {Qu}}]{Caputa:2024vrn}%
  \BibitemOpen
  \bibfield  {author} {\bibinfo {author} {\bibfnamefont {P.}~\bibnamefont {Caputa}}, \bibinfo {author} {\bibfnamefont {H.-S.}\ \bibnamefont {Jeong}}, \bibinfo {author} {\bibfnamefont {S.}~\bibnamefont {Liu}}, \bibinfo {author} {\bibfnamefont {J.~F.}\ \bibnamefont {Pedraza}},\ and\ \bibinfo {author} {\bibfnamefont {L.-C.}\ \bibnamefont {Qu}},\ }\bibfield  {title} {\bibinfo {title} {{Krylov complexity of density matrix operators}},\ }\href {https://doi.org/10.1007/JHEP05(2024)337} {\bibfield  {journal} {\bibinfo  {journal} {JHEP}\ }\textbf {\bibinfo {volume} {05}},\ \bibinfo {pages} {337}},\ \Eprint {https://arxiv.org/abs/2402.09522} {arXiv:2402.09522 [hep-th]} \BibitemShut {NoStop}%
\bibitem [{\citenamefont {Nandy}\ \emph {et~al.}(2025{\natexlab{b}})\citenamefont {Nandy}, \citenamefont {Pathak}, \citenamefont {Xian},\ and\ \citenamefont {Erdmenger}}]{Nandy:2024mml}%
  \BibitemOpen
  \bibfield  {author} {\bibinfo {author} {\bibfnamefont {P.}~\bibnamefont {Nandy}}, \bibinfo {author} {\bibfnamefont {T.}~\bibnamefont {Pathak}}, \bibinfo {author} {\bibfnamefont {Z.-Y.}\ \bibnamefont {Xian}},\ and\ \bibinfo {author} {\bibfnamefont {J.}~\bibnamefont {Erdmenger}},\ }\bibfield  {title} {\bibinfo {title} {{Krylov space approach to singular value decomposition in non-Hermitian systems}},\ }\href {https://doi.org/10.1103/PhysRevB.111.064203} {\bibfield  {journal} {\bibinfo  {journal} {Phys. Rev. B}\ }\textbf {\bibinfo {volume} {111}},\ \bibinfo {pages} {064203} (\bibinfo {year} {2025}{\natexlab{b}})},\ \Eprint {https://arxiv.org/abs/2411.09309} {arXiv:2411.09309 [quant-ph]} \BibitemShut {NoStop}%
\bibitem [{\citenamefont {Haake}(1991)}]{haake1991quantum}%
  \BibitemOpen
  \bibfield  {author} {\bibinfo {author} {\bibfnamefont {F.}~\bibnamefont {Haake}},\ }\href@noop {} {\emph {\bibinfo {title} {Quantum signatures of chaos}}}\ (\bibinfo  {publisher} {Springer},\ \bibinfo {year} {1991})\BibitemShut {NoStop}%
\bibitem [{\citenamefont {Bhattacharjee}\ and\ \citenamefont {Nandy}(2025)}]{Bhattacharjee:2024yxj}%
  \BibitemOpen
  \bibfield  {author} {\bibinfo {author} {\bibfnamefont {B.}~\bibnamefont {Bhattacharjee}}\ and\ \bibinfo {author} {\bibfnamefont {P.}~\bibnamefont {Nandy}},\ }\bibfield  {title} {\bibinfo {title} {{Krylov fractality and complexity in generic random matrix ensembles}},\ }\href {https://doi.org/10.1103/PhysRevB.111.L060202} {\bibfield  {journal} {\bibinfo  {journal} {Phys. Rev. B}\ }\textbf {\bibinfo {volume} {111}},\ \bibinfo {pages} {L060202} (\bibinfo {year} {2025})},\ \Eprint {https://arxiv.org/abs/2407.07399} {arXiv:2407.07399 [quant-ph]} \BibitemShut {NoStop}%
\bibitem [{\citenamefont {Verbaarschot}\ and\ \citenamefont {Zahed}(1993)}]{Verbaarschot93}%
  \BibitemOpen
  \bibfield  {author} {\bibinfo {author} {\bibfnamefont {J.~J.~M.}\ \bibnamefont {Verbaarschot}}\ and\ \bibinfo {author} {\bibfnamefont {I.}~\bibnamefont {Zahed}},\ }\bibfield  {title} {\bibinfo {title} {Spectral density of the qcd dirac operator near zero virtuality},\ }\href {https://doi.org/10.1103/PhysRevLett.70.3852} {\bibfield  {journal} {\bibinfo  {journal} {Phys. Rev. Lett.}\ }\textbf {\bibinfo {volume} {70}},\ \bibinfo {pages} {3852} (\bibinfo {year} {1993})}\BibitemShut {NoStop}%
\bibitem [{\citenamefont {Verbaarschot}(1994)}]{Verbaarschot94}%
  \BibitemOpen
  \bibfield  {author} {\bibinfo {author} {\bibfnamefont {J.}~\bibnamefont {Verbaarschot}},\ }\bibfield  {title} {\bibinfo {title} {Spectrum of the qcd dirac operator and chiral random matrix theory},\ }\href {https://doi.org/10.1103/PhysRevLett.72.2531} {\bibfield  {journal} {\bibinfo  {journal} {Phys. Rev. Lett.}\ }\textbf {\bibinfo {volume} {72}},\ \bibinfo {pages} {2531} (\bibinfo {year} {1994})}\BibitemShut {NoStop}%
\bibitem [{\citenamefont {Altland}\ and\ \citenamefont {Zirnbauer}(1997)}]{Altland97}%
  \BibitemOpen
  \bibfield  {author} {\bibinfo {author} {\bibfnamefont {A.}~\bibnamefont {Altland}}\ and\ \bibinfo {author} {\bibfnamefont {M.~R.}\ \bibnamefont {Zirnbauer}},\ }\bibfield  {title} {\bibinfo {title} {Nonstandard symmetry classes in mesoscopic normal-superconducting hybrid structures},\ }\href {https://doi.org/10.1103/PhysRevB.55.1142} {\bibfield  {journal} {\bibinfo  {journal} {Phys. Rev. B}\ }\textbf {\bibinfo {volume} {55}},\ \bibinfo {pages} {1142} (\bibinfo {year} {1997})}\BibitemShut {NoStop}%
\bibitem [{\citenamefont {Witten}(1981)}]{Witten81}%
  \BibitemOpen
  \bibfield  {author} {\bibinfo {author} {\bibfnamefont {E.}~\bibnamefont {Witten}},\ }\bibfield  {title} {\bibinfo {title} {Dynamical breaking of supersymmetry},\ }\href {https://doi.org/https://doi.org/10.1016/0550-3213(81)90006-7} {\bibfield  {journal} {\bibinfo  {journal} {Nuclear Physics B}\ }\textbf {\bibinfo {volume} {188}},\ \bibinfo {pages} {513} (\bibinfo {year} {1981})}\BibitemShut {NoStop}%
\bibitem [{\citenamefont {Cooper}\ and\ \citenamefont {Freedman}(1983)}]{Cooper83}%
  \BibitemOpen
  \bibfield  {author} {\bibinfo {author} {\bibfnamefont {F.}~\bibnamefont {Cooper}}\ and\ \bibinfo {author} {\bibfnamefont {B.}~\bibnamefont {Freedman}},\ }\bibfield  {title} {\bibinfo {title} {Aspects of supersymmetric quantum mechanics},\ }\href {https://doi.org/https://doi.org/10.1016/0003-4916(83)90034-9} {\bibfield  {journal} {\bibinfo  {journal} {Annals of Physics}\ }\textbf {\bibinfo {volume} {146}},\ \bibinfo {pages} {262} (\bibinfo {year} {1983})}\BibitemShut {NoStop}%
\bibitem [{\citenamefont {Cooper}\ \emph {et~al.}(1995)\citenamefont {Cooper}, \citenamefont {Khare},\ and\ \citenamefont {Sukhatme}}]{Cooper95}%
  \BibitemOpen
  \bibfield  {author} {\bibinfo {author} {\bibfnamefont {F.}~\bibnamefont {Cooper}}, \bibinfo {author} {\bibfnamefont {A.}~\bibnamefont {Khare}},\ and\ \bibinfo {author} {\bibfnamefont {U.}~\bibnamefont {Sukhatme}},\ }\bibfield  {title} {\bibinfo {title} {Supersymmetry and quantum mechanics},\ }\href {https://doi.org/https://doi.org/10.1016/0370-1573(94)00080-M} {\bibfield  {journal} {\bibinfo  {journal} {Physics Reports}\ }\textbf {\bibinfo {volume} {251}},\ \bibinfo {pages} {267} (\bibinfo {year} {1995})}\BibitemShut {NoStop}%
\bibitem [{\citenamefont {{Gendenshte{\"\i}n}}(1983)}]{Gendenshtein83}%
  \BibitemOpen
  \bibfield  {author} {\bibinfo {author} {\bibfnamefont {L.~{\'E}.}\ \bibnamefont {{Gendenshte{\"\i}n}}},\ }\bibfield  {title} {\bibinfo {title} {{Derivation of exact spectra of the Schr{\"o}dinger equation by means of supersymmetry}},\ }\href@noop {} {\bibfield  {journal} {\bibinfo  {journal} {Soviet Journal of Experimental and Theoretical Physics Letters}\ }\textbf {\bibinfo {volume} {38}},\ \bibinfo {pages} {356} (\bibinfo {year} {1983})}\BibitemShut {NoStop}%
\end{thebibliography}%
\end{document}